\documentclass[reprint,prx,superscriptaddress,longbibliography,amsmath,amssymb,aps] {revtex4-1}
\usepackage{graphicx}
\usepackage{dcolumn}
\usepackage{bm}

\usepackage{afterpage}
\usepackage{float}
\usepackage{placeins}
\usepackage{dsfont}
\usepackage[utf8]{inputenc}
\usepackage{bm}
\usepackage{amssymb}
\usepackage{amsmath}
\usepackage{amsfonts}
\usepackage{graphicx}
\usepackage[usenames,dvipsnames]{xcolor} 
\usepackage{dcolumn}
\usepackage{bm}
\usepackage[pdftex,
            pdftitle={1904.04266},
            pdfauthor={Pablo Sala},
            bookmarks,
            colorlinks,
            linkcolor=black,
            citecolor=magenta,
            menucolor=black,
            urlcolor=blue,
            plainpages=false,
            pdfpagelabels,
            hypertexnames=false]{hyperref}

\newcommand{\avg}[1]{\left< #1 \right>}

\newcommand{\ket}[1]{\left| #1 \right>}
\newcommand{\bra}[1]{\left< #1 \right|}
\newcommand{\braket}[1]{ \langle{{#1}}\rangle}

\let\baraccent=\= \renewcommand{\=}[1]{\stackrel{#1}{=}}

\usepackage[normalem]{ulem}

\begin{document}
\title{Ergodicity-breaking arising from Hilbert space fragmentation \\ in dipole-conserving Hamiltonians}

\author{Pablo Sala}
\email{pablo.sala@tum.de}
\affiliation{Department of Physics, Technical University of Munich, 85748 Garching, Germany}
\affiliation{Munich Center for Quantum Science and Technology (MCQST), Schellingstr. 4, D-80799 M\"unchen, Germany }

\author{Tibor Rakovszky }
\affiliation{Department of Physics, Technical University of Munich, 85748 Garching, Germany}
\affiliation{Munich Center for Quantum Science and Technology (MCQST), Schellingstr. 4, D-80799 M\"unchen, Germany }

\author{Ruben Verresen}
\affiliation{Department of Physics, Technical University of Munich, 85748 Garching, Germany}
\affiliation{Max-Planck-Institute for the Physics of Complex Systems, 01187 Dresden, Germany}

\author{Michael Knap}
\affiliation{Department of Physics, Technical University of Munich, 85748 Garching, Germany}
\affiliation{Munich Center for Quantum Science and Technology (MCQST), Schellingstr. 4, D-80799 M\"unchen, Germany }
\affiliation{Institute for Advanced Study, Technical University of Munich, 85748 Garching, Germany}

\author{Frank Pollmann}
\affiliation{Department of Physics, Technical University of Munich, 85748 Garching, Germany}
\affiliation{Munich Center for Quantum Science and Technology (MCQST), Schellingstr. 4, D-80799 M\"unchen, Germany }

\date{\today }

\begin{abstract}
We show that the combination of charge and dipole conservation---characteristic of fracton systems---leads to an extensive fragmentation of the Hilbert space, which in turn can lead to a breakdown of thermalization.
As a concrete example, we investigate the out-of-equilibrium dynamics of one-dimensional spin-1 models that conserve charge (total $S^z$) and its associated dipole moment.
First, we consider a minimal model including only three-site terms and find that the infinite temperature auto-correlation saturates to a finite value---showcasing non-thermal behavior.  
The absence of thermalization is identified as a consequence of the \emph{strong fragmentation} of the Hilbert space into exponentially many invariant subspaces in the local $S^z$ basis, arising from the interplay of dipole conservation and local interactions. 
Second, we extend the model by including four-site terms and find that this perturbation leads to a \emph{weak fragmentation}: the system still has exponentially many invariant subspaces, but they are no longer sufficient to avoid thermalization for typical initial states. 
More generally, for any finite range of interactions, the system still exhibits non-thermal eigenstates appearing throughout the entire spectrum.
We compare our results to charge and dipole moment conserving random unitary circuit models for which we reach identical conclusions. 

\end{abstract}

\maketitle

\section{Introduction}
Recent years have seen a great deal of effort---both theoretical and experimental---to understand \emph{quantum thermalization}: the question of how closed quantum systems, evolving under unitary dynamics, reach a state of thermal equilibrium~\cite{Gring1318,Hild2014,Brown540,Kaufman794,Yijun2018,Brydges2018,ETHreviewRigol16,GogolinReview,Meinert945}. Thermalization is believed to be characterized in terms of the Eigenstate Thermalization Hypothesis (ETH)~\cite{Deutsch91,Srednicki94,Rigol2008,ETHreviewRigol16}. According to this, each eigenstate of a thermalizing Hamiltonian essentially behaves like a thermal ensemble as far as expectation values of local observables are concerned. While no proof of ETH exists, there are many cases where it has been shown numerically that indeed all eigenstates satisfy this hypothesis~\cite{Rigol2008,Kim2014,ETHreviewRigol16}.

Given its supposed generality, there has been much interest in systems that violate ETH. Two well-known instances are integrable systems~\cite{Rigol2007,Kinoshita2006} and the many-body localized (MBL) phase~\cite{Basko06,Nandkishore14,AltmanVosk,Schreiber2015}, both of which avoid ETH due to the existence of extensively many conserved quantities~\cite{Essler16,Huse14,Serbyn13cons}. These conservation laws lead to non-ergodicity even at high energy densities. One important question concerns whether behavior similar to MBL can appear in systems without spatial disorder~\cite{Muller2015,Yao2016,PAPIC2015,Smith01,Smith02,Brenes18,Refael18,Schulz19}. 

\begin{figure}[t!]
	\centering
	\includegraphics[width=0.95\linewidth]{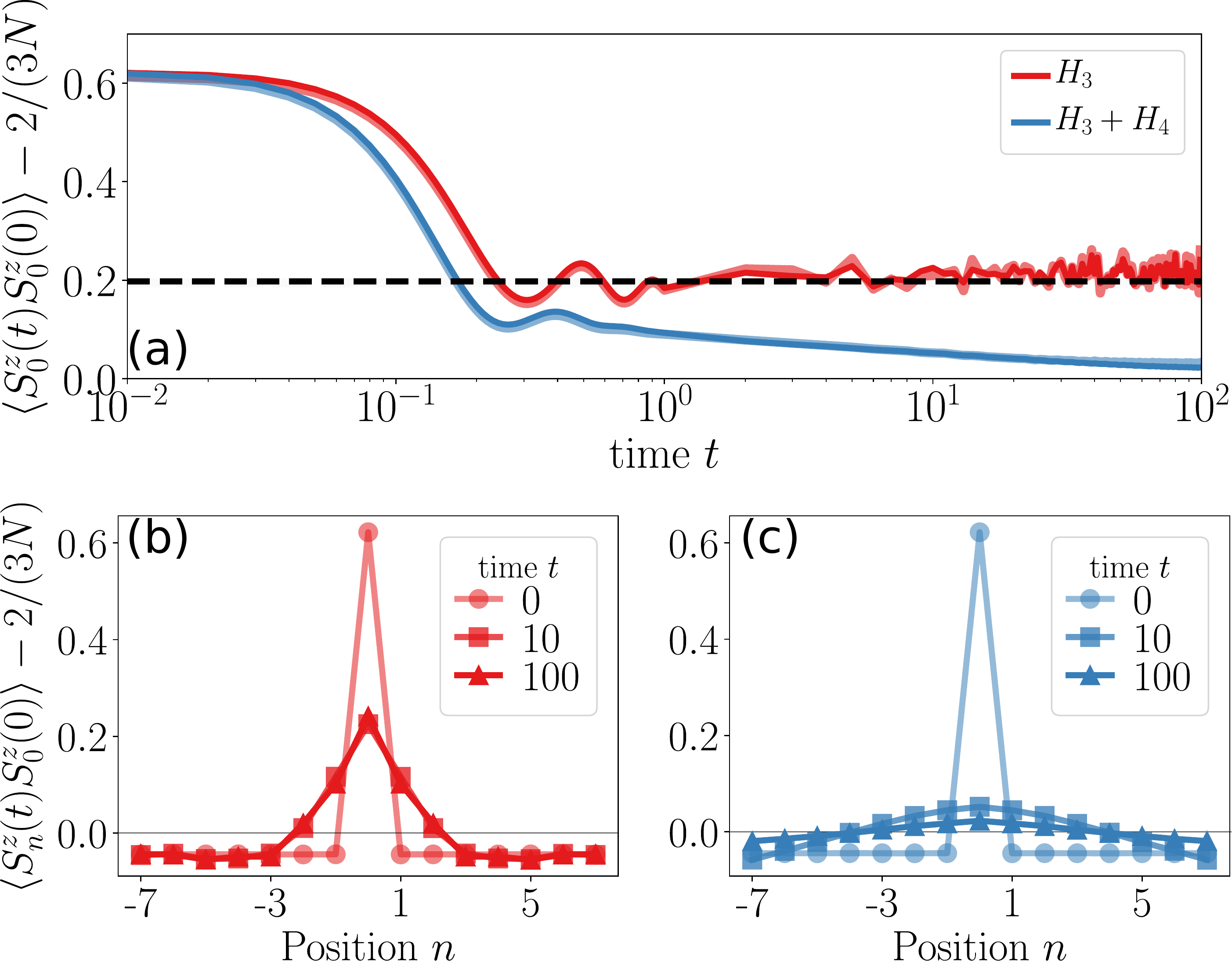}
	\caption{\textbf{Thermalization and its absence in the auto-correlation function.}  Panel (a) shows the auto-correlation function $C_0^z(t)\equiv\avg{S_0^z(t)S_0^z(0)}$ in the full Hilbert space at infinite temperature for $N=13$ (transparent curves) and $N=15$ (opaque curves) spins. For Hamiltonian $H_3$ in Eq.~\eqref{H_3}, $C_0^z(t)$ saturates to a finite value at long times, closely matching the lower bound in Eq.~\eqref{AnalPred} (dashed line). The auto-correlation function of the combined Hamiltonian $H_3+H_4$ decays to zero at long times. Panels (b) and (c) show the spatially resolved correlator $\avg{S_n^z(t)S_0^z(0)}$ for $H_3$ and $H_3+H_4$ respectively.}
	\label{fig:fig1}
\end{figure}

Another key question is about the possibility of systems that exhibit interesting intermediate behavior, neither localized, nor fully ergodic. In particular we can distinguish between \emph{strong} and \emph{weak} ETH: the former says that all eigenstates in the bulk of the spectrum become thermal in the thermodynamic limit, while the latter allows for the presence of outlying non-thermal states, as long as their ratio is vanishingly small at any given energy~\cite{Biroli2010,MoriWeakETH,ShiraishiMori}. It is important to stress that if only weak ETH is satisfied, then we can always find initial conditions which have narrow energy distributions but nevertheless fail to thermalize~\cite{GogolinReview}. Indeed, generic systems are expected to exhibit the strong version of ETH~\cite{Kim2014}. Recently, however, several exceptions have been discovered~\cite{Moudgalya01,Moudgalya02,Iadecola2018,Iadecola2019,Neupert2019}. Systems with constrained dynamics~\cite{Garrahan2015,Garrahan2018} are an especially promising avenue where non-thermal eigenstates, dubbed scar states, can occur~\cite{ShiraishiMori,TurnerNatPhys,TurnerPRB,Choi2018,Motrunich2018,Feldmeier19}. These are believed to be responsible for persistent oscillations observed in a recent Rydberg atom experiment~\cite{Bernien2017}.

Constrained dynamics occurs naturally in so-called \emph{fracton systems}, which are characterized by the existence of excitations that exhibit restricted mobility~\cite{Haah11,Vijay15}. On one hand, these systems have been studied in three-dimensional exactly-solvable lattice models with discrete symmetries, where fractons are created on the corners of a membrane or fractal operator \cite{Chamon05,Vijay15,Vijay16,Hsieh17,Gabor17,Kevin18}. On the other hand, different approaches for fractons with U(1) symmetry have shown that their mobility constraints are related to the conservation of the dipole moment which localizes isolated charged excitations. \cite{PretkoSub,PretkoPrin,Pretko17b,Williamson18,Bulmash18b, DimertoQLM, Gromov18} An analytical connection between the two approaches has been discussed in Refs.~\onlinecite{Bulmash18, Han18}. The exotic behavior of fractons also gave rise to the study of their non-equilibrium physics \cite{Abhinav17,Pretko17}, and it has been argued that fracton models with discrete symmetries show glassy dynamics~\cite{Chamon05,Castelnovo12, Abhinav17}.

In this paper, we study the consequences of dipole conservation associated with a global U(1) charge (i.e., the conservation of total spin $S^z$) in one-dimensional (1D) spin systems, for which a numerical study is feasible. Apart from fracton systems, such charge and dipole conserving Hamiltonians also occur naturally in other contexts, for example in the quantum Hall effect~\cite{PhysRevB.50.17199,Bergholtz08,BERGHOLTZ2011755,PhysRevB.86.155104,PhysRevLett.109.016401} and in systems of charged particles exposed to a strong electric field~\cite{Schulz19,Refael18}. Interestingly, a recent work~\cite{Pai18} has argued that random local unitary dynamics with such symmetries fails to thermalize. We find the same non-ergodic behavior in a minimal Hamiltonian that contains only three-site interactions. We discover that the source of non-ergodicity is an extensive fragmentation of the Hilbert space into exponentially many disconnected sectors in the local $z$-basis. In particular, based on the Hilbert space structure, we obtain a lower bound for the long-time auto-correlation, which remains finite in the thermodynamic limit. This is a novel type of non-ergodic behavior, arising in a translation invariant system, but nevertheless sharing certain features of MBL, which we denote by \emph{strong fragmentation} of the Hilbert space. 

However, we find that this strongly non-ergodic behavior disappears once we add longer-range interactions, such as a four-site term. In this case, the dipole constraint is no longer sufficient to violate ergodicity, and the infinite temperature autocorrelator decays to zero. Nevertheless, the model still violates the strong version of ETH and exhibits exponentially many non-thermal eigenstates, disconnected from the bulk of the spectrum, and co-existing with thermal eigenstates at the same energies. We term this behavior, which is reminiscent to quantum many-body scars, \emph{weak fragmentation} and give an analytical lower bound on the number of product eigenstates for arbitrary finite range of dipole-conserving interactions. We compare our results to random unitary circuit dynamics, and find the same behavior: while circuits constructed from three-site gates fail to thermalize, adding four-site gates is sufficient to delocalize the system and lead to thermalization for typical initial states. We numerically verify that the invariant subspaces for Hamiltonian and random circuit dynamics coincide exactly. 

The remainder of the paper is organized as follows. In Sec.~\ref{sec:model} we introduce the Hamiltonians we study, and describe their relevant symmetries. In Sec.~\ref{sec:H3} we investigate the minimal model containing only three-site interactions and show that it fails to thermalize. We prove that the Hilbert space fragments into exponentially many invariant subspaces, some of which we construct analytically, and connect these to the finite saturation value of the auto-correlation function. In Sec.~\ref{sec:H3H4} we extend the model by adding four-site interactions and argue that while these are sufficient to make the majority of eigenstates thermal---leading to ergodic behavior for typical initial states---the system still violates strong ETH. In Sec.~\ref{sec:RUC} we compare our results to random unitary circuit dynamics and find similar behavior. We conclude in Sec.~\ref{sec:summary} with a summary and outlook. The appendices provide further comparisons of our numerical results on auto-correlations for different system sizes, as well as other dynamical quantities, such as entanglement growth and operator spreading. App.~\ref{app:PXP} shows an explicit construction, relating the minimal Hamiltonian we consider to the PXP model~\cite{Lesanovsky12,TurnerNatPhys} that appears in the context of quantum many-body scars~\cite{Zlatko}.

\section{Model and symmetries}\label{sec:model}
We consider  two spin-$1$ Hamiltonians on a chain of length $N$ of the form  
\begin{align} 
\label{H_3}  H_3= &-\sum_{n}{\Big[ S_n^+\big(S_{n+1}^-\big)^2S_{n+2}^++\text{H.c.}\Big]}
\end{align}
and
\begin{align} 
\label{H_4}  H_4= &-\sum_{n}{\Big[ S_n^+S_{n+1}^-S_{n+2}^-S_{n+3}^++\text{H.c.}\Big]},
\end{align}
acting on three and four consecutive sites, respectively.  

Apart from being translation and inversion symmetric, both Hamiltonians share two additional global symmetries: they conserve a U(1) charge $Q$ and its associated dipole moment $P_{n_0}$:
\begin{align} 
\label{QP} Q\equiv\sum_{n}{S_n^z} \hspace{10pt} \textrm{and} & & P_{n_0}\equiv\sum_{n}{(n-n_0)S_n^z}, 
\end{align}
with respective eigenvalues $q$ and $p$ defining the symmetry sector $\mathcal{H}_{q,p}$~\cite{GralPcons}. 
Since $[Q,P_{n_0}]=0$, the local $S^z$-basis, denoted by $|+\rangle,|0\rangle,|-\rangle$, is a common eigenbasis of $Q$ and $P_{n_0}$. 
The definition of the dipole symmetry $P_{n_0}$ also depends on the reference position $n_0$, except when $Q=0$. 
Unless specified otherwise, we choose open boundary conditions~\cite{PBC_Note} and take $N=2m+1$ odd, labeling sites $n = -m,\dots,0,\dots,m$. 
We choose the reference site $n_0$ to be the center site, $n_0 = 0$, and denote $P\equiv P_{n_0=0}$.
The operator $P$ does not commute with spatial translations and changes sign under inversion; thus, it is not an internal symmetry~\cite{Gromov18}. 
Dipole conservation is the relevant global symmetry appearing in the description of fracton phases of matter with U(1) symmetry group~\cite{PretkoSub,PretkoPrin,Pretko17b,Williamson18,Bulmash18b, DimertoQLM, Gromov18}. Motivated by this, we use the following notations: we call the states $\ket{\pm}$ on a given site a \emph{fracton} with charge $q=\pm 1$, and a two-site configuration $\ket{+-}$ ($\ket{-+})$ a \emph{dipole} with zero charge and dipole moment $p=-1$ ($+1$). Notice that the dipole moment of a $(\pm)$-fracton on a site $n$ is $p=\pm n$. Thus, in order to conserve the total dipole moment, a fracton can only move by emitting dipoles~\cite{PretkoSub,Pai18}. 

There also exists an operator that anti-commutes with $H_3$, but commutes with $Q$ and $P$ (see App.~\ref{app:Symmetries} for details). Consequently, the spectrum of $H_3$ is symmetric around zero in every $(q,p)$-sector. The same is also true when $H_4$ is considered separately, but not for the combined Hamiltonian $H_3 + H_4$. These anti-commuting symmetries can also be broken by adding terms diagonal in the $S^z$ basis, which would not change any of the physics observed in the following.

We note in passing that similar charge and dipole conserving Hamiltonians can be written for any spin representation, in any spatial dimension, as well as for fermionic systems. 
For the latter, the dipole symmetry becomes the center of mass of the particle number operator and the corresponding Hamiltonian consists of a symmetric redistribution of charges with respect to the center sites. 
A similar fermionic Hamiltonian appears in the study of fractional quantum Hall on a torus in the Tao-Thouless limit \cite{PhysRevB.50.17199,Bergholtz08,BERGHOLTZ2011755,PhysRevB.86.155104,PhysRevLett.109.016401}.
In addition, such dipole-conserving chains can arise naturally in the presence of strong electric fields, as we discuss in the outlook.

\section{Hamiltonian $H_3$}\label{sec:H3}
We start by investigating the three-site Hamiltonian $H_3$ in Eq.~\eqref{H_3}, as a minimal model that conserves both the total charge $Q$ and the dipole moment $P$. 
We detail its unusual non-ergodic dynamics and identify it as a consequence of extensive fragmentation of the Hilbert space into invariant subspaces.
In Sec.~\ref{sec:H3H4} we will add longer-range terms to this minimal model and describe their effect on the dynamics.

\subsection{Lack of thermalization}
We first investigate the behavior of the auto-correlation function $C^z_{0}(t)\equiv \avg{S_{0}^z(t)S_{0}^z(0)}$ at infinite temperature. 
Relying on quantum typicality~\cite{Reimann07,QuantumThermo,Steinigeweg15}, we compute $C^z_0(t)$ for a  random state on the full Hilbert space.
For thermalizing and translational invariant spin-$1$ systems, $C^z_{0}(t)$ is expected to decay to $2/(3N)$ for a chain of length $N$, up to potential boundary contributions~\cite{Czth_Note}.
In Figure~\ref{fig:fig1}(a) we show $C^z_{0}(t)$, obtained via an iterative Krylov space based algorithm~\cite{Saad92}, for system sizes $N=13,15$. 
Instead of relaxing to the thermal expectation value, the auto-correlation saturates to a finite value $C^z_{0}(t)-2/(3N)\sim 0.2$ at long times. 
In App.~\ref{app:scaling} we confirm that this finite saturation value persists up to long times, $t \sim 10^{10}$, with no sign of decay. 
Moreover, the long-time values appear to be largely independent of $N$, indicating truly localized behavior that persists even in the thermodynamic limit.
Figure~\ref{fig:fig1}(b) shows the spatially resolved correlation function $ \avg{S_n^z(t)S_{0}^z(0)}$, which exhibits a peak in the center site at all times. 
We conclude that the system exhibits non-ergodic behavior. 
This is also supported by calculating the growth of entanglement starting from a random product state, which saturates to a sub-thermal von Neumann entropy density, as we show in App.~\ref{app:EE} 

\subsection{Hilbert space fragmentation}\label{sec:strong_fragment}
In this section, we demonstrate that the constrained dynamics of $H_3$ leads to a \emph{fragmentation} of the many-body Hilbert space: most $(q,p)$ symmetry sectors split into many smaller invariant subspaces in the local $S^z$-basis, such that the total number of such subspaces grows exponentially with system size.  
These disconnected sectors come in a variety of different sizes; they include `frozen' states (product eigenstates of $H_3$) and finite dimensional subspaces, where the chain splits into spatially disconnected regions.

\subsubsection{Frozen states}\label{sec:frozen}

We begin by constructing a family of exponentially many exact eigenstates of the Hamiltonian, which are all product states in the local $z$-basis. We will refer to these as \emph{frozen states}.
The simplest example is the \textit{vacuum} state $\ket{\bm{0}}\equiv\ket{\cdots 0000\cdots}$, which is annihilated by all terms in $H_3$, due to $(S^\pm)^2\ket{0} = 0$.
We can easily construct other frozen states by adding blocks of at least two contiguous charges of equal sign on top of the vacuum, e.g., $\ket{0\cdots 0++0\cdots 0---0\cdots}$. 
These are  annihilated by all terms, since $S_n^+S_{n+1}^-\ket{\pm\pm} = 0$. 
We conclude that any configuration where charges always occur in blocks of at least two consecutive sites are zero energy (mid-spectrum) eigenstates of $H_3$.
It is clear from the construction that their number is exponentially large in system size. 

\begin{figure}
	\centering
	\includegraphics[width=1.0\linewidth]{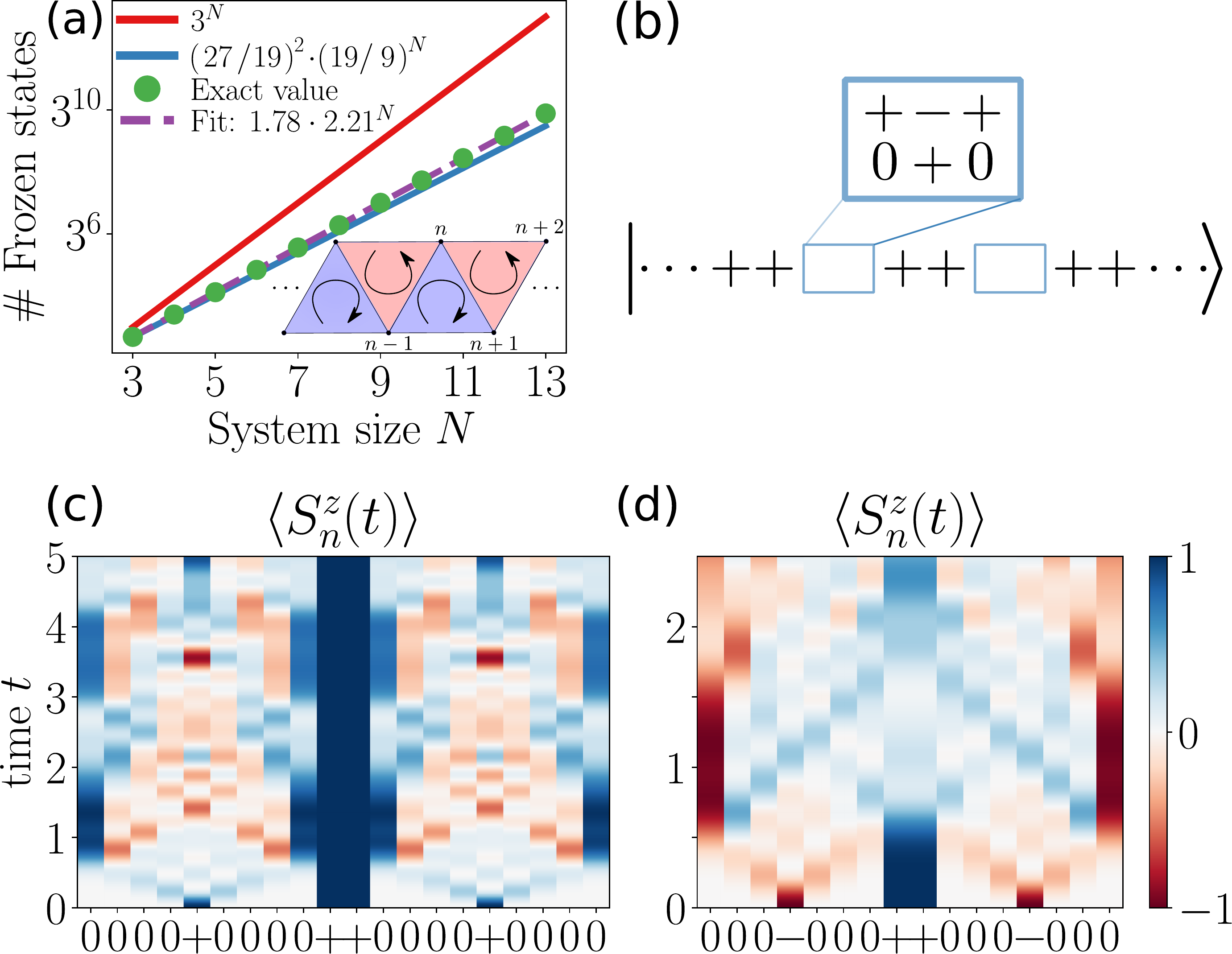}
	
	\caption{\textbf{Fragmentation of the Hilbert space into smaller subspaces.} (a) Exponential scaling of frozen states, which correspond to invariant subspaces of dimension $D=1$, and comparison to the Pauling estimate, (b) example of higher dimensional sectors, in the form of spatially separated 2-level `bubbles'. (c-d) Time evolved charge density $\avg{S_n^z(t)}$ for the two initial states indicated under each figure. (c) The $++$ block in the middle cuts the system in half when $(+)$-fractons are placed on each side. (d) When they are replaced by $(-)$-fractons, the block melts and the two halves become connected.  }
	\label{fig:fig2}
\end{figure}

We can follow Pauling~\cite{Pauling, Dupuis59} to estimate the total number of frozen states. 
To do this, we map the spin chain of length $N$ to a triangular ladder with spins placed on the vertices as shown in the inset of Fig.~\ref{fig:fig2}(a).
Treating the constraints on each triangle as independent, and using that there are $N-2$ triangles and 19 frozen states per triangle, we estimate their total number to be $3^N \times \big( 19/27)^{N-2}\approx 2.02 \times 2.11^N$. 
In Fig.~\ref{fig:fig2}(a), we numerically verify that the estimate is quite close to the actual number of frozen states, as obtained by exact diagonalization.
The numerical results together with an explicit computation for small system sizes, suggest that this estimate gives a lower bound of the actual number of frozen states.
However, we do not have a proof of this general statement.
Nevertheless, in Sec.~\ref{sec:recursion}  we provide an analytical lower bound for arbitrary finite-range dipole-conserving Hamiltonians.

\subsubsection{Larger dimensional sectors}\label{sec:bubbles}

Above we saw that blocks of two or more consecutive charges of equal sign are annihilated by the local terms in $H_3$ that act on them.
Let us now consider the empty region ($\ket{00\cdots0}$) between two such frozen blocks and fill it with a random configuration of charges.
These charges can now move around and potentially destroy the blocks on the two sides. 
However, we argue that there are initial configurations where this \emph{cannot} happen: when the sign of the rightmost charge within the region matches the charge of the frozen block to its right, then this block remains inert at all times. 
The same holds for the frozen block on the left when its charge is of the same sign as the leftmost charge within the region. 
When the charges match on both sides, then both blocks are stable and the charges in the middle bounce back-and-forth between them, disconnected from the rest of the chain. 
This appears as a direct consequence of the general rule: \emph{For a region surrounded by empty sites, the signs of the left- and rightmost charges are invariant under the dynamics generated by $H_3$}. 

The simplest example where we can observe this behavior, is as a 2-level system shown in Fig.~\ref{fig:fig2}(b), defined by the states $\ket{++{0+0}++}$ and $\ket{++{+-+}++}$.
We can check that these two states can only evolve to each other under $H_3$, defining a small invariant subspace.
More generally we can consider states of the form $\ket{++0\cdots 0{+}0 \cdots 0++}$: an isolated fracton surrounded by two `walls’ of positive charge.
Acting on this state with $H_3$, maps the configuration $00{+}00$ in the middle to $0{+-+}0$, showing that the $(+)$-fracton can move by emitting a dipole $+-$ (or $-+$) in the opposite direction~\cite{PretkoSub,Pai18}.
The fracton can then move forward by emitting further dipoles, until it reaches one of the walls.
However, when it eventually hits the wall, we end up with the configuration $+++$, which is annihilated by $H_3$; the wall therefore remains intact and the fracton bounces back harmlessly. 
Consequently, if the fractons on both sides of a $++$ block have positive charge, the chain is cut into two disconnected halves, as shown in Fig.~\ref{fig:fig2}(c). 
To destroy the wall, we would need to flip the charge of the isolated fracton to get a $(-)$-fracton: the resulting $-++$ configuration can then peel off a freely-moving $-+$ dipole, eventually melting the walls that surround it as shown in Fig.~\ref{fig:fig2}(d).

A similar situation occurs for the initial configuration $\ket{--0\cdots 0{-+}0 \cdots 0++}$.
In this case the walls on the two sides have opposite charges and a single dipole is placed between them. 
For a single dipole surrounded by empty sites, the Hamiltonian $H_3$ acts as a free hopping term, moving the dipole from site $n$ to $n\pm1$~\cite{Diffusion_Note}.
Eventually it reaches one of the surrounding walls, but since the charges in the dipole are aligned with those of the walls, it always bounces back, effectively defining a single particle hopping problem on a finite region. 
If, on the other hand, the initial dipole in the middle was of the form $+-$ it could again peel off charges from the two walls, eventually melting them.

The previously stated general rule, together with the fact that blocks with a given charge are frozen, allow us to construct more general spatially disconnected regions in the chain: take an arbitrary configuration in some finite interval and surround it with walls that have the same charge as the one closest to them on the inside.
One can then cover the entire chain with such regions, each of which has its own conserved charge and dipole moment, giving rise to many invariant subspaces within each global $(q,p)$ symmetry sector.
The resulting eigenstates clearly break translation invariance and have small amounts of entanglement, limited by the size of the largest connected spatial region.

These constructions highlight the intertwined relation between dipole conservation, spatial translations, and locality.
After the dipole quantum number is fixed, translation/inversion symmetry is generically broken, which allows us to derive conservation laws within different spatially disconnected regions.
Our construction also shows that in order to determine which invariant subspace a given initial configuration belongs to, one has to consider it on the entire chain: even if a certain region looks initially frozen, it can eventually be melted by additional charges coming from the outside.
This indicates that it might not be possible to systematically label all invariant subspaces in terms of quantum numbers of \emph{local} conserved quantities.

\begin{figure}
	\centering
	\includegraphics[width=1.0\linewidth]{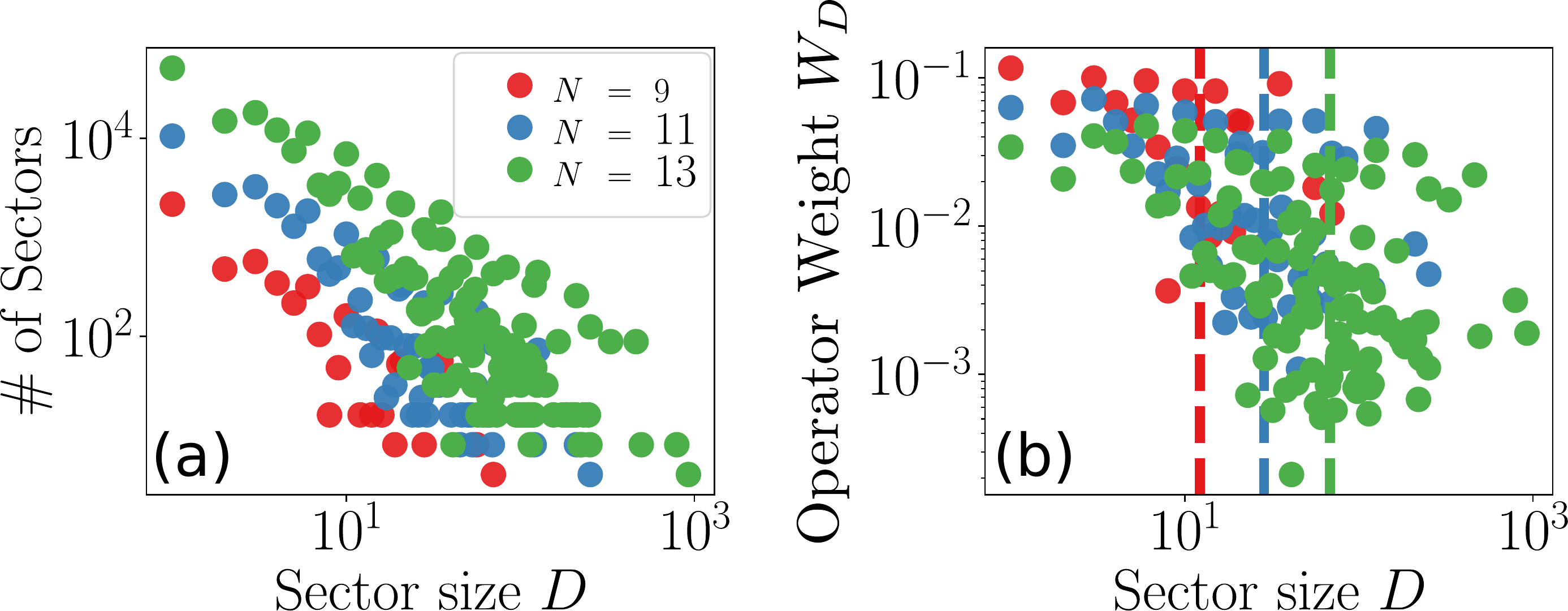}
	\caption{\textbf{Sector size and weight distributions.} (a) Distribution of invariant subspaces of size $D$ and (b) operator weight $W_D$ (see text for its definition) of the operator $S_0^z$ in each invariant subspace $\mathcal{H}_i$ of dimension $D$ in the full Hilbert space. The vertical dashed lines indicate the averaged sector size of the distribution, which is exponentially smaller than the largest sector.}
	\label{fig:fig3}
\end{figure}

\subsubsection{Distribution of dimensions of invariant subspaces}

Above we explicitly constructed invariant subspaces of $H_3$ of various dimensions within given $(q,p)$ symmetry sectors.
The distribution of these invariant subspaces can be studied by numerically identifying the connected components of the Hamiltonian written in the $S^z$ basis. 
The resulting distribution is plotted in Fig.~\ref{fig:fig3}(a), showing exponentially many sectors with a broad distribution.
We point out that since the sectors are obtained in the local $z$-basis, they remain invariant under any perturbation that is diagonal in this basis. 
However, such diagonal perturbations would have the effect of changing the energy of the different frozen states, moving them away from zero energy, and distributing them throughout the entire spectrum.

Based on the constructions in the previous section, we infer that the existence of these invariant subspaces is a consequence of the interplay between the conservation of dipole moment (which fails to commute with translation and inversion) and the locality of interactions.
In particular, in Sec.~\ref{sec:recursion} we prove that exponentially many invariant subspaces exist for any extension of the model that only involves dipole-conserving interactions with finite range.

We close this section by noting that, apart from the overall fragmented structure of the Hilbert space, which is our main concern in this paper, there is also the possibility of interesting dynamics \emph{within} certain connected components. 
For example, as we show in App.~\ref{app:PXP}, there are particular subspaces where the Hamiltonian $H_3$ maps exactly to the so-called PXP Hamiltonian~\cite{Zlatko}, studied in the context of quantum many-body scars~\cite{TurnerNatPhys,TurnerPRB}. 
A similar mapping has been uncovered in a spin-1/2 version of this model in a recent preprint~\cite{Sanjay19}.

\subsection{Saturation value of $C_0^z(t)$} \label{sec:Pred}
Equipped with the knowledge of the fragmented Hilbert space structure, we are now able to explain the long-time value of the auto-correlation function observed in Fig.~\ref{fig:fig1}(a).
To this end, let us define $\mathcal{P}_i$ as the projection onto the connected subspaces $\mathcal{H}_i$.
These projectors form an orthogonal set of conserved quantities ($\mathcal{P}_i \mathcal{P}_j = \delta_{ij}\mathcal{P}_i$), such that one can use Mazur's inequality~\cite{Mazur69,Suzuki71,Caux10} to lower bound the infinite time average of the charge auto-correlator as
\begin{align} \label{AnalPred} \lim_{T\to\infty}\frac{1}{T}\int_0^Tdt\,\avg{S_0^z( t)S_0^z(0)}&\geq \sum_{i}{\frac{\left[\mathrm{tr}\big(Z_i\big)\right]^2}{3^N \, D_i}} \equiv C_0^z(\infty),
\end{align}
where $Z_i \equiv \mathcal{P}_i S_0^z \mathcal{P}_i = \mathcal{P}_i S_0^z$ is the projection of $S_0^z$ onto $\mathcal{H}_i$, and $D_i = \text{tr}\left(\mathcal{P}_i\right)$ is the dimension of the subspace.
\begin{figure}
	\centering
	\includegraphics[width=0.8\linewidth]{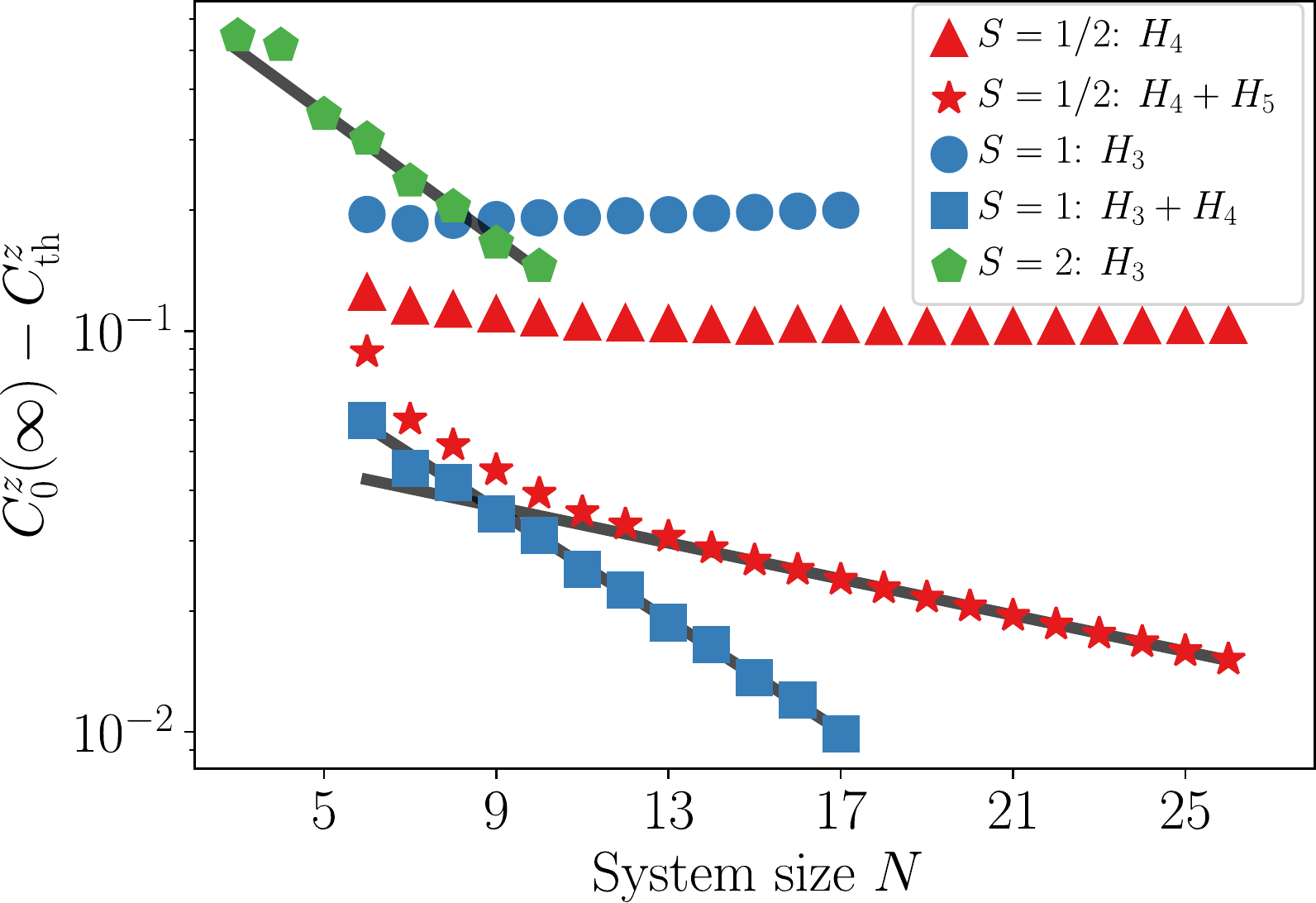}
	\caption{\textbf{Saturation value of the autocorrelator.} Finite-size study of the lower bound in Eq.~\eqref{AnalPred} for the time-averaged correlation function $C_0^z(t)$ as a function of system size. We have substracted the thermal value $C_{\textrm{th}}^z\equiv \frac{1}{N} \textrm{tr}\big[\big( S_0^z\big)^2\big]/((2S+1)^N)$ for a general spin $S$ and a chain of length $N$. For the minimal spin-1 model $H_3$ (blue dots) the lower bound is slightly increasing with system size. On the other hand, it decays to zero exponentially for the combined Hamiltonian $H_3+H_4$ (blue squares). For comparison, we also show results for other local spin $S$: the larger the on-site Hilbert space dimension, the easier it is for the system to thermalize~\cite{H5}.}
	\label{fig:fig4}
\end{figure}
The bound $ C_0^z(\infty)$ is shown in Figure~\ref{fig:fig1}(a) for $N=15$ by the dashed horizontal line; we observe that it is close to being tight, indicating that the main cause of the lack of ergodicity is indeed the fragmentation of the Hilbert space. 
We computed the estimated value for $C_0^z(\infty)-2/(3N)$ for a variety of different system sizes, and found that the result appears to remain finite in the thermodynamic limit, even increasing slightly with $N$ for the system sizes available in our numerics (blue dots in Fig.~\ref{fig:fig4}).

Since the $\mathcal{H}_i$'s are invariant and disjoint subspaces, the weight of the operator $S_0^z$ within a given sector, $\mathrm{tr}\big(Z_i^2\big)$, remains constant under time evolution. 
Therefore, we introduce the operator weight $W_D \equiv \sum_{D_i=D}\mathrm{tr}\big(Z_i^2\big)/\mathrm{tr}\big[\big(S_0^z\big)^2\big]$ as a function of the sector size $D$ for all invariant subspaces $\mathcal{H}_i$. This defines a probability distribution, shown in Fig.~\ref{fig:fig3}(b).  
We find a wide distribution with significant weight on small sectors.
While the number of frozen states scales as $\sim 2.2^N$, the size of the largest sector in the entire Hilbert space scales as $\sim 1.9^N$, both much smaller than the total dimension $3^N$.
This suggests that sectors of all sizes have significant contributions to the evolution of $S_0^z(t)$, even in the thermodynamic limit.
We also confirm the same behavior when considering only the largest symmetry sector $(q,p) = (0,0)$ (see Appendices \ref{app:Symmetries} and \ref{app:scaling}); this emphasizes the relevance of the fragmentation within each $(q,p)$-sector.

\section{Combined Hamiltonian $H_3 + H_4$}\label{sec:H3H4}

So far we have only considered the `minimal model', defined by the Hamiltonian $H_3$ in Eq.~\eqref{H_3}. 
We will now investigate to which extent the features found above are robust against local perturbations that preserve the symmetries $Q$ and $P$. 

\subsection{Thermalization for $H_3 + H_4$}

In the following, we add the four-site terms defined in Eq.~\eqref{H_4} and consider the combined Hamiltonian $H_3 + H_4$.
We find that, while this Hamiltonian shares certain features with $H_3$---in particular, it has exponentially many invariant subspaces---it nevertheless thermalizes at infinite temperature.
Indeed, the auto-correlation function $C_0^z(t)$ for the Hamiltonian $H_3+H_4$ decays to zero at long times, in contrast to the dynamics governed by $H_3$ alone; see Fig.~\ref{fig:fig1} for a comparison. 
This is accompanied by the spatially resolved correlation function, $\avg{S_n^z(t)S_0^z(0)}$, becoming approximately homogeneous at long times, as shown in Fig.~\ref{fig:fig1}(c). 
The remaining small peak is due to finite size effects, as we show in App.~\ref{app:scaling}.
Moreover, as we discuss in App.~\ref{app:EE}, for a random product state evolving under $H_3+H_4$, the entanglement entropy approaches its thermal value at long times, providing an additional indication that the system thermalizes.

\begin{figure}
	\centering
	\includegraphics[width=1.\linewidth]{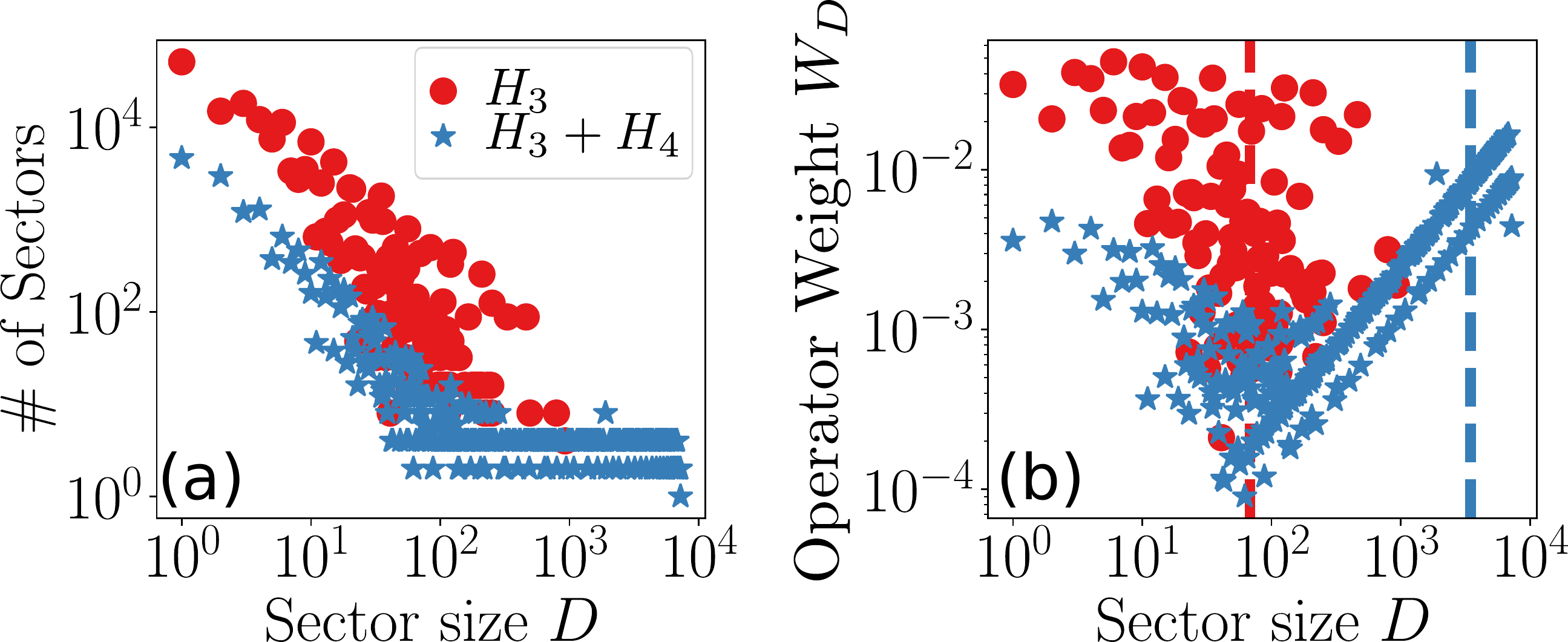}
	\caption{\textbf{Comparison of the Hilbert space connectivity.} (a) Sector size distribution for $H_3$ (red dots) and $H_3+H_4$ (blue stars). (b) The operator weight $W_D$ distribution for $S_0^z$ is qualitatively different, dominated by large sectors in the latter case. }
	\label{fig:fig5}
\end{figure}

This qualitative difference suggests that the Hilbert space structure uncovered in Sec.~\ref{sec:strong_fragment} should also be modified by adding $H_4$ to the Hamiltonian.
Figure~\ref{fig:fig5}(a) compares the distribution of sector sizes $D$ for $H_3+H_4$ (blue stars) with the minimal Hamiltonian $H_3$ (red dots).
While exponentially many invariant subspaces still exist, their total number is drastically reduced, as many previously disconnected sectors are coupled to each other by the perturbation $H_4$. 
Thus the number of sectors of small dimension $D$ decreases and there are new larger blocks appearing; in fact, the largest global symmetry sector, $q=p=0$, becomes almost (but not exactly) fully connected, as we discuss in Sec.~\ref{sec:StrVSWeak}.
This effect is even more apparent in the distribution of the operator weight $W_D$ (defined in Sec.~\ref{sec:Pred}) for the operator $S_0^z$, which we show in Fig.~\ref{fig:fig5}(b). 
Most of the weight is now concentrated around the largest sector, similarly to the case of a single global U$(1)$ symmetry. 
Thus, even though invariant subspaces within symmetry sectors still exist, they do not appear to be sufficiently relevant to make the system non-ergodic.
This is also reflected in the long-time value of the auto-correlation function as predicted in Eq.~\eqref{AnalPred}: plugging in the invariant subspaces of $H_3 + H_4$ we find that $C_0^z(\infty)$  approaches the thermal value, $2/(3N)$, exponentially in the thermodynamic limit, as shown in Fig.~\ref{fig:fig4}.

From these results we infer that including longer-range interactions makes the system sufficiently ergodic to thermalize. 
One possible reason for this qualitative difference is that the 4-site terms break the rule stated at the beginning of Sec.~\ref{sec:bubbles}, thus allowing for the desctruction of blocks of charges that would be inert under the dynamics of $H_3$.
A different path to break the non-ergodicity of $H_3$ would be to increase the local Hilbert space dimension, making the dynamics less constrained.
Consequently, we expect that for larger spin, even a three-site Hamiltonian of the form~\eqref{H_3} would lead to thermalization. 
Indeed, computing the lower bound $C_0^z(\infty)$ for the charge autocorrelator using Eq.~\eqref{AnalPred} for $H_3$ acting on a spin-2 chain, we find that it decays to zero in the thermodynamic limit, as shown in Fig.~\ref{fig:fig4}.
Similarly, if we consider spin-1/2 chains, the shortest range non-trivial model is $H_4$, which appears to be non-ergodic, while adding 5-site interactions restores ergodicity.

\subsection{Constructing frozen states}\label{sec:recursion}

While the combined Hamiltonian $H_3+H_4$ appears thermalizing at infinite temperature, it nonetheless violates the strong version of the Eigenstate Thermalization Hypothesis~\cite{Deutsch91,Srednicki94,Rigol2008,Kim2014}. 
In particular, certain frozen states continue to exist not only for $H_3+H_4$, but even in the presence of longer-range local interactions.
In fact, as we now prove, for a spin-1 chain that conserves charge and dipole, and involves only local terms with range at most $\ell$, there exist at least $2\cdot 5^{N/\ell}$ frozen states. 
While for $\ell=3$ this lower bound is not as tight as the Pauling estimate discussed in Sec.\ref{sec:frozen}, it provides useful insight into generic longer-range Hamiltonians and can be generalized to any spin representation.

\begin{figure}
	\centering
	\includegraphics[width=0.9\linewidth]{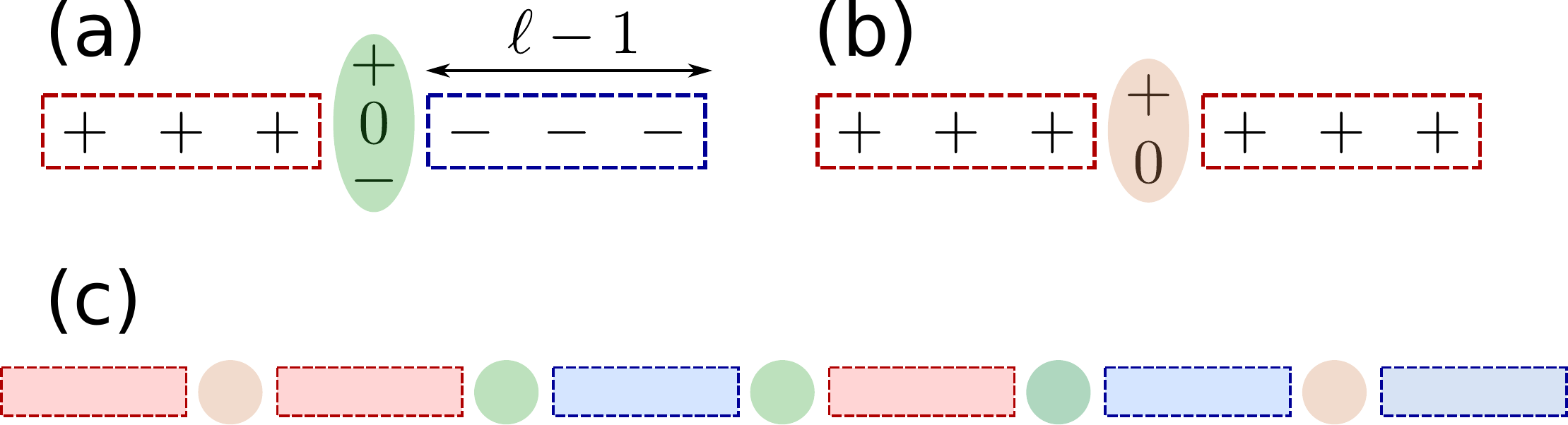}
	\caption{\textbf{Constructing frozen states for arbitrary finite-range interactions.} For interactions of maximal range $\ell$, one can create frozen patches of $2\ell-1$ sites with a flippable spin surrounded by domains of (a) opposite or (b) the same charges (shown here for $\ell=4$, relevant for $H_3+H_4$). These can then be combined to cover the entire chain, resulting in exponentially many frozen states, such as the one in panel (c).}
	\label{fig:recusrion}
\end{figure}

We begin our construction by considering the configuration shown in Fig.~\ref{fig:recusrion}(a), with a center site surrounded by a block of $\ell-1$ ($+$)-fractons on one side and $\ell-1$ ($-$)-fractons on the other. We now prove that this configuration is an eigenstate of any dipole-conserving term with range at most $\ell$, where without loss of generality we can measure the dipole moment relative to the center site. It is sufficient to consider off-diagonal terms (in the $z$ basis), consisting of spin raising and lowering operators. Due to the way we constructed the state, the only such terms that do not annihilate it are those that have only $S^-$ on one side and $S^+$ on the other. However, any such term would lead to a change in the dipole moment and is thus prohibited. Terms only acting on the center site do not change $P$ but they are also excluded due to charge-conservation. We conclude that this configuration is frozen, independently of the state of the center spin, as promised.

Next, we consider a similar configuration, but one where the center spin is surrounded by blocks of the same, rather than opposite, charges, as shown in Fig.~\ref{fig:recusrion}(b). Let these blocks be made out of ($+$)-fractons. Then the only off-diagonal operators that can act on them are powers of $S^-$, decreasing the total charge $Q$. One has to compensate for these charges by adding additional charges on the center site. Therefore the only allowed terms that could change this configuration are of the form $S^-_{-n}(S^+_0)^2S^-_{n}$, and only when the central spin is occupied by a ($-$)-fracton. When it is either $0$ or $+$, the state is frozen. 

One can combine these two types of `frozen patches' we constructed above to cover the entire 1D chain, resulting in a globally frozen state. These states are made up by blocks of $+$ or $-$ charges, with a single site between any two consecutive blocks, as shown in Fig.~\ref{fig:recusrion}(c). As we showed above, these sites host flippable spins: the ones separating blocks of equal charge can take two values (e.g $+$ or $0$ between blocks of $+$ charge), while those that separate blocks of opposite charge can be in any of the 3 possible spin states. This construction then results in exponentially many frozen states, coming from both the possible arrangements of $\pm$ blocks and from flipping the spins between blocks within a given arrangement. 

We can count the total number of frozen states resulting from this construction iteratively, starting from the left edge of the system (assuming open boundaries). We cut the systems into blocks of $\ell$ sites, consisting of a wall of $\ell-1$ positive/negative charges, followed by a flippable spin. Let $F^\pm_k$ denote the number of different such configurations to the left of the $k$-th wall (but before the flippable spin), ending in a $(\pm)$-block. Then the considerations outlined above lead to the following recursion formula:
\begin{equation}
\begin{pmatrix}
F^+_{k+1} \\ F^-_{k+1} 
\end{pmatrix}
= \begin{pmatrix}
2 & 3 \\ 3 & 2
\end{pmatrix}
\begin{pmatrix}
F^+_{k} \\ F^-_{k} 
\end{pmatrix}
= \begin{pmatrix}
2 & 3 \\ 3 & 2
\end{pmatrix}^k
\begin{pmatrix}
1 \\ 1
\end{pmatrix}
= 5^k
\begin{pmatrix}
1 \\ 1
\end{pmatrix},
\end{equation}
where we have used that $F^+_{1} = F^-_{1} = 1$. Since each step $k\to k+1$ corresponds to a shift by $\ell$ sites, we conclude that the number of frozen states we constructed scales as $2 \cdot 5^{N/\ell}$. This is only a lower bound on the total number of frozen states, which can include other configurations not captured by this construction. In particular one could
systematically improve the bound by allowing blocks to be separated by more then one site.

\begin{figure}
	\centering
	\includegraphics[width=0.75\linewidth]{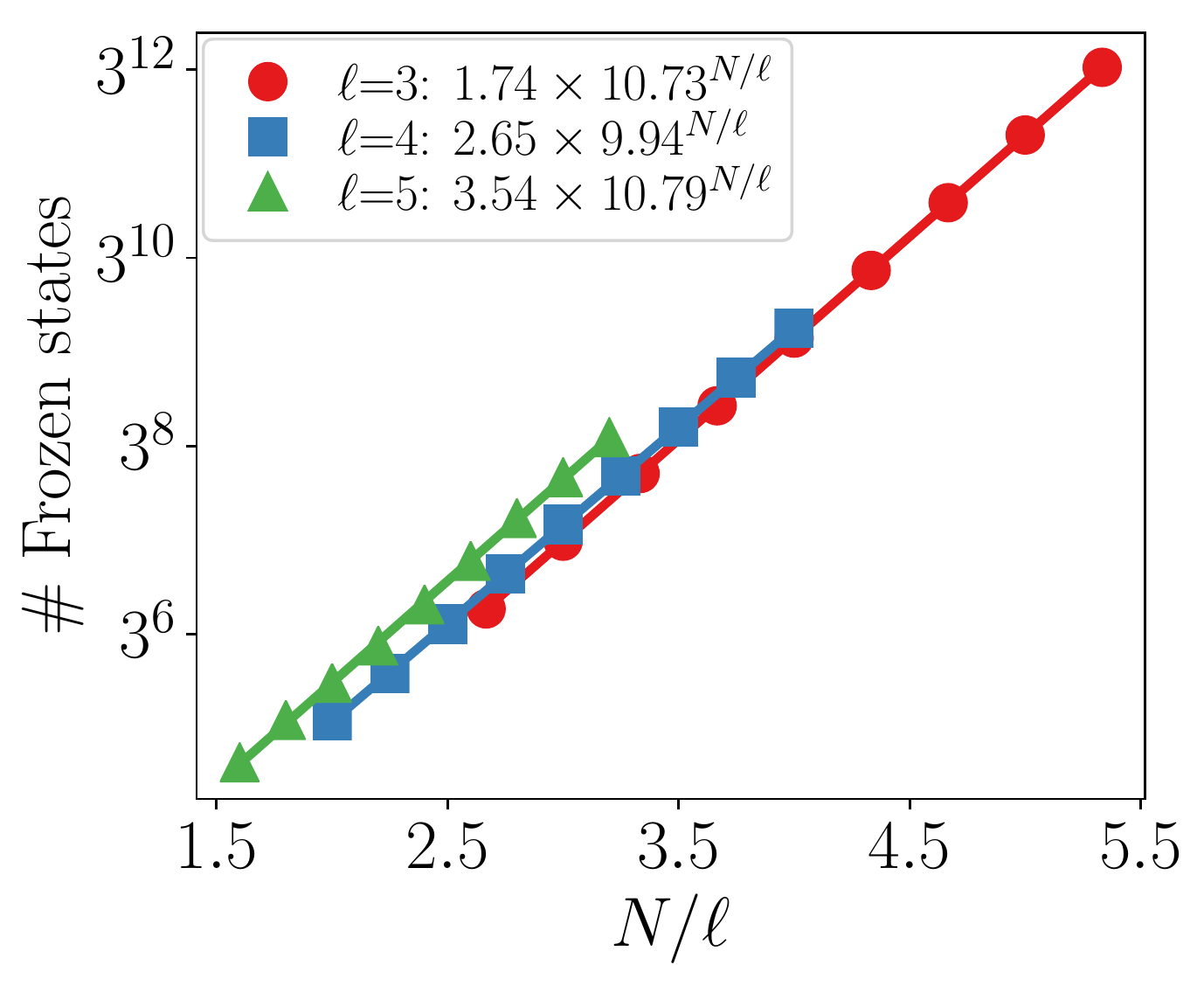}
	\caption{ \textbf{Scaling of the number of frozen states for Hamiltonians with at most range $\boldsymbol \ell$ terms.} We consider Hamiltonians with all possible combinations of charge and dipole conserving terms, quartic in spin operators of range at most $\ell$, for $\ell=3,4,5$. The number of frozen states grows exponentially with system size $N$, with an exponent that decreases with $\ell$, but is larger than the analytical lower bound $2\cdot 5^{N/l}$. }
	\label{fig:ScalFrozen}
\end{figure}

We compare the lower bound $\sim  5^{N/\ell}$ to the numerical results on the number of frozen states for Hamiltonians with interactions of range at most $\ell=3,4,5$ in Fig.~\ref{fig:ScalFrozen}, where we extract the asymptotic scaling. The comparison to numerical data in Fig.~\ref{fig:ScalFrozen} shows that the scaling is relatively close to $\sim 10^{N/\ell}$~\cite{ThirdRef}: $10.73^{N/3}$ ($\ell=3$), $9.94^{N/4}$ ($\ell=4$) and $10.79^{N/5}$ ($\ell=5$). Thus, the lower bound is not tight but it proves the exponential scaling of frozen states.

We conclude this section with some comments about the construction we presented. First, while above we did not distinguish between different overall $(q,p)$-sectors, one could similarly construct frozen states with a given $q$ and $p$. For example one can apply the construction on only the left half of the chain and for each state repeat the same configuration on the right half to obtain a state with $p=0$. Second, the bound can be easily extended to chains with local spin $S > 1$. For example one can consider blocks that have maximal positive/negative charge; repeating the same arguments then gives a scaling~\cite{Dimratio_Note} $(2S+3)^{N/\ell}$. Last, we note that in the limit $\ell\to\infty$ the lower bound tends to one, consistent with the fact that for all possible charge and dipole conserving infinite-range interactions every $(q,p)$ sector becomes completely connected.

\subsection{Strong vs. weak fragmentation} \label{sec:StrVSWeak}

\begin{figure}[t!]
	\centering
	\includegraphics[width=1.\linewidth]{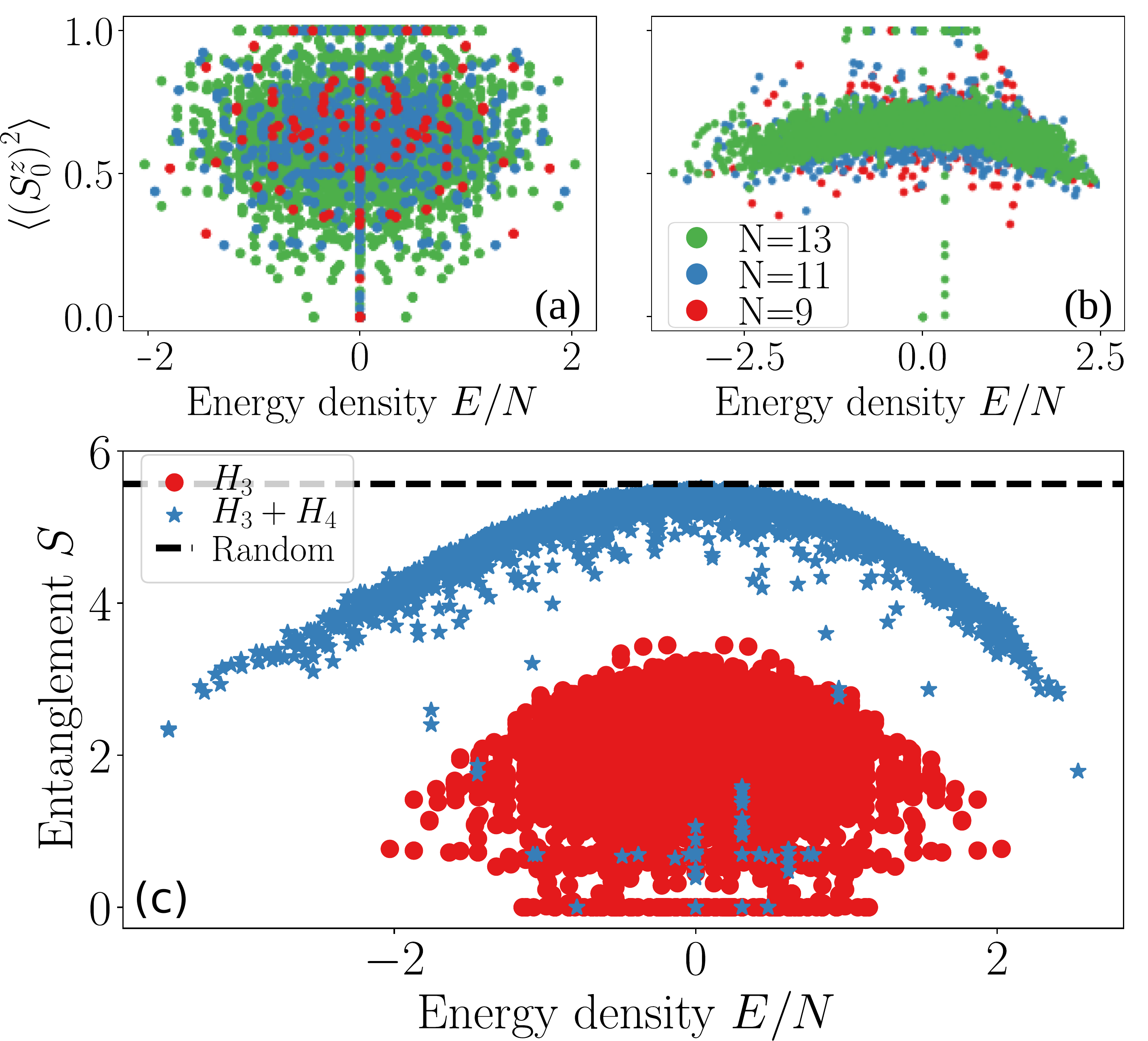}
	\caption{\textbf{Ergodicity breaking due to strong and weak fragmentation.} Expectation value of the local operator $(S_0^z)^2$ for eigenstates within the $(q,p)=(0,0)$ sector as a function of energy for different system sizes. (a) \emph{Strong fragmentation:} for the minimal Hamiltonian $H_3$, the width of the distribution does not decrease with $N$, violating the Eigenstate Thermalization Hypothesis. (b) \emph{Weak fragmentation:} for $H_3 + H_4$ most eigenstates appear thermal, and the bulk of the distribution narrows with $N$, but outlyers remain, showing that the system obeys weak, but not strong ETH. (c) Half-chain entanglement entropy of the eigenstates for $H_3$ (red dots) and $H_3+H_4$ (blue stars), for $N=13$, leads to the same conclusion. The black dashed line shows the entanglement entropy of a random state in the $(q,p)=(0,0)$ sector.}
	\label{fig:eigstates}
\end{figure}

As the previous section shows, the combination of dipole conservation and strictly local interactions is sufficient to lead to an emergence of exponentially many dynamically disconnected sectors in the many-body Hilbert space, even after fixing $q$ and $p$.
While we only showed this rigorously for the case of one-dimensional sectors, we find numerically that others with larger dimension also exist (see Fig~\ref{fig:fig5}).
While both $H_3$ and $H_3+H_4$ share this feature, their behavior with respect to thermalization appears to be quite different, as we already observed in Fig~\ref{fig:fig1}.
This motivates us to distinguish two cases, dubbed \emph{weak} and \emph{strong fragmentation}, which violate strong and weak ETH, respectively.

Let us first make precise what we mean by violation of ETH. 
In defining ETH we consider expectation values of few-body observables for all eigentates of the Hamiltonian within a fixed global $(q,p)$ symmetry sector (we do not consider off-diagonal matrix elements here).
By strong ETH we then mean the statement that the expectation values are the same for \emph{all} eigenstates at the same energy density in the thermodynamic limit.
Weak ETH, on the other hand, means that this statement only holds up to a small number of outlying states, where `small' means here `measure zero in the thermodynamic limit'.
Here we take the point of view of fixing only local symmetries, as non-local ones usually do not lead to distinct distributions for the diagonal matrix elements~\cite{Santos2010,Sorg2014,Rigol2DIsing}.
In our case, this means fixing $Q$ and $P$, but not the additional symmetries that correspond to the invariant subspaces, since we expect these to be non-local~\cite{ETHnote}.

Our construction in the previous section then proves that any dipole-conserving, strictly local Hamiltonian has weak fragmentation in the above sense, i.e., non-thermal eigenstates are present in the middle of the spectrum.
Apart from the aforementioned frozen states, these also include other low entanglement eigenstates, stemming from small invariant subspaces, analogous to the ones discussed in Sec.~\ref{sec:bubbles}.
Generically, however, their ratio compared to thermal ones is vanishingly small within any energy shell in the thermodynamic limit; this is the case of $H_3 + H_4$ as we argue below.
Thus the weak version of ETH~\cite{Biroli2010,MoriWeakETH,ShiraishiMori} is still obeyed, and the system thermalizes for typical initial states, provided they have narrow energy distributions.
On the other hand, we argue that the Hamiltonian $H_3$, discussed in Sec.~\ref{sec:H3}, has strong fragmentation in the sense that at least a finite fraction of the eigenstates is non-thermal, leading to the manifestly non-thermalizing behavior we observed. 

The difference is illustrated in Figs.~\ref{fig:eigstates}(a,b), which shows the expectation value of a simple observable ($(S_0^z)^2$, where $0$ is the central spin) for all energy eigenstates within the $q=p=0$ symmetry sector, for $H_3$ and $H_3+H_4$.
For the combined Hamiltonian, $H_3+H_4$, the majority of eigenstates, which all belong to the same invariant subspace, behave as predicted by ETH: $\braket{(S_0^z)^2}$ takes similar values for states within a narrow energy shell, with the width of its distribution decreasing with system size.
Nevertheless, we also observe outlying eigenstates, stemming from small invariant subspaces, that \emph{do not} approach this line, violating strong ETH.
The minimal Hamiltonian, $H_3$, on the other hand, violates even the weak version of ETH: the distribution of $\braket{(S_0^z)^2}$ does not become narrower with increasing $N$, as shown in Fig.~\ref{fig:eigstates}(a). 
This is in contradiction with ETH, which predicts a vanishing width in the thermodynamic limit. 
Similar behavior occurs in the half-chain entanglement entropy of the eigenstates, shown in Fig.~\ref{fig:eigstates}(c): the non-thermalizing nature of $H_3$ is reflected by the fact that the entropies of its eigenstates do not fall on a line when plotted as a function of the energy, instead being distributed over values much smaller than what is predicted at infinite temperature, as realized by a random state in the $(0,0)$ sector.

\begin{figure}
	\centering
	\includegraphics[width=0.65\linewidth]{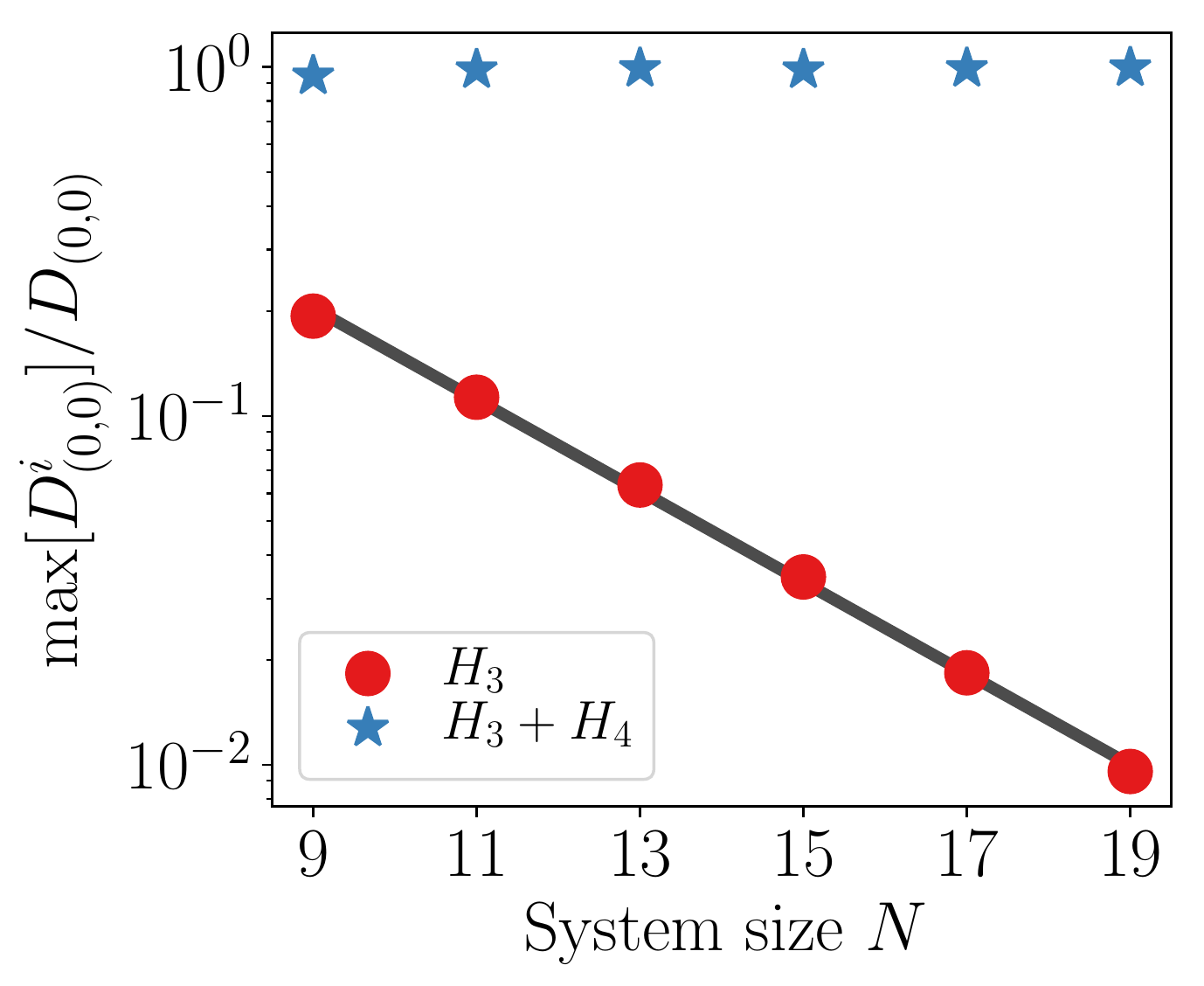}

	\caption{\textbf{Diagnosing strong and weak fragmentation.} Ratio between the dimension of the largest invariant subspace $\mathcal{H}^i$ within the $(0,0)$ symmetry sector, and the total dimension of the $(0,0)$ sector $D_{(0,0)}$. For $H_3$ (red dots), this ratio vanishes exponentially fast with system size while it approaches one for $H_3+H_4$ (blue stars).}
	\label{fig:ratio}
\end{figure}

The above discussion suggest that the difference between strong and weak fragmentation can be diagnosed by considering the sizes of the connected subspaces, in comparison with the size of the global $(q,p)$ symmetry sector they belong to.
In the strongly fragmented case of $H_3$ studied above, for a typical $(q,p)$ symmetry sector, the dimension of the largest connected subspace is exponentially smaller than the dimension of the full symmetry sector, i.e., $\mathrm{max}[D_{(q,p)}^i]/D_{(q,p)} \propto \exp(-\alpha N)$ for some $\alpha>0$.  
In Fig.~\ref{fig:ratio} we verify that this  is indeed the case for the largest symmetry sector $(0,0)$ of $H_3$.
We propose that this decay indicates strong fragmentation, naturally leading to the absence of thermalization for physical observables such as the auto-correlation function considered above.

In the weakly fragmented case, the symmetry sectors can still split into many subspaces.
However, the largest of these spans almost the entire $(q,p)$-sector: $\mathrm{max}[D_{(q,p)}^i] \approx D_{(q,p)}$, with the ratio approaching $1$ in the thermodynamic limit.
Figure~\ref{fig:ratio} shows that this is the case for $H_3+H_4$.
Consequently, the vast majority of eigenstates within any energy shell in a given $(q,p)$ symmetry sector belong to the same \emph{large} invariant subspace, and look thermal as a consequence.
Thus, while weakly fragmented systems violate strong ETH---due to outlying non-thermal eigenstates---they nevertheless thermalize for typical (but not all) initial states.
This weak fragmentation is reminiscent to what has been observed in other models in the context of \emph{many-body quantum scars}: although the majority of the eigenstates obey ETH, non-thermal eigenstates exist even in the bulk of the spectrum~\cite{ShiraishiMori,Moudgalya01,Moudgalya02,TurnerNatPhys,TurnerPRB}.
However, while the number of these `scarred' states is usually $\mathcal O(N)$, in our case we find exponentially many such states.
Note that the non-thermal eigenstates belonging to low-dimensional invariant sectors in our system have finite overlap with simple product states which can potentially be prepared in experimental settings.
This implies that a lack of thermalization up to \emph{infinite times} could be observed, even in the weakly fragmented case, for appropriately chosen initial states.

In the above we observed that the ratio $\mathrm{max}[D_{(q,p)}^i]/D_{(q,p)}$ either decays (exponentially) to zero or approaches unity. It is an interesting and open question, whether systems with intermediate behavior---with either slower than exponential decay or convergence to a finite fraction---can exist, and whether they exhibit strong or weak fragmentation.

\section{Comparison to random unitary circuits}\label{sec:RUC}
\begin{figure}
	\centering
	\includegraphics[width=1.\linewidth]{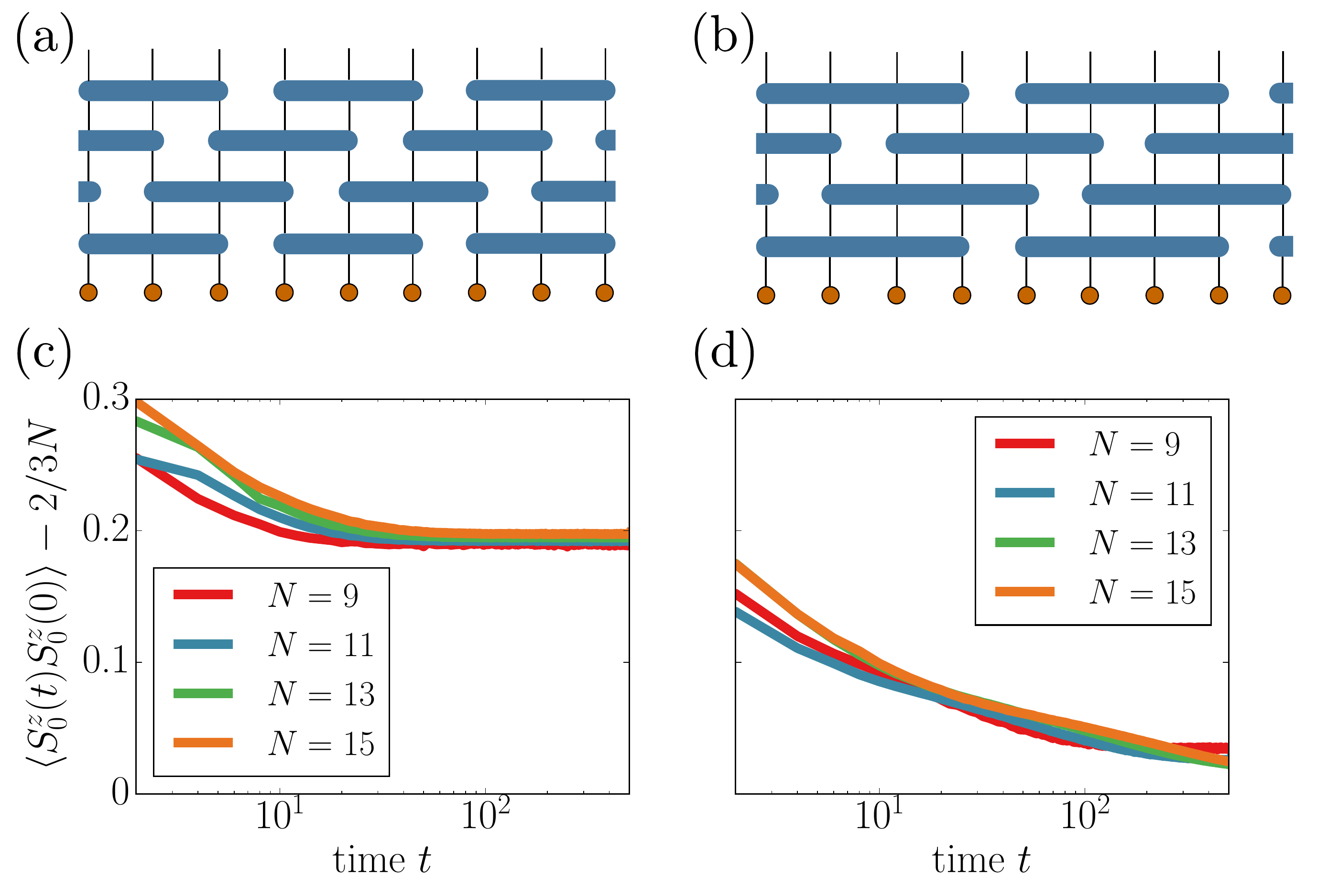}
	\caption{\textbf{Thermalization in charge and dipole conserving random circuit models.} The two versions of the circuit, with (a) three-site (b) four-site gates resemble the Hamiltonians $H_3$ and $H_3 + H_4$ respectively. Consequently, (c) the auto-correlator obtained for the three-site circuit has a finite long-time value, while (d) in the four-site circuit it slowly decays to zero. The curves correspond to the infinite temperature correlator, averaged over 50 random state and circuit realizations for $N \leq 13$ and 20 realizations of $N=15$.}
	\label{fig:RCFig}
\end{figure}
We now argue that our findings are not specific to the Hamiltonians we considered so far, and generalize to arbitrary systems with the same global symmetries and a fixed range of interactions. In particular, we compare with random unitary circuits of the form originally introduced in Ref. \onlinecite{Pai18}. These define a discrete time evolution, the building blocks of which are unitary gates acting on $\ell$ sites, each of which is required to be block diagonal in $Q$ and $P$, but is otherwise chosen randomly (i.e. every block is independently Haar random). In Ref. \onlinecite{Pai18} it was argued that such circuits always lead to localized behavior. Here we argue that this is in fact only the case for gates with $\ell=3$, where the circuit exhibits exactly the same Hilbert space structure as the Hamiltonian $H_3$ above, and is therefore indeed similarly localized. When introducing larger gates of size $\ell=4$, we find that the system thermalizes, also in complete agreement with our results on the Hamiltonian $H_3 + H_4$. 

The two circuit geometries, with gates of size $\ell=3$ and $4$, are shown in Fig.~\ref{fig:RCFig}. In both cases we compute the connectivity of the Hilbert space. Instead of the Hamiltonian, we consider the unitary operator defined by the first $\ell$ layers of the circuit. This is a matrix with random entries, but its connected components are independent of the particular realizations. We find numerically that the connected components for $\ell=3$ ($\ell=4$) coincide \emph{exactly} with those of the Hamiltonians $H_3$ ($H_3 + H_4$), shown previously in Fig.~\ref{fig:fig5}. This follows from the fact that the allowed local transitions are the same in the Hamiltonian and the random unitary circuit.   The fact that the invariant subspaces coincide supports the idea that the additional invariant subspaces are a consequence of dipole conservation and locality alone, and do not depend on any additional structure that might be present in the Hamiltonian case. Based on our previous analysis, we therefore expect that the three-site circuit does not thermalize, but the four-site circuit does. This is confirmed by calculating the autocorrelator $C_0^z(t)$, which (after subtracting its thermal value) goes to a constant in the former case, while it decays to zero in the latter, as shown in Fig.~\ref{fig:RCFig}. In App.~\ref{app:rhoR} we also consider the spatial spreading of an initial $S_n^z$ operator and similarly find that for $\ell=4$ the operator is delocalized at long times. We therefore conclude that the localized behavior observed in Ref. \onlinecite{Pai18} is particular to the case of the circuit with three-site gates, contrary to what is suggested there.

The fact that the Hilbert space fragmentation coincides exactly between the random circuit and Hamiltonian cases also means that the conclusions we drew regarding non-thermal eigenstates in Sec.~\ref{sec:StrVSWeak} also generalize to time-periodic (Floquet) models built out of similar local gates.   In particular this implies the presence of exponentially many frozen eigenstates for such models, especially for the $\ell=3$ case where we predict that the majority of eigenstates should be nonthermal.

\section{Summary and Outlook}\label{sec:summary}
In this work, we studied the out-of-equilibrium dynamics of spin chains conserving a charge and its associated dipole moment. For the minimal spin-1 Hamiltonian which is restricted to only three-site interactions, we found non-ergodic behavior in the charge-charge auto-correlation function. We explained this finding in terms of a \emph{strong fragmentation} of the Hilbert space into exponentially many disconnected sectors which all contribute significantly to the dynamics even at infinite temperature. We found that a weaker form of fragmentation survives for more general, longer-range Hamiltonians, and while it is no longer sufficient to make the infinite-temperature dynamics non-ergodic, it nevertheless results in exponentially many non-thermal eigenstates. Furthermore, we showed numerically that the fragmentation of the Hilbert space exactly matches that of random circuit dynamics with the same range of interactions, giving rise to similar dynamical behavior.

The observed fragmentation lies in-between the known cases of systems with a few global symmetries and that of integrable or many-body localized systems. 
The former have at most polynomially many symmetry sectors ---most of which are exponentially large---while the latter have $\sim N$ independent local conserved quantities.
In our case, however, sectors of all sizes co-exist and in the case of strong fragmentation they all are relevant for the dynamics, even at infinite temperature.
Understanding how these different sector can be consistently labeled and what the corresponding conserved operators look like is an interesting open problem.
 
Another interesting problem for future work is to investigate the equilibrium properties and dynamics of the system at finite or zero temperature, both within the whole Hilbert space and within individual subspaces, as well as to clarify the requirements for strong and weak fragmentation. It would be also worth investigating the relationship between the phenomenon of \emph{Hilbert space fragmentation} and the concept of \emph{reducibility} in classical constrained models~\cite{Ritort03}. Moreover, the extension of the current analysis to higher spatial dimensions seems promising, where disconnected sectors also appear but different roads to thermalization can be present. It is also worth exploring further the connections to: many-body localization~\cite{BASKO20061126,Serbyn13cons}, quantum scars~\cite{Sanjay18,TurnerNatPhys}, gauged  formulations of the considered systems~\cite{PretkoPrin,PretkoMach} as well as Stark localization~\cite{Schulz19,Refael18}. The latter also provides a potential experimental platform for realizing dipole-conserving Hamiltonians of the type studied here in the limit of strong electric fields~\cite{vanNieuwenburg19}, which perturbatively lead to the kind of terms we have considered above. Finally, it would be interesting to search for connections between the models discussed in this work and type I fracton models \cite{Vijay16,Vijay15}, as well as for the possible connection between the discussed fragmentation of the Hilbert space and the emergence of superselection rules in the cubic code model~\cite{Kim16}.

\section{Acknowledgments}
The authors thank Mari Carmen Bañuls, Claudio Castelnovo, Juan Garrahan, Johannes Hauschild, Daniel Hetterich, Vedika Khemani, Sanjay Moudgalya,  Rahul M. Nandkishore, Shriya Pai, Michael Pretko, Nic Shannon, John Sous, D\'avid Horv\'ath and Giuseppe de Tomasi for discussions. 
P.S. specially thanks María del Carmen Giménez Aurusa, and acknowledges the Boulder school 2018, which inspired part of this work.
We acknowledge support from “la Caixa” Foundation (ID 100010434) fellowship grant for post-graduate studies (P.S.), the Technical University of Munich - Institute for Advanced Study, funded by the German Excellence Initiative and the European Union FP7 under grant agreement 291763 (M.K.), the Deutsche Forschungsgemeinschaft (DFG, German
Research Foundation) under Germany's Excellence Strategy -- EXC-2111-390814868 (F.P., M.K.), from the DFG grant No. KN 1254/1-1 (M.K.), the DFG TRR80, Project number 107745057 (F.P., M.K.), the DFG Research Unit FOR 1807 through grant no. PO 1370/2- 1 (F.P.), the DFG Collaborative Research Center SFB 1143 (R.V.), the Nanosystems Initiative Munich (NIM) by the German Excellence Initiative (F.P.), and the European Research Council (ERC) under the European Unions Horizon 2020 research and innovation program grant agreement No. 771537 (F.P.) and No. 851161 (M.K.). This research was conducted in part at the KITP, which is supported by NSF Grant No. NSF PHY-1748958 and the Heising-Simons Foundation.

\textbf{Note added.}  While we were finalizing our manuscript, we have learned about related work by V. Khemani and R. Nandkishore which will appear in the same arXiv posting.

\onecolumngrid
\appendix 
\section{Symmetries of the `minimal' Hamiltonian $H_3$ \label{app:Symmetries}}

Here we discuss some additional symmetries possessed by the 'minimal model' represented by the Hamiltonian $H_3$ in Eq.~\eqref{H_3}. 
One of these is the sublattice parity symmetry $\Pi^z_\text{odd}$. However, one can easily prove that this quantity is fixed by the dipole moment as
\begin{align} \Pi^z_\text{odd}=\exp\big({i\pi\sum_{n\,\mathrm{odd}}{S_n^z}}\big)=\exp\big({i\pi\sum_{n}{nS_n^z}}\big)= \exp\big({i\pi P}\big),
\end{align}
where in the second step we have used that for spin-1, a $2\pi$ rotation is equal to the identity. 
From this it is clear that the total parity $\Pi^z=\exp\big({i\pi\sum_{n}{S_n^z}}\big) $ is obtained as $\Pi^z=\Pi_\text{odd}^z \Pi^z_\text{even}$ and is related to the total charge as $\Pi^z=\exp\big(i\pi Q\big)$. 
In general, the terms in $H_3$ are also invariant under the parity transformations given by $\Pi^x=\exp\big({i\pi\sum_{n}{S_n^x}}\big)$ and $\Pi^y=\exp\big({i\pi\sum_{n}{S_n^y}}\big)$, which map 
\begin{equation} 
S_{n_1}^+S_{n_2}^-S_{n_3}^-S_{n_4}^+\overset{\Pi^x,\Pi^y}{\longleftrightarrow} S_{n_1}^-S_{n_2}^+S_{n_3}^+S_{n_4}^- \end{equation}
for all $n_1, n_2,n_3,n_4$. 
Note that $\Pi^x$ and $\Pi^y$ do not commute with $Q$ or $P$. 

Moreover, as stated in the main text, there exists an operator
$ \mathcal{C}=\prod_{n}{e^{i\pi\big( S_{4n}^z + S_{4n+1}^z \big)}} $ which anti-commutes with $H_3$. Since $\mathcal{C}$ commutes with $Q$ and $P$, the spectrum of $H_3$ is symmetric around zero in every $(q,p)$-sector. However, when additional terms diagonal in the $S^z$ basis are considered, which by construction do not change the fragmentation of the Hilbert space as we discussed in the main text, $\mathcal{C}$ does not anti-commute anymore with the resulting Hamiltonian.
There is also at least one additional anti-commuting operator $\tilde{\mathcal{C}}=\prod_{n}{e^{i\pi\big( S_{4n+2}^z + S_{4n+3}^z \big)}}$ but since $\mathcal{C}\tilde{\mathcal{C}}=\Pi^z$, they are not independent. Note that since $\mathcal{C}$ commutes rather than anti-commutes with $H_4$, the spectrum of $H_3 + H_4$ is no longer symmetric as can be seen e.g., in Fig.~\ref{fig:eigstates}.
Nevertheless, there also exists a separate anti-commuting operator for $H_4$\eqref{H_4} taking the form $\mathcal{C}_4=\prod_n e^{\pi S_{4n}^z}$, which does not anti-commute with $H_3$.

\begin{figure}[b!]
	\centering
	\includegraphics[width=0.7\linewidth]{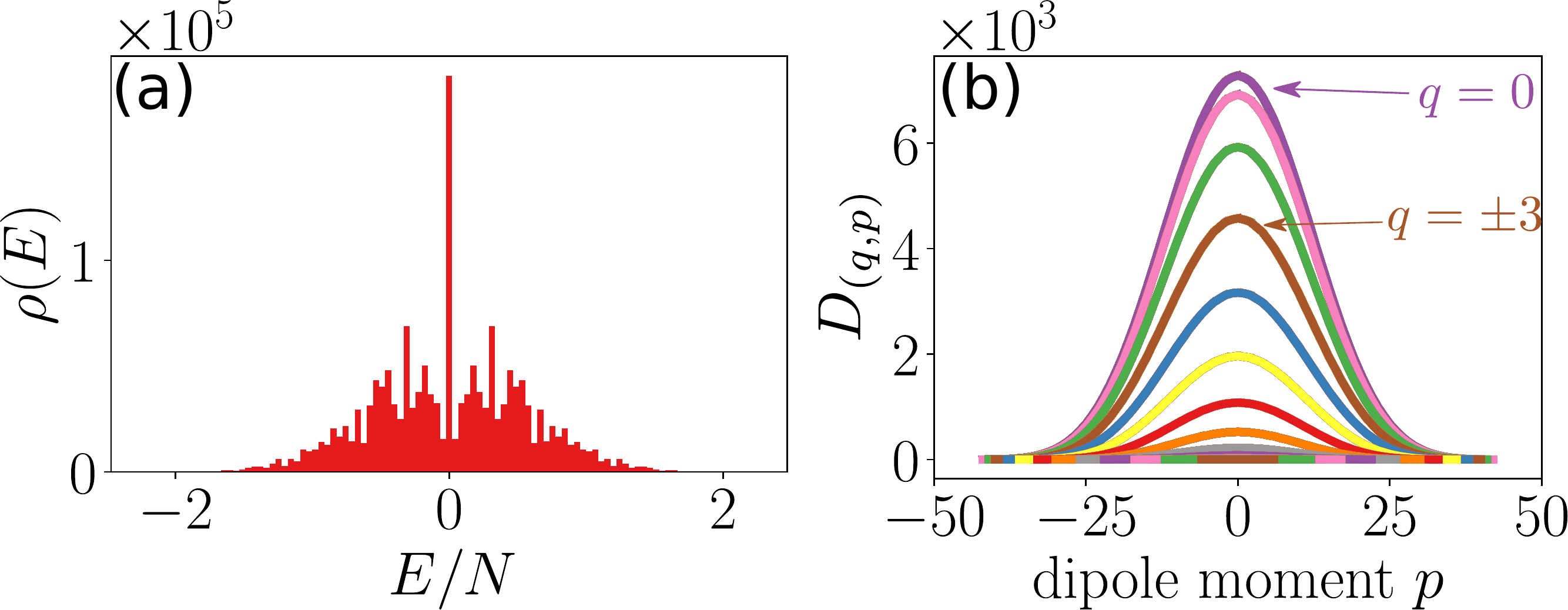}
	\caption{ (a) Density of states (DOS) for the Hamiltonian $H_3$ in Eq.~\eqref{H_3} for system size $N=13$. (b) Distribution of dimensions $D_{(q,p)}$ for the $\mathcal{H}_{(q,p)}$ invariant subspaces. Each curve corresponds to a subspace with fixed charge $q$.  }
	\label{fig:App1}
\end{figure}

In Fig.~\ref{fig:App1}(a) we show the density of states, $\rho(E)$, of $H_3$ for a chain of length $N=13$, which has a divergent delta peak at zero energy.
This peak contains all the frozen states described in the main text, among other zero energy eigenstates that arise as a consequence of the aforementioned anti-commuting symmetry.
One could remove the peak at zero energy by adding e.g. a finite mass term of the form $m\sum_n(S_n^z)^2$ to $H_3$, which breaks the anti-commuting symmetry.
This term also has the effect of shifting the energy of the frozen states to finite values and distributing them throughout the spectrum.

In Fig.~\ref{fig:App1}(b) we show the size of the symmetry sectors with different global quantum numbers $q$ and $p$.
Note that this distribution is independent of the specific Hamiltonian under study. 
Each curve corresponds to a fixed value of the charge quantum number $q$. The dimension $D_{(q,p)}$ decreases with increasing absolute value of the charge. 
The distributions for $+q$ and $-q$ coincide due to time reversal invariance, the way we have chosen the reference site $n_0$, and labeling the sites in the chain. 
A different labeling of sites, would simply shift the mean value of both distributions symmetrically with respect $p=0$. 
We also observe that the distribution attains a maximum at the $(0,0)$-sector, as claimed in the main text. 
In addition, we obtain symmetric distributions because $P$ changes sign under inversion, while $Q$ is invariant.

\begin{figure}
	\centering
	\includegraphics[width=0.7\linewidth]{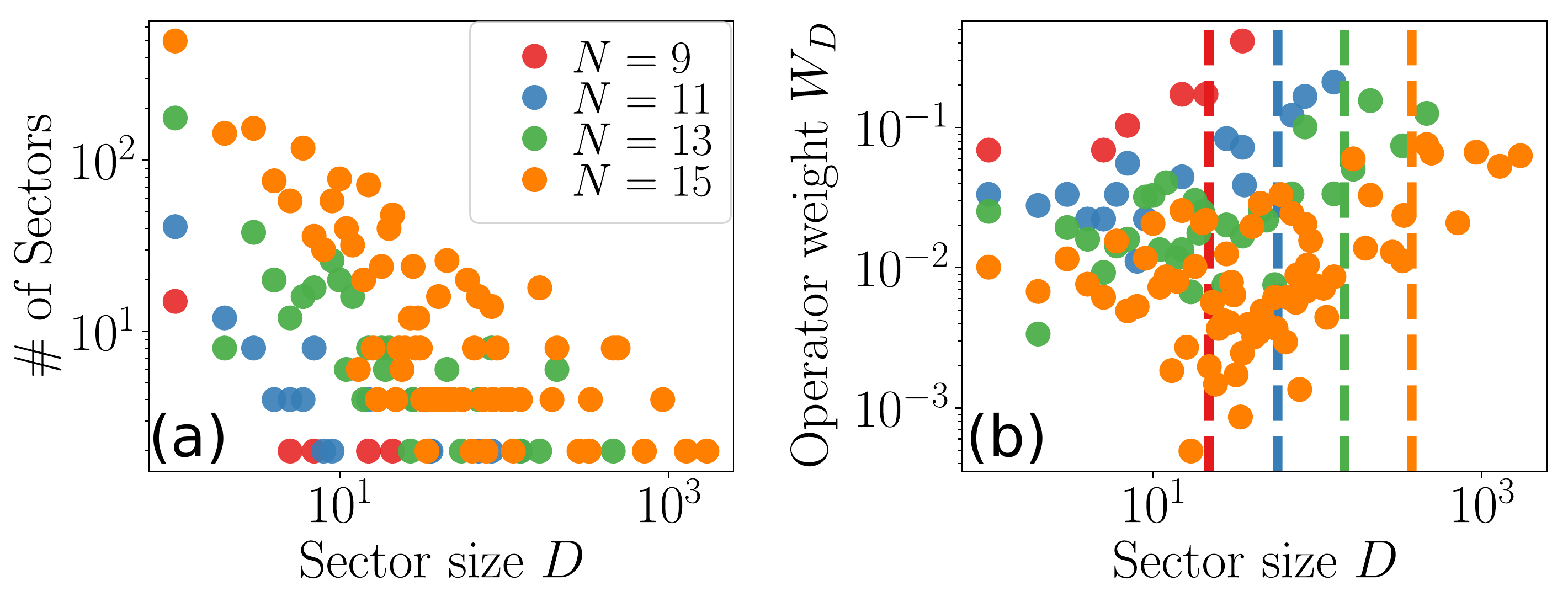}
	\caption{(a) Distribution of sector sizes D with $\mathcal{H}_i\subset \mathcal{H}_{(0,0)}$ for the Hamiltonian $H_3$. (b) operator weight distribution $W_D$. Both plots are similar to the full distributions shown in the main text. The vertical dashed lines in (b) indicates the average sector size, which grows exponentially in system size, but is nevertheless exponentially smaller than the largest sector. }
	\label{fig:App1b}
\end{figure}

Finally, we show the sector size and the operator weight distributions with $W_D\equiv\sum_{D_i=D}\mathrm{tr}\big(Z_i^2\big)/\Big(\sum_{D}\sum_{D_i=D}\mathrm{tr}\big(Z_i^2\big)\Big)$ for invariant subspaces within the largest $(q,p)$-sector, $q=p=0$. Fig.~\ref{fig:App1b}(a) shows qualitatively the same sector size distribution as in Fig.~\ref{fig:fig2}(c) in the full Hilbert space. Fig.~\ref{fig:App1b}(b) also reflects the main properties of the operator weight distribution, featuring a wide distribution with significant weight on small sectors.

\section{Finite-size scaling of the autocorrelator\label{app:scaling}}

In this section we present in more detail the finite size scaling of the auto-correlation function and its lower bound.

\begin{figure}
	\centering
	\includegraphics[width=0.7\linewidth]{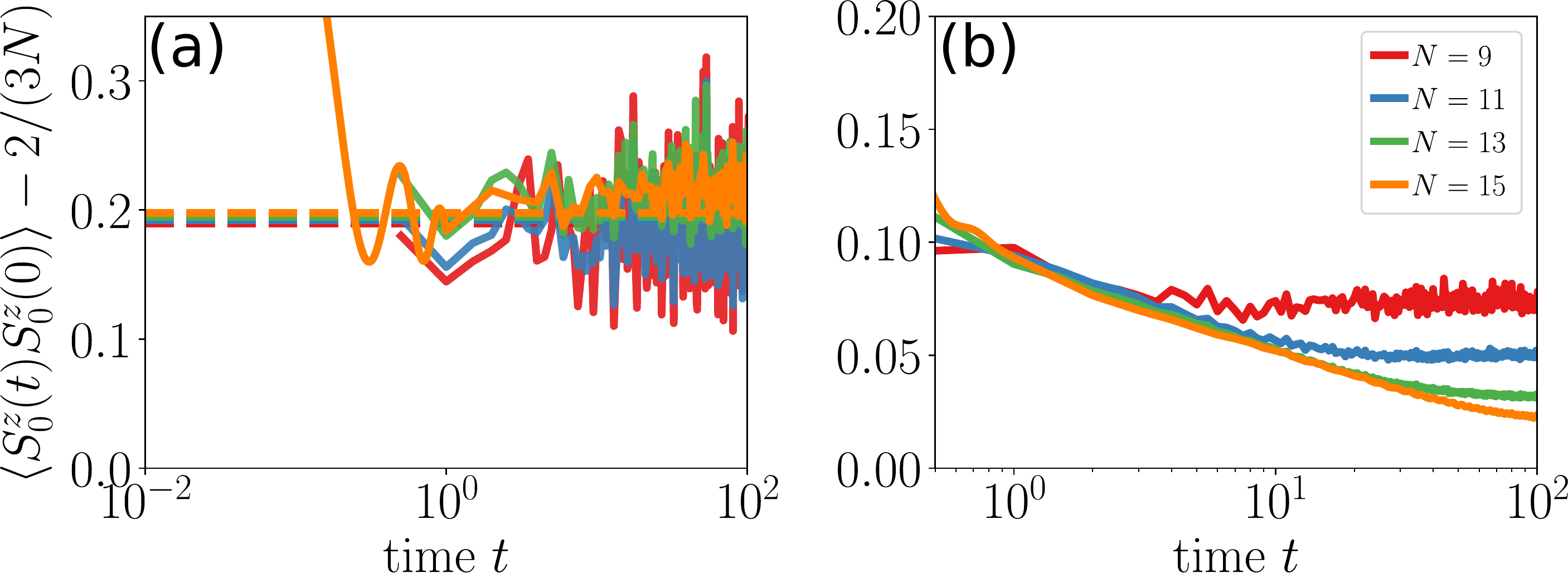}
	
	\caption{Finite size scaling for the auto-correlation function $\avg{S_0^z(t)S_0^z}$ at infinite temperature in the full Hilbert space after substracting the thermal value. Panel (a) shows a finite value for the auto-correlation under the evolution of $H_3$ in Eq.~\eqref{H_3}. The dashed lines show the lower bound in Eq.~\eqref{AnalPredqp}.  (b) The auto-correlation function decays to zero with system size once the longer range-interaction $H_4$ in Eq.~\eqref{H_4} is added to $H_3$. }
	\label{fig:App2a}
\end{figure}

First, we discuss the scaling of the auto-correlation function $\avg{S_0^z(t)S_0^z}$ at infinite temperature in the full Hilbert space in Fig.~\ref{fig:App2a} for both the minimal model $H_3$ in Eq.~\eqref{H_3} and the combined Hamiltonian $H_3+H_4$. On the one hand, the minimal model realizes a finite saturation value at long times which slightly grows with system size as can be seen in Fig.~\ref{fig:App2a}(a). On the other hand, when the combined Hamiltonian $H_3+H_4$ is considered, the auto-correlation decays to zero with system size. This agrees with the discussion in the main text, where it was argued that for longer range Hamiltonians, the system thermalizes and the correlation decays to zero at long times in the thermodynamic limit.

\begin{figure}
	\centering
	\includegraphics[width=0.6\linewidth]{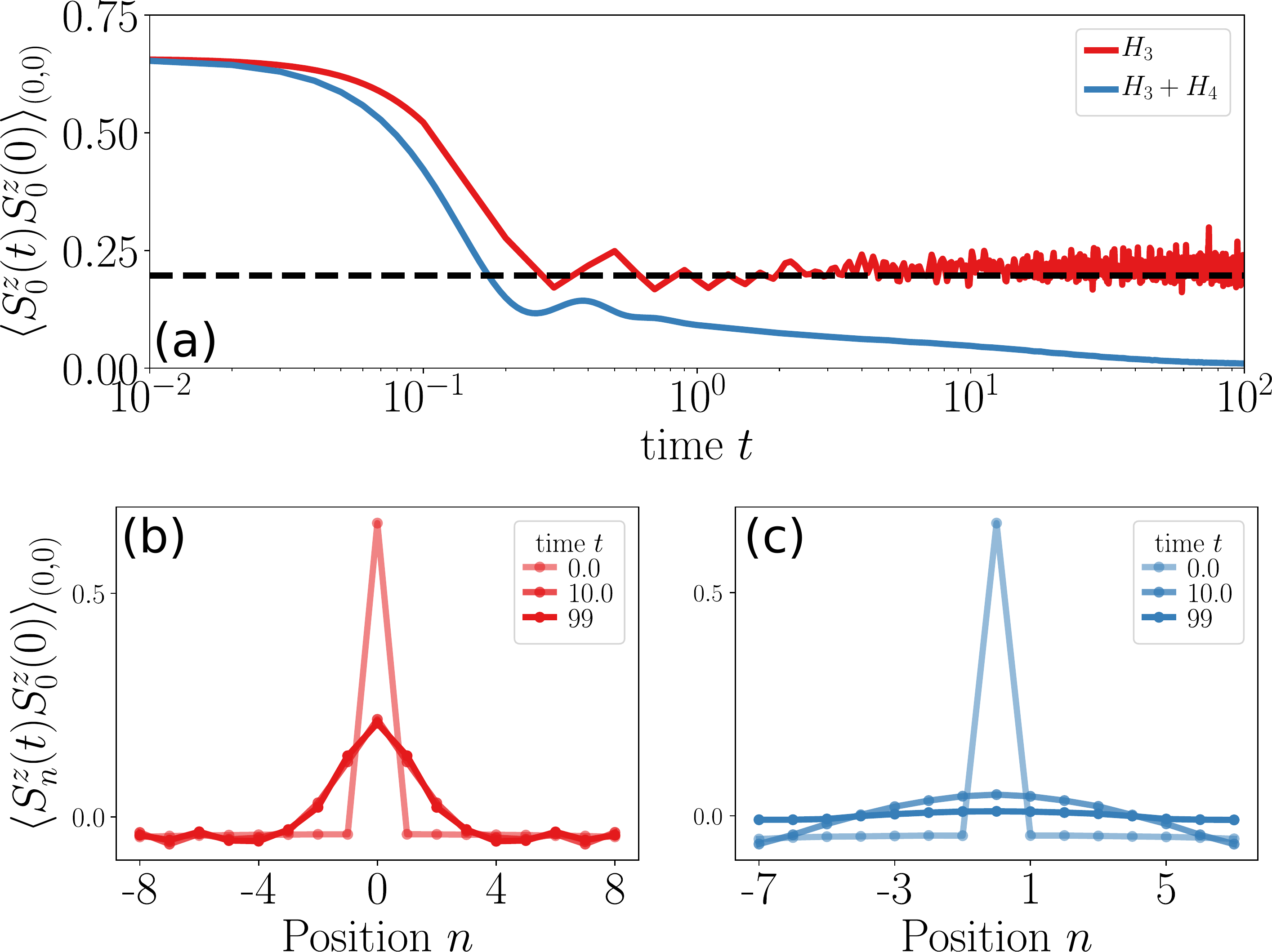}
	
	\caption{ (a) Auto-correlation function (upper panel) $\avg{S_0^z(t)S_0^z}_{(0,0)}$ in the symmetry sector $q=p=0$ at infinite temperature for $H_3$ (red curve) and $H_3+H_4$ (blue curve) for $N=15$. Spatially resolved correlation functions for (b) $H_3$ and (c)$H_3+H_4$. }
	\label{fig:App2aa}
\end{figure}

Moreover, as we discussed in the main text, not only the auto-correlation function in the full Hilbert space shows a non-thermal (thermal) behavior for $H_3$ ($H_3+H_4$). We can also realize this behavior within a specific restricted symmetry sector. In Figs.~\ref{fig:App2aa}(a) we show the behavior of the auto-correlation function $\avg{S_0^z(t)S_0^z}_{(0,0)}$ in the largest $(q,p)$-symmetry sector, $q=p=0$, and size $N=15$ showing the same qualitative behavior: a finite saturation value at long times for $H_3$ (panel (a)) and thermalization for the combined Hamiltonian $H_3+H_4$ (panel (b)). Note that since charge is conserved and we evaluate the correlation within the $q=0$ sector, $\sum_n\avg{S_n^z(t)S_0^z(0)}=0$ at all times and thus the surface under the peak must add up to zero.

\begin{figure}
	\centering
	\includegraphics[width=0.45\linewidth]{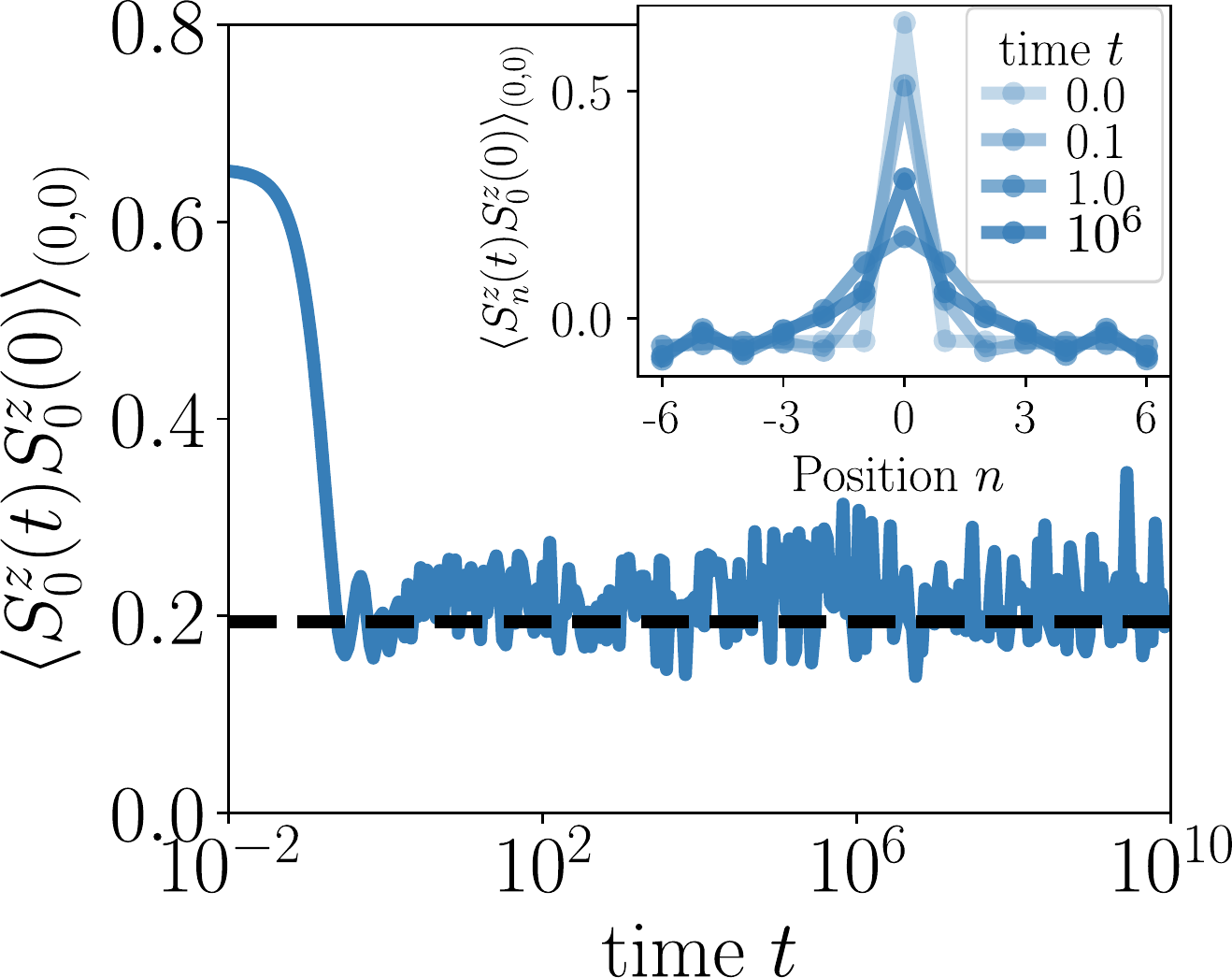}
	
	\caption{Evolution of the auto-correlation function $\avg{S_0^z(t)S_0^z}_{(0,0)}$ for the Hamiltonian $H_3$ and system size $N=13$, where longer time scales $t\sim 10^{10}$ can be numerically reached. We observe the same qualitative behavior as in Fig. \ref{fig:fig1}. }
	\label{fig:App2L}
\end{figure}

Moreover, in Fig.~\ref{fig:App2L}, we show the persistence of the non-thermalizing behavior for $H_3$ at longer times $t=10^{10}$ for smaller system size $N=13$ and within the $(0,0)$-sector. The space resolved correlation function is also shown in the inset showing the absence of thermalization even at long time scales.

\begin{figure}
	\centering
	\includegraphics[width=.8\linewidth]{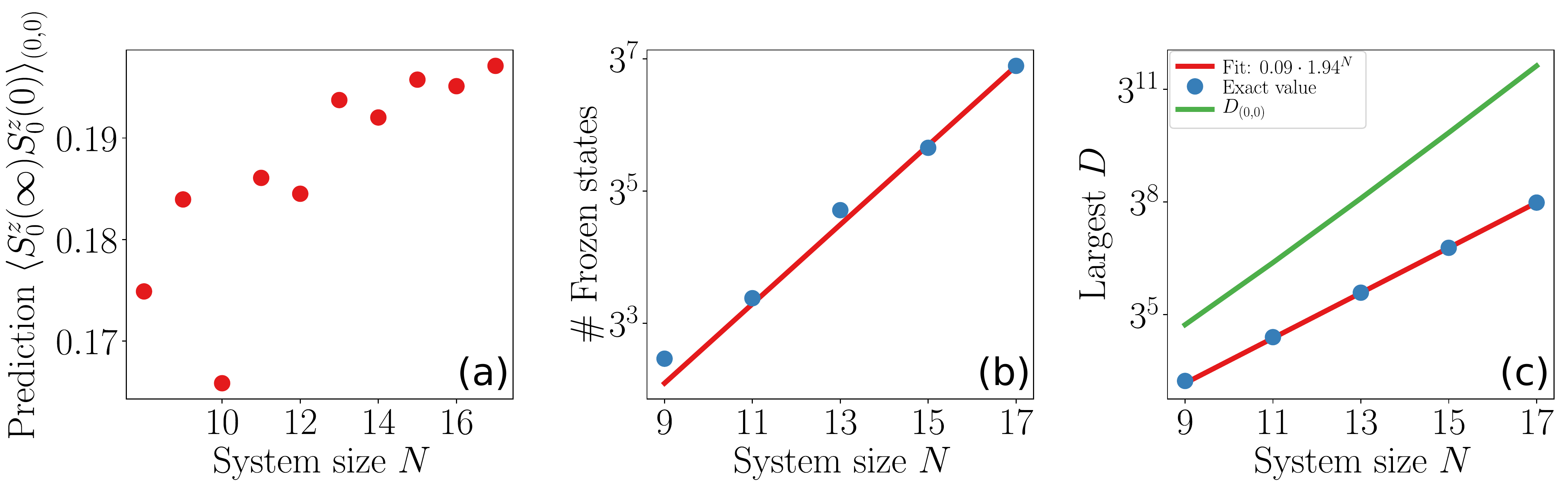}
	\caption{Scaling within the $q=p=0$ sector for Hamiltonian $H_3$. (a) Scaling of the lower bound $\lim_{T\to\infty}1/T\int_0^Tdt\,\avg{S_0^z(t)S_0^z(0)}_{(0,0)}$ with system size. (b) Scaling of the number of frozen states. (c) Scaling of the largest sector dimension (blue dots) in comparison to the dimension of the $(0,0)$ sector (green line).}
	\label{fig:App2S}
\end{figure}

In Fig.~\ref{fig:App2S}(a) we show the scaling of the lower bound $C_0^z(\infty)$ in Eq.~\eqref{AnalPred} with system size $N$ restricted to the $(0,0)$ symmetry sector of $H_3$. In this case the lower bound takes the form
\begin{align} \label{AnalPredqp} \lim_{T\to\infty}\frac{1}{T}\int_0^Tdt\,\avg{S_0^z( t)S_0^z(0)}_{(0,0)}\geq \frac{1}{D_{(0,0)}}\sum_{\mathcal{H}_i\subset \mathcal{H}_{(q,p)}}{\frac{1}{D_i}\left[\mathrm{tr}\big(Z_i\big)\right]^2}.
\end{align}

We observe that the value increases with $N$ and realize an even-odd dependence on $N$ decreasing with system size. In addition, Figs.~\ref{fig:App2S}(b-c) show, respectively, how the number of frozen states and the size of the largest invariant subspace within the $(0,0)$ symmetry sector grow with system size. Since the largest sector does not scale with the size of the entire Hilbert space, the lower dimensional sectors become thermodynamically important. Compare for example with a spin 1/2 chain with charge conservation only. The dimension of the full Hilbert space is $2^N$ and the largest (zero charge) subspace scales as $ \sqrt{1/ N}\cdot 2^N $; hence, the exponents are the same up to logarithmic corrections. 

\section{Entanglement growth from random product states}\label{app:EE}

In this appendix we complement our results on auto-correlations with a different measure of thermalization: entanglement growth from a (random) product state. By choosing the initial state Haar randomly on each site and averaging, we ensure that the dynamics explores all $(q,p)$ symmetry sectors. For an ergodic system, the long-time state is then expected to resemble a global random state in the entire Hilbert space. In particular, the entanglement between two halves of a bi-partition is expected to be given by the Page formula, which in our case (maximal bi-partition of a spin-1 chain with odd lengths) reads \cite{Page93} $S_\text{Page} = \frac{N-1}{2} \log{3} - \frac{1}{6}$. 

\begin{figure}
	\centering
	\includegraphics[width=0.95\linewidth]{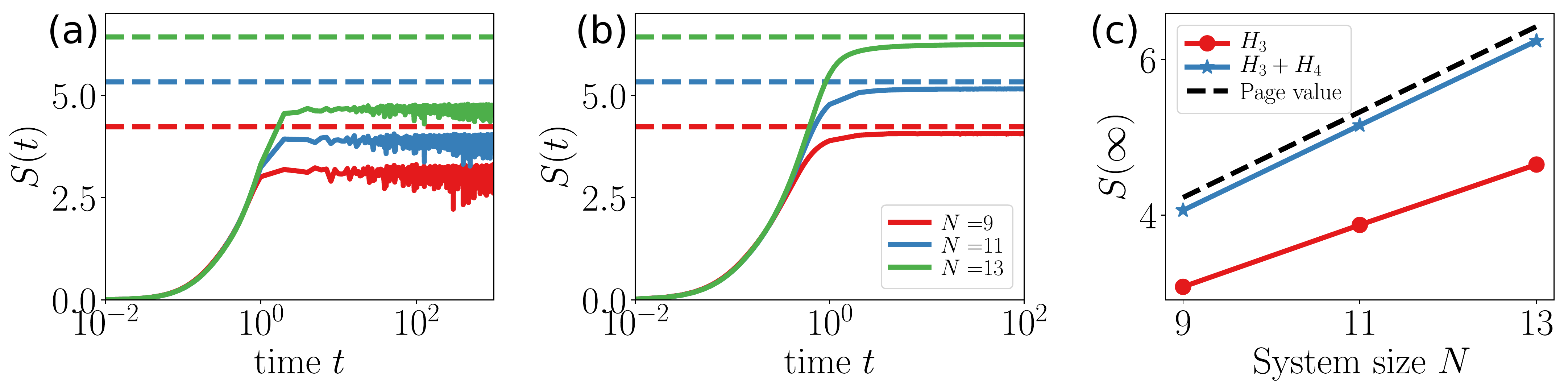}
	\caption{Half chain entanglement entropy (EE) growth for an initial random product state for different system sizes. The dashed line signals the Page value~\cite{Page93}. Panel (a) shows the behavior of the EE for the minimal model $H_3$. The entanglement reaches a size dependent saturation value below the Page value. This can be understood from the exponential fragmentation of the Hilbert space. However, when the combined Hamiltonian $H_3+H_4$ is considered in panel (b), the EE almost reaches the Page value. We associate the offset between them to the existence of still exponentially many invariant subspaces. (c) Scaling of the time-averaged saturation value for EE reached at long times. While for $H_3$ the slope is different from that of the Page value, signalling a sub-thermal entropy density in the steady state, the off-set for the combined Hamiltonian appears to be constant. }  
	\label{fig:EntEnt}
\end{figure}

We evaluate the time evolution starting from the aforementioned random product states exactly, for both the minimal Hamiltonian $H_3$ and the combined Hamiltonian $H_3 + H_4$. In the former case, shown in Fig.~\ref{fig:EntEnt}(a), we find that while the entanglement quickly saturates to a volume law, the associated entropy \emph{density} is smaller than the expected Page value, indicating a non-thermal state. This is consistent with our results on auto-correlation functions in Fig.~\ref{fig:fig1}, as well the entanglement of eigenstates in Fig.~\ref{fig:eigstates}, all consistent with non-ergodic behavior. The entanglement growth for $H_3 + H_4$ is shown in~\ref{fig:EntEnt}(b) where we observe that the entanglement saturates to a value close to $S_\text{Page}$. There is still a constant offset, which we associate to the influence of the remaining non-thermal eigenstates, but this does not affect the entropy density in the thermodynamic limit.

\section{Mapping to the PXP model}\label{app:PXP}
In this appendix we explain the relation between the dipole-conserving Hamiltonian $H_3$ introduced in Eq.~\eqref{H_3} and the PXP model~\cite{Zlatko}, that appears in the context of quantum many-body scars~\cite{TurnerNatPhys,TurnerPRB,Choi2018,Shiraishi19}. Such relation has been already obtained in Ref.~\cite{Sanjay19} in the spin-$1/2$ version of the models we study (see e.g., Fig.~\ref{fig:fig4}) in the context of fractional quantum Hall.  The PXP model describes a chain of interacting Rydberg atoms~\cite{Bernien2017}, which in the limit of strong nearest-neighbor interactions is effectively described  by the spin-$1/2$ Hamiltonian
\begin{equation}\label{eq:PXP}
H_{{\text{\tiny PXP}}}=\sum_{n=1}^L P_{n-1}\sigma_n^xP_{n+1},
\end{equation}
where the projectors $P_n\equiv (1-\sigma_n^z)/2$, ensure that not two adjacent Rydberg atoms become simultaneously excited into the $\ket{\uparrow}$ state, a phenomenon known as Rydberg blockade.
%
%
Restricted to the lowest energy subspace with no adjacent excited states, the dimension of the constrained Hilbert space can be shown to be~\cite{TurnerPRB} $d_L=F_{L+2}$ for open (OBC) ---in the presence of additional boundary terms~\cite{OBCPXP}--- and $d_L=F_{L-1}+F_{L+1}$ for periodic boundary conditions (PBC), where $F_n$ is the $n$th Fibonacci number.
Note that in particular, this subspace contains the Néel states $\ket{\mathbb{Z}_2}=\ket{\downarrow \uparrow \downarrow \uparrow  \cdots}$ and $\ket{\mathbb{Z}'_2}=\ket{\uparrow \downarrow \uparrow \downarrow \cdots} $, whose atypical real-time dynamics has been experimentally realized~\cite{Bernien2017} and has been identified as a probe of the existence of quantum many-body scars~\cite{TurnerPRB}.

In the following we show that the dynamics of certain connected subspaces $\mathcal{H}_i$ discussed in the main text, are governed by the PXP Hamiltonian in Eq.~\eqref{eq:PXP} and identify the analogs to the Néel states in the fractonic language.
Let us consider states of the form
\begin{equation} \label{eq:Step0}
\ket{\mathbb{F}_2}=\ket{\dots \boxed{-\bm{+}-}\bm{-}\overset{2(k-1)}{\boxed{-\bm{+}-}}\overset{2k}{\bm{-}}\overset{2(k+1)}{\boxed{-\bm{+}-}}\bm{-}\boxed{-\bm{+}-}\dots},
\end{equation}
 with a $\ket{+}$ state on every fourth site separated by three $\ket{-}$'s.
In the following we fix the length of the chain to be a multiple of four, such that we contain an integer number of unit cells.
For OBC the dipole moment is given by $p(n_+)=L/2(1-L/2 +n_+)$, where $n_+$ is the location of the first $\ket{+}$ state starting from the left boundary.
Thus, the location $n_+$ labels different symmetry sectors containing the same spin pattern.
However, due to the periodicity of the configuration, there only exist four different dipole moments $p(n_+)$ containing such configuration.
When considering PBC, the dipole is defined modulo $L$.
Recalling that every local term $h_n$ in the Hamiltonian $H_3=\sum_nh_n$ takes the form
\begin{equation}\label{}
h_n\equiv S_{n-1}^+\big(S_{n}^-\big)^2S_{n+1}^++\text{H.c.},
\end{equation}
it is clear that the \emph{only non-trivial} local actions of $H_3$ on the state $\ket{\mathbb{F}_2}$, are those contained within the blocks shown in Eq.~\eqref{eq:Step0}.
After applying two local terms $h_n$ centered around the location of $\ket{+}$ states at sites $n=2(k\pm 1)$, $\ket{\mathbb{F}_2}$ becomes
\begin{equation} \label{eq:Step1}
\ket{\mathbb{F}_2}\longrightarrow h_{2(k-1)}h_{2(k+1)}\ket{\mathbb{F}_2}=\ket{\dots \boxed{-\bm{+}-}\bm{-} 0\overset{2(k-1)}{\bm{-}}\overset{2k}{\boxed{0\bm{-}0}}\overset{2(k+1)}{\bm{-}}0\bm{-}\boxed{-\bm{+}-}\dots}.
\end{equation}

Now the action on the intermediate site $2k$ becomes non-trivial
\begin{equation} \label{eq:Step2}h_{2(k-1)}h_{2(k+1)}\ket{\mathbb{F}_2} \longrightarrow h_{2k}h_{2(k-1)}h_{2(k+1)}\ket{\mathbb{F}_2}=\ket{\dots \boxed{-\bm{+}-}\bm{-} 0\overset{2(k-1)}{\bm{-}}\overset{2k}{\boxed{-\bm{+}-}}\overset{2(k+1)}{\bm{-}}0\bm{-}\boxed{-\bm{+}-}\dots}.
\end{equation}

One can then realize that only terms $h_{2k}$ centered around even sites generate non-trivial dynamics conditioned to the states on odd near sites, such that the only allowed local transition is $\ket{-+-} \leftrightarrow \ket{0-0}$.
Then, the restriction of the Hamiltonian $H_3$ to subspaces containing configurations of the form given by Eq.~\eqref{eq:Step0} becomes
\begin{equation} \label{eq:res1}
\left. H_{3}\right|_{\text{\tiny  PXP}}=\sum_{k=1}^{L/2}h_{2k}=4\sum_{k=1}^{L/2} \ket{-\bm{+}_{2k}-}\bra{0\bm{-}_{2k}0} +\text{H.c.}.
\end{equation}

Note also that there are never two $\ket{+}$ states in adjacent even sites, i.e., the local configuration $\ket{+}_{2k}\ket{+}_{2(k+1)}$ is not generated under the evolution of $H_3$.
This effectively implements the Rydberg blockade as imposed by the projectors in Eq.~\eqref{eq:PXP}.
With these observations in mind, we can construct a reversible map relating local spin-$1$ configurations centered around even sites $\{2k-1,2k,2k+1\} $, to spin-$1/2$ configurations on even sites $\{2(k-1),2k,2(k+1)\} $ in the PXP model via
\begin{align}
&\ket{0-0}  \leftrightarrow \ket{\downarrow \downarrow \downarrow},    \hspace{25pt} \ket{-+-} \leftrightarrow \ket{\downarrow \uparrow \downarrow}, \hspace{25pt} \ket{0--}  \leftrightarrow \ket{\downarrow \downarrow \uparrow}, 
\\  &\ket{--0}  \leftrightarrow \ket{\uparrow \downarrow \downarrow},    \hspace{25pt} \ket{---}  \leftrightarrow \ket{\uparrow \downarrow \uparrow},
\end{align}
such that Eq.~\eqref{eq:res1} becomes
\begin{equation}
\left. H_{3}\right|_{\text{\tiny  PXP}}=4\sum_{k=1}^{L/2}\ket{\downarrow \uparrow \downarrow}\bra{\downarrow \downarrow\downarrow} +H.c.= 4\sum_{k=1}^{L/2}\ket{\downarrow}\bra{\downarrow}_{2(k-1)}\otimes \ket{\uparrow}\bra{\downarrow}_{2k}\otimes\ket{\downarrow}\bra{\downarrow}_{2(k+1)} +H.c.=4H_{\text{\tiny PXP}},
\end{equation}
i.e., the restriction of $H_3$ into this family of connected subspaces becomes equivalent to a PXP model on a chain of length $L/2$ up to a factor of 4.
Thus, there exist eight different symmetry sectors (the other four subspaces are obtained applying the $\Pi^x$ parity symmetry (see App.~\ref{app:Symmetries}) to the configuration $\ket{\mathbb{F}_2}$), whose evolution is governed by the PXP Hamiltonian.
This explicitly shows that quantum many-body scars appear in the dipole conserving Hamiltonian $H_3$, similarly to Ref.~\cite{Sanjay19}.

\section{Operator spreading of $S_0^z(t)$}\label{app:rhoR}

Here we consider another measure of localization, that contains complementary information about the Heisenberg picture evolution of the charge density operator $S_0^z(t)$ compared to its auto-correlation function. In particular we look at how $S_0^z$ spreads out in the space of all possible operators, becoming a complicated superposition of many operators, and how its spatial support grows in time.

In order to do this, we first need to introduce a local basis in the space of operators acting on a single site of the spin chain. For the spin-$1$ models we consider, such a basis constists of 9 linearly independent operators that span the entire space of on-site operators. A possible choice is given by the 8 Gell-Mann matrices, together with the identity $1\!\!1$. Let us denote these as $\lambda^a$ for $a=0,\ldots,8$, where $\lambda^0 \equiv 1\!\!1$. A basis of operators on the entire chain is then given by products of such local basis elements of the form $\lambda^{\vec{a}}\equiv \bigotimes_{n=-N/2}^{N/2}\lambda_{n}^{a_n}$, labeled by a list of $N$ indices $\{a_n\}$. These operator string form an orthonormal basis in the Hilbert space of operators with respect to the Frobenius inner product $\langle A,B \rangle \equiv \text{tr}(A^\dagger B) / 3^L$ where $A$ and $B$ are two arbitrary operators. 

Given such a basis, one can always expand the time evolved operator as 
\begin{equation}
S_0^z(t) = \sum_{\vec{a}} c_{\vec{a}}(t) \lambda^{\vec{a}}.
\end{equation}
The coefficients $c_{\vec{a}}(t)$ characterize how $S_0^z(t)$ spreads out in the space of all possible operators. In particular, focusing on spatial spreading, it is useful to classify the basis strings $\lambda^{\vec{a}}$ according to their right endpoints (assuming open boundary conditions), i.e., the rightmost site $n$ such that $\lambda^{a_n} \neq 1\!\!1$ but $\lambda^{a_{m>n}} = 1\!\!1$. Denoting this site by $\text{RHS}(\vec{a})$ we can define the \emph{right endpoint density} of $S_0^z$ at time $t$ as~\cite{Roberts2015,Keyserlingk2018,Nahum2018}
\begin{equation}
\rho_R(n,t) \equiv \sum_{\text{RHS}(\vec{a})=n} |c_{\vec{a}}(t)|^2.
\end{equation}
At time $t=0$ this is a delta function at the initial position of the operator, $\rho_R(n,0) = \delta_{n0}$. During time evolution, as the support of $S_0^z(t)$ increases, $\rho_R(n,t)$ moves to the right, ballistically for generic clean systems. At the same time, its value near the origin decays to zero, exponentially when symmetries are not present~\cite{Keyserlingk2018}, and as a power law when the operator is a conserved density~\cite{Rakovszky2018,Khemani2018}. A possible alternative measure of localized behavior is therfore to look at the spreading of the right endpoint density and look for a finite weight remaining near the origin at infinite times, even in the thermodynamic limit.

We first consider the evolution of $\rho_R(t)$ in random circuits, first  with $3$- and then with $4$-site gates. In order to evaluate $\rho_R(n,t)$ we represent $S_0^z(t)$ as a matrix product operator~\cite{MurgReview} (MPO) and apply the random gates to that to evolve it in time. In order to simplify the calculations, we consider slightly modified circuit geometries, which allow us to use the well known time-evolving block decimation (TEBD) algorithm, after blocking pairs of sites together~\cite{VidalTEBD}. 

\begin{figure}
	\centering
	\includegraphics[width=1.0\linewidth]{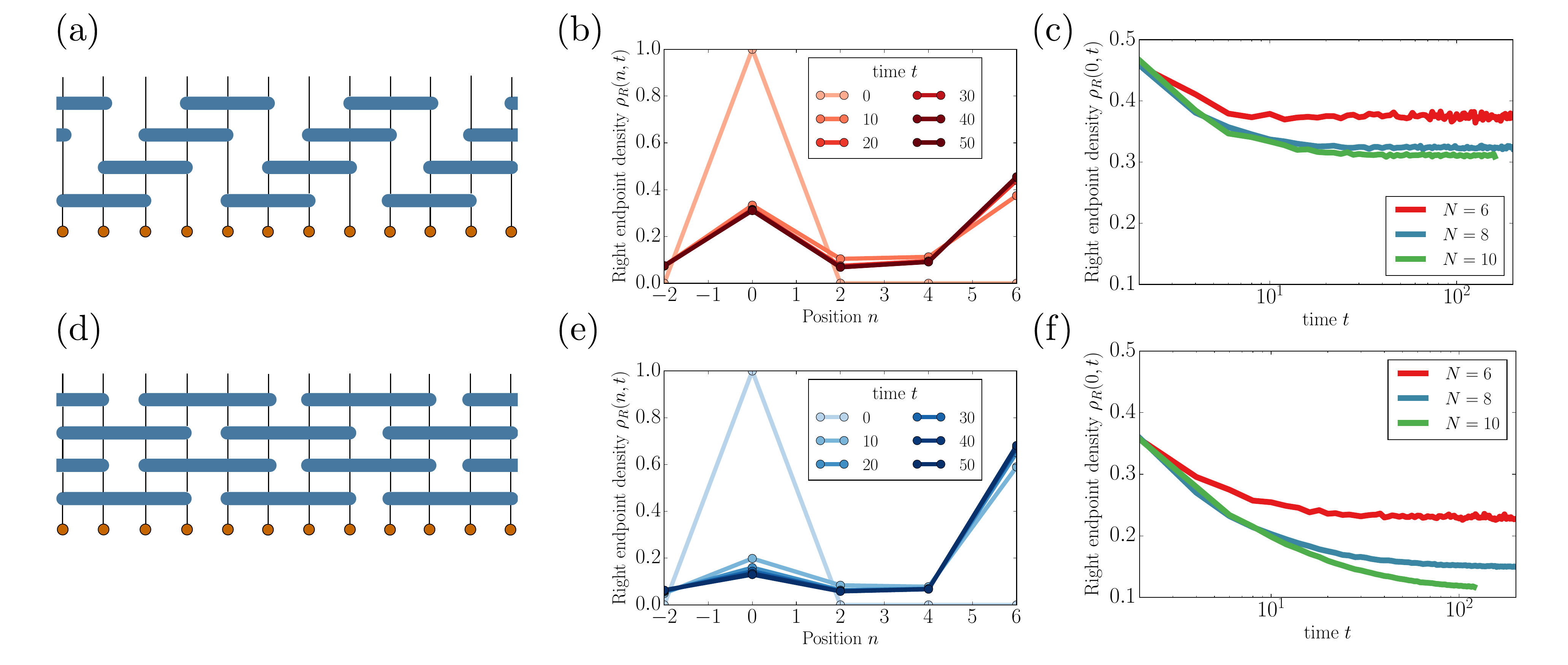}
	\caption{Operator spreading of $S_0^z(t)$ for random circuits with gate sizes $l=3$ and $4$. Panels (a) and (d) show the circuit geometries, slightly modified from the ones in the main text in order to ease numerical calculations. (b) and (e) show the profile of right endpoint weights $\rho_R(n,t)$ at different times for a 10-site chain, for $l=3$ and $l=4$ respectively. Both have a peak near the origin, but in the former case it is much larger and stops decaying after a few time steps, while in the latter case it keeps decaying to longer times. Finite size flow of the size of the peaks as a function of time, shown in (c) and (f) indicates that while for $\ell=3$ the system saturates to a finite value, this is not the case for $\ell=4$, where the long-time value scales to zero for large system sizes. }
	\label{fig:fig2RUC}
\end{figure}

Our numerics only allow us to access small systems of size $N=6,8,10$. To compute the spreading of $\rho_R(n,t)$, we place an operator $S^z$ on the third site from the left end of the system and calculate $\rho_R(n,t)$ at different positions and times. For a circuit made out of 3-site gates, we find a persistent peak near the original position, whose size decays only slightly with system size (Fig.~\ref{fig:fig2RUC}(a)-(c)). For the circuit with gate-size $l=4$ on the other hand, we observe a much smaller peak, which keeps decreasing until finite size effects kick in, similar to the behavior observed for the autocorrelator in the main text, and consistent with the perdiction that in the thermodynamic limit the peak would eventually disappear (Fig.~\ref{fig:fig2RUC}(d)-(f)). We also observe a larger peak at the rightmost site, where most of the operator weight accumulates at long times. 

\begin{figure}
	\centering
	\includegraphics[width=0.75\linewidth]{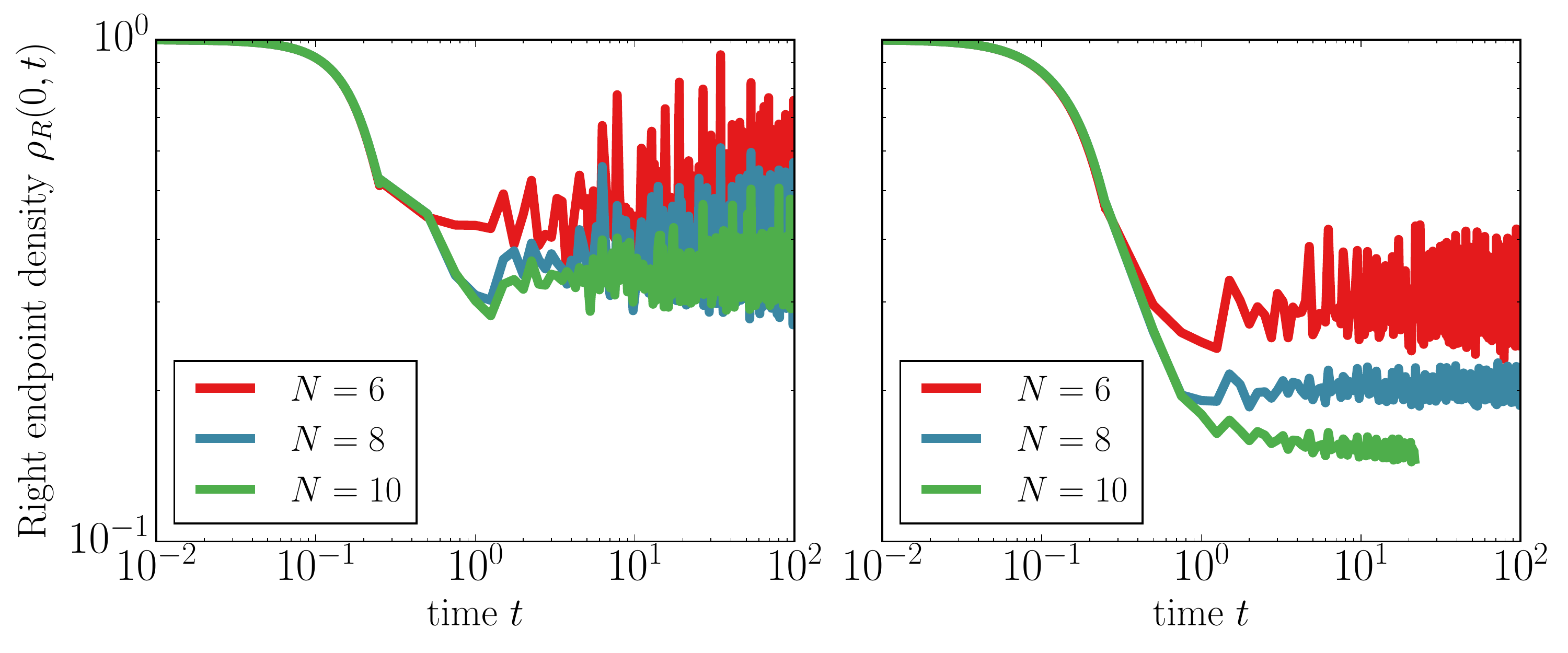}
	\caption{Height of the peak in $\rho_R$ of $S_0^z(t)$ obtained for Hamiltonians $H_3$ (left) and $H_3 + H_4'$ (right).}
	\label{fig:rhoRHam}
\end{figure}

The same difference in behavior between 3-site and 4-site interactions is also present in the Hamiltonian case. For $H_3$ we find that the peak in $\rho_R(0,t)$ is almost independent of system size, in agreement with the non-ergodic behavior observed in the autocorrelator in the main text. This is shown in the left panel of Fig.~\ref{fig:rhoRHam}. This behavior changes, however, once we add 4-site terms to the Hamiltonian. In particular we consider the perturbation
\begin{align} 
\label{H_4p}  H_4'= &-\sum_{n\in 2\mathbb{Z}}{\Big[ S_n^+S_{n+1}^-S_{n+2}^-S_{n+3}^++H.c.\Big]}.
\end{align}
This is the same as in Eq.~\eqref{H_4}, except that only terms with even $n$ are present. This is done in order to simplify numerical calculations (making the Hamiltonian nearest neighbor after blocking pairs of neighboring sites together). We expect that if $H_3 + H_4'$ does not exhibit a presistent peak in $\rho_R$, then neither should $H_3 + H_4$, therefore it is enough to show its absence in the former case. This is indeed what we find as shown in the right panel of Fig.~\ref{fig:rhoRHam}

\vspace{1cm}
\twocolumngrid
\bibliography{1DfractonsRef}

\begin{thebibliography}{115}%
\makeatletter
\providecommand \@ifxundefined [1]{%
 \@ifx{#1\undefined}
}%
\providecommand \@ifnum [1]{%
 \ifnum #1\expandafter \@firstoftwo
 \else \expandafter \@secondoftwo
 \fi
}%
\providecommand \@ifx [1]{%
 \ifx #1\expandafter \@firstoftwo
 \else \expandafter \@secondoftwo
 \fi
}%
\providecommand \natexlab [1]{#1}%
\providecommand \enquote  [1]{``#1''}%
\providecommand \bibnamefont  [1]{#1}%
\providecommand \bibfnamefont [1]{#1}%
\providecommand \citenamefont [1]{#1}%
\providecommand \href@noop [0]{\@secondoftwo}%
\providecommand \href [0]{\begingroup \@sanitize@url \@href}%
\providecommand \@href[1]{\@@startlink{#1}\@@href}%
\providecommand \@@href[1]{\endgroup#1\@@endlink}%
\providecommand \@sanitize@url [0]{\catcode `\\12\catcode `\$12\catcode
  `\&12\catcode `\#12\catcode `\^12\catcode `\_12\catcode `\%12\relax}%
\providecommand \@@startlink[1]{}%
\providecommand \@@endlink[0]{}%
\providecommand \url  [0]{\begingroup\@sanitize@url \@url }%
\providecommand \@url [1]{\endgroup\@href {#1}{\urlprefix }}%
\providecommand \urlprefix  [0]{URL }%
\providecommand \Eprint [0]{\href }%
\providecommand \doibase [0]{http://dx.doi.org/}%
\providecommand \selectlanguage [0]{\@gobble}%
\providecommand \bibinfo  [0]{\@secondoftwo}%
\providecommand \bibfield  [0]{\@secondoftwo}%
\providecommand \translation [1]{[#1]}%
\providecommand \BibitemOpen [0]{}%
\providecommand \bibitemStop [0]{}%
\providecommand \bibitemNoStop [0]{.\EOS\space}%
\providecommand \EOS [0]{\spacefactor3000\relax}%
\providecommand \BibitemShut  [1]{\csname bibitem#1\endcsname}%
\let\auto@bib@innerbib\@empty
\bibitem [{\citenamefont {Gring}\ \emph {et~al.}(2012)\citenamefont {Gring},
  \citenamefont {Kuhnert}, \citenamefont {Langen}, \citenamefont {Kitagawa},
  \citenamefont {Rauer}, \citenamefont {Schreitl}, \citenamefont {Mazets},
  \citenamefont {Smith}, \citenamefont {Demler},\ and\ \citenamefont
  {Schmiedmayer}}]{Gring1318}%
  \BibitemOpen
  \bibfield  {author} {\bibinfo {author} {\bibfnamefont {M.}~\bibnamefont
  {Gring}}, \bibinfo {author} {\bibfnamefont {M.}~\bibnamefont {Kuhnert}},
  \bibinfo {author} {\bibfnamefont {T.}~\bibnamefont {Langen}}, \bibinfo
  {author} {\bibfnamefont {T.}~\bibnamefont {Kitagawa}}, \bibinfo {author}
  {\bibfnamefont {B.}~\bibnamefont {Rauer}}, \bibinfo {author} {\bibfnamefont
  {M.}~\bibnamefont {Schreitl}}, \bibinfo {author} {\bibfnamefont
  {I.}~\bibnamefont {Mazets}}, \bibinfo {author} {\bibfnamefont {D.~Adu}\
  \bibnamefont {Smith}}, \bibinfo {author} {\bibfnamefont {E.}~\bibnamefont
  {Demler}}, \ and\ \bibinfo {author} {\bibfnamefont {J.}~\bibnamefont
  {Schmiedmayer}},\ }\bibfield  {title} {\enquote {\bibinfo {title} {Relaxation
  and prethermalization in an isolated quantum system},}\ }\href {\doibase
  10.1126/science.1224953} {\bibfield  {journal} {\bibinfo  {journal}
  {Science}\ }\textbf {\bibinfo {volume} {337}},\ \bibinfo {pages} {1318--1322}
  (\bibinfo {year} {2012})},\ \Eprint
  {http://arxiv.org/abs/http://science.sciencemag.org/content/337/6100/1318.full.pdf}
  {http://science.sciencemag.org/content/337/6100/1318.full.pdf} \BibitemShut
  {NoStop}%
\bibitem [{\citenamefont {Hild}\ \emph {et~al.}(2014)\citenamefont {Hild},
  \citenamefont {Fukuhara}, \citenamefont {Schau\ss{}}, \citenamefont {Zeiher},
  \citenamefont {Knap}, \citenamefont {Demler}, \citenamefont {Bloch},\ and\
  \citenamefont {Gross}}]{Hild2014}%
  \BibitemOpen
  \bibfield  {author} {\bibinfo {author} {\bibfnamefont {Sebastian}\
  \bibnamefont {Hild}}, \bibinfo {author} {\bibfnamefont {Takeshi}\
  \bibnamefont {Fukuhara}}, \bibinfo {author} {\bibfnamefont {Peter}\
  \bibnamefont {Schau\ss{}}}, \bibinfo {author} {\bibfnamefont {Johannes}\
  \bibnamefont {Zeiher}}, \bibinfo {author} {\bibfnamefont {Michael}\
  \bibnamefont {Knap}}, \bibinfo {author} {\bibfnamefont {Eugene}\ \bibnamefont
  {Demler}}, \bibinfo {author} {\bibfnamefont {Immanuel}\ \bibnamefont
  {Bloch}}, \ and\ \bibinfo {author} {\bibfnamefont {Christian}\ \bibnamefont
  {Gross}},\ }\bibfield  {title} {\enquote {\bibinfo {title}
  {Far-from-equilibrium spin transport in heisenberg quantum magnets},}\ }\href
  {\doibase 10.1103/PhysRevLett.113.147205} {\bibfield  {journal} {\bibinfo
  {journal} {Phys. Rev. Lett.}\ }\textbf {\bibinfo {volume} {113}},\ \bibinfo
  {pages} {147205} (\bibinfo {year} {2014})}\BibitemShut {NoStop}%
\bibitem [{\citenamefont {Brown}\ \emph {et~al.}(2015)\citenamefont {Brown},
  \citenamefont {Wyllie}, \citenamefont {Koller}, \citenamefont {Goldschmidt},
  \citenamefont {Foss-Feig},\ and\ \citenamefont {Porto}}]{Brown540}%
  \BibitemOpen
  \bibfield  {author} {\bibinfo {author} {\bibfnamefont {R.~C.}\ \bibnamefont
  {Brown}}, \bibinfo {author} {\bibfnamefont {R.}~\bibnamefont {Wyllie}},
  \bibinfo {author} {\bibfnamefont {S.~B.}\ \bibnamefont {Koller}}, \bibinfo
  {author} {\bibfnamefont {E.~A.}\ \bibnamefont {Goldschmidt}}, \bibinfo
  {author} {\bibfnamefont {M.}~\bibnamefont {Foss-Feig}}, \ and\ \bibinfo
  {author} {\bibfnamefont {J.~V.}\ \bibnamefont {Porto}},\ }\bibfield  {title}
  {\enquote {\bibinfo {title} {Two-dimensional superexchange-mediated
  magnetization dynamics in an optical lattice},}\ }\href {\doibase
  10.1126/science.aaa1385} {\bibfield  {journal} {\bibinfo  {journal}
  {Science}\ }\textbf {\bibinfo {volume} {348}},\ \bibinfo {pages} {540--544}
  (\bibinfo {year} {2015})},\ \Eprint
  {http://arxiv.org/abs/http://science.sciencemag.org/content/348/6234/540.full.pdf}
  {http://science.sciencemag.org/content/348/6234/540.full.pdf} \BibitemShut
  {NoStop}%
\bibitem [{\citenamefont {Kaufman}\ \emph {et~al.}(2016)\citenamefont
  {Kaufman}, \citenamefont {Tai}, \citenamefont {Lukin}, \citenamefont
  {Rispoli}, \citenamefont {Schittko}, \citenamefont {Preiss},\ and\
  \citenamefont {Greiner}}]{Kaufman794}%
  \BibitemOpen
  \bibfield  {author} {\bibinfo {author} {\bibfnamefont {Adam~M.}\ \bibnamefont
  {Kaufman}}, \bibinfo {author} {\bibfnamefont {M.~Eric}\ \bibnamefont {Tai}},
  \bibinfo {author} {\bibfnamefont {Alexander}\ \bibnamefont {Lukin}}, \bibinfo
  {author} {\bibfnamefont {Matthew}\ \bibnamefont {Rispoli}}, \bibinfo {author}
  {\bibfnamefont {Robert}\ \bibnamefont {Schittko}}, \bibinfo {author}
  {\bibfnamefont {Philipp~M.}\ \bibnamefont {Preiss}}, \ and\ \bibinfo {author}
  {\bibfnamefont {Markus}\ \bibnamefont {Greiner}},\ }\bibfield  {title}
  {\enquote {\bibinfo {title} {Quantum thermalization through entanglement in
  an isolated many-body system},}\ }\href {\doibase 10.1126/science.aaf6725}
  {\bibfield  {journal} {\bibinfo  {journal} {Science}\ }\textbf {\bibinfo
  {volume} {353}},\ \bibinfo {pages} {794--800} (\bibinfo {year}
  {2016})}\BibitemShut {NoStop}%
\bibitem [{\citenamefont {Tang}\ \emph {et~al.}(2018)\citenamefont {Tang},
  \citenamefont {Kao}, \citenamefont {Li}, \citenamefont {Seo}, \citenamefont
  {Mallayya}, \citenamefont {Rigol}, \citenamefont {Gopalakrishnan},\ and\
  \citenamefont {Lev}}]{Yijun2018}%
  \BibitemOpen
  \bibfield  {author} {\bibinfo {author} {\bibfnamefont {Yijun}\ \bibnamefont
  {Tang}}, \bibinfo {author} {\bibfnamefont {Wil}\ \bibnamefont {Kao}},
  \bibinfo {author} {\bibfnamefont {Kuan-Yu}\ \bibnamefont {Li}}, \bibinfo
  {author} {\bibfnamefont {Sangwon}\ \bibnamefont {Seo}}, \bibinfo {author}
  {\bibfnamefont {Krishnanand}\ \bibnamefont {Mallayya}}, \bibinfo {author}
  {\bibfnamefont {Marcos}\ \bibnamefont {Rigol}}, \bibinfo {author}
  {\bibfnamefont {Sarang}\ \bibnamefont {Gopalakrishnan}}, \ and\ \bibinfo
  {author} {\bibfnamefont {Benjamin~L.}\ \bibnamefont {Lev}},\ }\bibfield
  {title} {\enquote {\bibinfo {title} {Thermalization near integrability in a
  dipolar quantum newton's cradle},}\ }\href {\doibase
  10.1103/PhysRevX.8.021030} {\bibfield  {journal} {\bibinfo  {journal} {Phys.
  Rev. X}\ }\textbf {\bibinfo {volume} {8}},\ \bibinfo {pages} {021030}
  (\bibinfo {year} {2018})}\BibitemShut {NoStop}%
\bibitem [{\citenamefont {Brydges}\ \emph {et~al.}(2018)\citenamefont
  {Brydges}, \citenamefont {Elben}, \citenamefont {Jurcevic}, \citenamefont
  {Vermersch}, \citenamefont {Maier}, \citenamefont {Lanyon}, \citenamefont
  {Zoller}, \citenamefont {Blatt},\ and\ \citenamefont {Roos}}]{Brydges2018}%
  \BibitemOpen
  \bibfield  {author} {\bibinfo {author} {\bibfnamefont {Tiff}\ \bibnamefont
  {Brydges}}, \bibinfo {author} {\bibfnamefont {Andreas}\ \bibnamefont
  {Elben}}, \bibinfo {author} {\bibfnamefont {Petar}\ \bibnamefont {Jurcevic}},
  \bibinfo {author} {\bibfnamefont {Benoît}\ \bibnamefont {Vermersch}},
  \bibinfo {author} {\bibfnamefont {Christine}\ \bibnamefont {Maier}}, \bibinfo
  {author} {\bibfnamefont {Ben~P.}\ \bibnamefont {Lanyon}}, \bibinfo {author}
  {\bibfnamefont {Peter}\ \bibnamefont {Zoller}}, \bibinfo {author}
  {\bibfnamefont {Rainer}\ \bibnamefont {Blatt}}, \ and\ \bibinfo {author}
  {\bibfnamefont {Christian~F.}\ \bibnamefont {Roos}},\ }\href@noop {}
  {\enquote {\bibinfo {title} {Probing entanglement entropy via randomized
  measurements},}\ } (\bibinfo {year} {2018}),\ \Eprint
  {http://arxiv.org/abs/arXiv:1806.05747} {arXiv:1806.05747} \BibitemShut
  {NoStop}%
\bibitem [{\citenamefont {{D'Alessio}}\ \emph {et~al.}(2016)\citenamefont
  {{D'Alessio}}, \citenamefont {{Kafri}}, \citenamefont {{Polkovnikov}},\ and\
  \citenamefont {{Rigol}}}]{ETHreviewRigol16}%
  \BibitemOpen
  \bibfield  {author} {\bibinfo {author} {\bibfnamefont {L.}~\bibnamefont
  {{D'Alessio}}}, \bibinfo {author} {\bibfnamefont {Y.}~\bibnamefont
  {{Kafri}}}, \bibinfo {author} {\bibfnamefont {A.}~\bibnamefont
  {{Polkovnikov}}}, \ and\ \bibinfo {author} {\bibfnamefont {M.}~\bibnamefont
  {{Rigol}}},\ }\bibfield  {title} {\enquote {\bibinfo {title} {{From quantum
  chaos and eigenstate thermalization to statistical mechanics and
  thermodynamics}},}\ }\href {\doibase 10.1080/00018732.2016.1198134}
  {\bibfield  {journal} {\bibinfo  {journal} {Advances in Physics}\ }\textbf
  {\bibinfo {volume} {65}},\ \bibinfo {pages} {239--362} (\bibinfo {year}
  {2016})},\ \Eprint {http://arxiv.org/abs/1509.06411} {arXiv:1509.06411
  [cond-mat.stat-mech]} \BibitemShut {NoStop}%
\bibitem [{\citenamefont {Gogolin}\ and\ \citenamefont
  {Eisert}(2016)}]{GogolinReview}%
  \BibitemOpen
  \bibfield  {author} {\bibinfo {author} {\bibfnamefont {Christian}\
  \bibnamefont {Gogolin}}\ and\ \bibinfo {author} {\bibfnamefont {Jens}\
  \bibnamefont {Eisert}},\ }\bibfield  {title} {\enquote {\bibinfo {title}
  {Equilibration, thermalisation, and the emergence of statistical mechanics in
  closed quantum systems},}\ }\href
  {http://stacks.iop.org/0034-4885/79/i=5/a=056001} {\bibfield  {journal}
  {\bibinfo  {journal} {Reports on Progress in Physics}\ }\textbf {\bibinfo
  {volume} {79}},\ \bibinfo {pages} {056001} (\bibinfo {year}
  {2016})}\BibitemShut {NoStop}%
\bibitem [{\citenamefont {Meinert}\ \emph {et~al.}(2017)\citenamefont
  {Meinert}, \citenamefont {Knap}, \citenamefont {Kirilov}, \citenamefont
  {Jag-Lauber}, \citenamefont {Zvonarev}, \citenamefont {Demler},\ and\
  \citenamefont {N{\"a}gerl}}]{Meinert945}%
  \BibitemOpen
  \bibfield  {author} {\bibinfo {author} {\bibfnamefont {Florian}\ \bibnamefont
  {Meinert}}, \bibinfo {author} {\bibfnamefont {Michael}\ \bibnamefont {Knap}},
  \bibinfo {author} {\bibfnamefont {Emil}\ \bibnamefont {Kirilov}}, \bibinfo
  {author} {\bibfnamefont {Katharina}\ \bibnamefont {Jag-Lauber}}, \bibinfo
  {author} {\bibfnamefont {Mikhail~B.}\ \bibnamefont {Zvonarev}}, \bibinfo
  {author} {\bibfnamefont {Eugene}\ \bibnamefont {Demler}}, \ and\ \bibinfo
  {author} {\bibfnamefont {Hanns-Christoph}\ \bibnamefont {N{\"a}gerl}},\
  }\bibfield  {title} {\enquote {\bibinfo {title} {Bloch oscillations in the
  absence of a lattice},}\ }\href {\doibase 10.1126/science.aah6616} {\bibfield
   {journal} {\bibinfo  {journal} {Science}\ }\textbf {\bibinfo {volume}
  {356}},\ \bibinfo {pages} {945--948} (\bibinfo {year} {2017})},\ \Eprint
  {http://arxiv.org/abs/https://science.sciencemag.org/content/356/6341/945.full.pdf}
  {https://science.sciencemag.org/content/356/6341/945.full.pdf} \BibitemShut
  {NoStop}%
\bibitem [{\citenamefont {Deutsch}(1991)}]{Deutsch91}%
  \BibitemOpen
  \bibfield  {author} {\bibinfo {author} {\bibfnamefont {J.~M.}\ \bibnamefont
  {Deutsch}},\ }\bibfield  {title} {\enquote {\bibinfo {title} {Quantum
  statistical mechanics in a closed system},}\ }\href {\doibase
  10.1103/PhysRevA.43.2046} {\bibfield  {journal} {\bibinfo  {journal} {Phys.
  Rev. A}\ }\textbf {\bibinfo {volume} {43}},\ \bibinfo {pages} {2046--2049}
  (\bibinfo {year} {1991})}\BibitemShut {NoStop}%
\bibitem [{\citenamefont {Srednicki}(1994)}]{Srednicki94}%
  \BibitemOpen
  \bibfield  {author} {\bibinfo {author} {\bibfnamefont {Mark}\ \bibnamefont
  {Srednicki}},\ }\bibfield  {title} {\enquote {\bibinfo {title} {Chaos and
  quantum thermalization},}\ }\href {\doibase 10.1103/PhysRevE.50.888}
  {\bibfield  {journal} {\bibinfo  {journal} {Phys. Rev. E}\ }\textbf {\bibinfo
  {volume} {50}},\ \bibinfo {pages} {888--901} (\bibinfo {year}
  {1994})}\BibitemShut {NoStop}%
\bibitem [{\citenamefont {Rigol}\ \emph {et~al.}(2008)\citenamefont {Rigol},
  \citenamefont {Dunjko},\ and\ \citenamefont {Olshanii}}]{Rigol2008}%
  \BibitemOpen
  \bibfield  {author} {\bibinfo {author} {\bibfnamefont {Marcos}\ \bibnamefont
  {Rigol}}, \bibinfo {author} {\bibfnamefont {Vanja}\ \bibnamefont {Dunjko}}, \
  and\ \bibinfo {author} {\bibfnamefont {Maxim}\ \bibnamefont {Olshanii}},\
  }\bibfield  {title} {\enquote {\bibinfo {title} {Thermalization and its
  mechanism for generic isolated quantum systems},}\ }\href {\doibase
  10.1038/nature06838} {\bibfield  {journal} {\bibinfo  {journal} {Nature}\
  }\textbf {\bibinfo {volume} {452}},\ \bibinfo {pages} {854--8} (\bibinfo
  {year} {2008})}\BibitemShut {NoStop}%
\bibitem [{\citenamefont {Kim}\ \emph {et~al.}(2014)\citenamefont {Kim},
  \citenamefont {Ikeda},\ and\ \citenamefont {Huse}}]{Kim2014}%
  \BibitemOpen
  \bibfield  {author} {\bibinfo {author} {\bibfnamefont {Hyungwon}\
  \bibnamefont {Kim}}, \bibinfo {author} {\bibfnamefont {Tatsuhiko~N.}\
  \bibnamefont {Ikeda}}, \ and\ \bibinfo {author} {\bibfnamefont {David~A.}\
  \bibnamefont {Huse}},\ }\bibfield  {title} {\enquote {\bibinfo {title}
  {Testing whether all eigenstates obey the eigenstate thermalization
  hypothesis},}\ }\href {\doibase 10.1103/PhysRevE.90.052105} {\bibfield
  {journal} {\bibinfo  {journal} {Phys. Rev. E}\ }\textbf {\bibinfo {volume}
  {90}},\ \bibinfo {pages} {052105} (\bibinfo {year} {2014})}\BibitemShut
  {NoStop}%
\bibitem [{\citenamefont {Rigol}\ \emph {et~al.}(2007)\citenamefont {Rigol},
  \citenamefont {Dunjko}, \citenamefont {Yurovsky},\ and\ \citenamefont
  {Olshanii}}]{Rigol2007}%
  \BibitemOpen
  \bibfield  {author} {\bibinfo {author} {\bibfnamefont {Marcos}\ \bibnamefont
  {Rigol}}, \bibinfo {author} {\bibfnamefont {Vanja}\ \bibnamefont {Dunjko}},
  \bibinfo {author} {\bibfnamefont {Vladimir}\ \bibnamefont {Yurovsky}}, \ and\
  \bibinfo {author} {\bibfnamefont {Maxim}\ \bibnamefont {Olshanii}},\
  }\bibfield  {title} {\enquote {\bibinfo {title} {Relaxation in a completely
  integrable many-body quantum system: An ab initio study of the dynamics of
  the highly excited states of 1d lattice hard-core bosons},}\ }\href {\doibase
  10.1103/PhysRevLett.98.050405} {\bibfield  {journal} {\bibinfo  {journal}
  {Phys. Rev. Lett.}\ }\textbf {\bibinfo {volume} {98}},\ \bibinfo {pages}
  {050405} (\bibinfo {year} {2007})}\BibitemShut {NoStop}%
\bibitem [{\citenamefont {Kinoshita}\ \emph {et~al.}(2006)\citenamefont
  {Kinoshita}, \citenamefont {Wenger},\ and\ \citenamefont
  {S~Weiss}}]{Kinoshita2006}%
  \BibitemOpen
  \bibfield  {author} {\bibinfo {author} {\bibfnamefont {Toshiya}\ \bibnamefont
  {Kinoshita}}, \bibinfo {author} {\bibfnamefont {Trevor}\ \bibnamefont
  {Wenger}}, \ and\ \bibinfo {author} {\bibfnamefont {David}\ \bibnamefont
  {S~Weiss}},\ }\bibfield  {title} {\enquote {\bibinfo {title} {A quantum
  newton's cradle},}\ }\href {\doibase 10.1038/nature04693} {\bibfield
  {journal} {\bibinfo  {journal} {Nature}\ }\textbf {\bibinfo {volume} {440}},\
  \bibinfo {pages} {900--3} (\bibinfo {year} {2006})}\BibitemShut {NoStop}%
\bibitem [{\citenamefont {Basko}\ \emph
  {et~al.}(2006{\natexlab{a}})\citenamefont {Basko}, \citenamefont {Aleiner},\
  and\ \citenamefont {Altshuler}}]{Basko06}%
  \BibitemOpen
  \bibfield  {author} {\bibinfo {author} {\bibfnamefont {D.~M.}\ \bibnamefont
  {Basko}}, \bibinfo {author} {\bibfnamefont {I.~L.}\ \bibnamefont {Aleiner}},
  \ and\ \bibinfo {author} {\bibfnamefont {B.~L.}\ \bibnamefont {Altshuler}},\
  }\bibfield  {title} {\enquote {\bibinfo {title} {Metal-insulator transition
  in a weakly interacting many-electron system with localized single-particle
  states},}\ }\href {\doibase 10.1016/j.aop.2005.11.014} {\bibfield  {journal}
  {\bibinfo  {journal} {Annals of Physics}\ }\textbf {\bibinfo {volume}
  {321}},\ \bibinfo {pages} {1126--1205} (\bibinfo {year}
  {2006}{\natexlab{a}})}\BibitemShut {NoStop}%
\bibitem [{\citenamefont {Nandkishore}\ and\ \citenamefont
  {Huse}(2015)}]{Nandkishore14}%
  \BibitemOpen
  \bibfield  {author} {\bibinfo {author} {\bibfnamefont {Rahul}\ \bibnamefont
  {Nandkishore}}\ and\ \bibinfo {author} {\bibfnamefont {David~A.}\
  \bibnamefont {Huse}},\ }\bibfield  {title} {\enquote {\bibinfo {title}
  {Many-body localization and thermalization in quantum statistical
  mechanics},}\ }\href {\doibase 10.1146/annurev-conmatphys-031214-014726}
  {\bibfield  {journal} {\bibinfo  {journal} {Annual Review of Condensed Matter
  Physics}\ }\textbf {\bibinfo {volume} {6}},\ \bibinfo {pages} {15--38}
  (\bibinfo {year} {2015})},\ \Eprint
  {http://arxiv.org/abs/http://dx.doi.org/10.1146/annurev-conmatphys-031214-014726}
  {http://dx.doi.org/10.1146/annurev-conmatphys-031214-014726} \BibitemShut
  {NoStop}%
\bibitem [{\citenamefont {Altman}\ and\ \citenamefont
  {Vosk}(2015)}]{AltmanVosk}%
  \BibitemOpen
  \bibfield  {author} {\bibinfo {author} {\bibfnamefont {Ehud}\ \bibnamefont
  {Altman}}\ and\ \bibinfo {author} {\bibfnamefont {Ronen}\ \bibnamefont
  {Vosk}},\ }\bibfield  {title} {\enquote {\bibinfo {title} {Universal dynamics
  and renormalization in many-body-localized systems},}\ }\href {\doibase
  10.1146/annurev-conmatphys-031214-014701} {\bibfield  {journal} {\bibinfo
  {journal} {Annual Review of Condensed Matter Physics}\ }\textbf {\bibinfo
  {volume} {6}},\ \bibinfo {pages} {383--409} (\bibinfo {year} {2015})},\
  \Eprint
  {http://arxiv.org/abs/https://doi.org/10.1146/annurev-conmatphys-031214-014701}
  {https://doi.org/10.1146/annurev-conmatphys-031214-014701} \BibitemShut
  {NoStop}%
\bibitem [{\citenamefont {Schreiber}\ \emph {et~al.}(2015)\citenamefont
  {Schreiber}, \citenamefont {Hodgman}, \citenamefont {Bordia}, \citenamefont
  {L{\"u}schen}, \citenamefont {Fischer}, \citenamefont {Vosk}, \citenamefont
  {Altman}, \citenamefont {Schneider},\ and\ \citenamefont
  {Bloch}}]{Schreiber2015}%
  \BibitemOpen
  \bibfield  {author} {\bibinfo {author} {\bibfnamefont {Michael}\ \bibnamefont
  {Schreiber}}, \bibinfo {author} {\bibfnamefont {Sean~S.}\ \bibnamefont
  {Hodgman}}, \bibinfo {author} {\bibfnamefont {Pranjal}\ \bibnamefont
  {Bordia}}, \bibinfo {author} {\bibfnamefont {Henrik~P.}\ \bibnamefont
  {L{\"u}schen}}, \bibinfo {author} {\bibfnamefont {Mark~H.}\ \bibnamefont
  {Fischer}}, \bibinfo {author} {\bibfnamefont {Ronen}\ \bibnamefont {Vosk}},
  \bibinfo {author} {\bibfnamefont {Ehud}\ \bibnamefont {Altman}}, \bibinfo
  {author} {\bibfnamefont {Ulrich}\ \bibnamefont {Schneider}}, \ and\ \bibinfo
  {author} {\bibfnamefont {Immanuel}\ \bibnamefont {Bloch}},\ }\bibfield
  {title} {\enquote {\bibinfo {title} {Observation of many-body localization of
  interacting fermions in a quasirandom optical lattice},}\ }\href {\doibase
  10.1126/science.aaa7432} {\bibfield  {journal} {\bibinfo  {journal}
  {Science}\ }\textbf {\bibinfo {volume} {349}},\ \bibinfo {pages} {842--845}
  (\bibinfo {year} {2015})},\ \Eprint
  {http://arxiv.org/abs/http://science.sciencemag.org/content/349/6250/842.full.pdf}
  {http://science.sciencemag.org/content/349/6250/842.full.pdf} \BibitemShut
  {NoStop}%
\bibitem [{\citenamefont {Essler}\ and\ \citenamefont
  {Fagotti}(2016)}]{Essler16}%
  \BibitemOpen
  \bibfield  {author} {\bibinfo {author} {\bibfnamefont {Fabian H~L}\
  \bibnamefont {Essler}}\ and\ \bibinfo {author} {\bibfnamefont {Maurizio}\
  \bibnamefont {Fagotti}},\ }\bibfield  {title} {\enquote {\bibinfo {title}
  {Quench dynamics and relaxation in isolated integrable quantum spin
  chains},}\ }\href {http://stacks.iop.org/1742-5468/2016/i=6/a=064002}
  {\bibfield  {journal} {\bibinfo  {journal} {Journal of Statistical Mechanics:
  Theory and Experiment}\ }\textbf {\bibinfo {volume} {2016}},\ \bibinfo
  {pages} {064002} (\bibinfo {year} {2016})}\BibitemShut {NoStop}%
\bibitem [{\citenamefont {Huse}\ \emph {et~al.}(2014)\citenamefont {Huse},
  \citenamefont {Nandkishore},\ and\ \citenamefont {Oganesyan}}]{Huse14}%
  \BibitemOpen
  \bibfield  {author} {\bibinfo {author} {\bibfnamefont {David~A.}\
  \bibnamefont {Huse}}, \bibinfo {author} {\bibfnamefont {Rahul}\ \bibnamefont
  {Nandkishore}}, \ and\ \bibinfo {author} {\bibfnamefont {Vadim}\ \bibnamefont
  {Oganesyan}},\ }\bibfield  {title} {\enquote {\bibinfo {title} {Phenomenology
  of fully many-body-localized systems},}\ }\href {\doibase
  10.1103/PhysRevB.90.174202} {\bibfield  {journal} {\bibinfo  {journal} {Phys.
  Rev. B}\ }\textbf {\bibinfo {volume} {90}},\ \bibinfo {pages} {174202}
  (\bibinfo {year} {2014})}\BibitemShut {NoStop}%
\bibitem [{\citenamefont {Serbyn}\ \emph {et~al.}(2013)\citenamefont {Serbyn},
  \citenamefont {Papi\ifmmode~\acute{c}\else \'{c}\fi{}},\ and\ \citenamefont
  {Abanin}}]{Serbyn13cons}%
  \BibitemOpen
  \bibfield  {author} {\bibinfo {author} {\bibfnamefont {Maksym}\ \bibnamefont
  {Serbyn}}, \bibinfo {author} {\bibfnamefont {Z.}~\bibnamefont
  {Papi\ifmmode~\acute{c}\else \'{c}\fi{}}}, \ and\ \bibinfo {author}
  {\bibfnamefont {Dmitry~A.}\ \bibnamefont {Abanin}},\ }\bibfield  {title}
  {\enquote {\bibinfo {title} {Local conservation laws and the structure of the
  many-body localized states},}\ }\href {\doibase
  10.1103/PhysRevLett.111.127201} {\bibfield  {journal} {\bibinfo  {journal}
  {Phys. Rev. Lett.}\ }\textbf {\bibinfo {volume} {111}},\ \bibinfo {pages}
  {127201} (\bibinfo {year} {2013})}\BibitemShut {NoStop}%
\bibitem [{\citenamefont {Schiulaz}\ \emph {et~al.}(2015)\citenamefont
  {Schiulaz}, \citenamefont {Silva},\ and\ \citenamefont
  {M\"uller}}]{Muller2015}%
  \BibitemOpen
  \bibfield  {author} {\bibinfo {author} {\bibfnamefont {Mauro}\ \bibnamefont
  {Schiulaz}}, \bibinfo {author} {\bibfnamefont {Alessandro}\ \bibnamefont
  {Silva}}, \ and\ \bibinfo {author} {\bibfnamefont {Markus}\ \bibnamefont
  {M\"uller}},\ }\bibfield  {title} {\enquote {\bibinfo {title} {Dynamics in
  many-body localized quantum systems without disorder},}\ }\href {\doibase
  10.1103/PhysRevB.91.184202} {\bibfield  {journal} {\bibinfo  {journal} {Phys.
  Rev. B}\ }\textbf {\bibinfo {volume} {91}},\ \bibinfo {pages} {184202}
  (\bibinfo {year} {2015})}\BibitemShut {NoStop}%
\bibitem [{\citenamefont {Yao}\ \emph {et~al.}(2016)\citenamefont {Yao},
  \citenamefont {Laumann}, \citenamefont {Cirac}, \citenamefont {Lukin},\ and\
  \citenamefont {Moore}}]{Yao2016}%
  \BibitemOpen
  \bibfield  {author} {\bibinfo {author} {\bibfnamefont {N.~Y.}\ \bibnamefont
  {Yao}}, \bibinfo {author} {\bibfnamefont {C.~R.}\ \bibnamefont {Laumann}},
  \bibinfo {author} {\bibfnamefont {J.~I.}\ \bibnamefont {Cirac}}, \bibinfo
  {author} {\bibfnamefont {M.~D.}\ \bibnamefont {Lukin}}, \ and\ \bibinfo
  {author} {\bibfnamefont {J.~E.}\ \bibnamefont {Moore}},\ }\bibfield  {title}
  {\enquote {\bibinfo {title} {Quasi-many-body localization in
  translation-invariant systems},}\ }\href {\doibase
  10.1103/PhysRevLett.117.240601} {\bibfield  {journal} {\bibinfo  {journal}
  {Phys. Rev. Lett.}\ }\textbf {\bibinfo {volume} {117}},\ \bibinfo {pages}
  {240601} (\bibinfo {year} {2016})}\BibitemShut {NoStop}%
\bibitem [{\citenamefont {Papić}\ \emph {et~al.}(2015)\citenamefont {Papić},
  \citenamefont {Stoudenmire},\ and\ \citenamefont {Abanin}}]{PAPIC2015}%
  \BibitemOpen
  \bibfield  {author} {\bibinfo {author} {\bibfnamefont {Z.}~\bibnamefont
  {Papić}}, \bibinfo {author} {\bibfnamefont {E.~Miles}\ \bibnamefont
  {Stoudenmire}}, \ and\ \bibinfo {author} {\bibfnamefont {Dmitry~A.}\
  \bibnamefont {Abanin}},\ }\bibfield  {title} {\enquote {\bibinfo {title}
  {Many-body localization in disorder-free systems: The importance of
  finite-size constraints},}\ }\href {\doibase
  https://doi.org/10.1016/j.aop.2015.08.024} {\bibfield  {journal} {\bibinfo
  {journal} {Annals of Physics}\ }\textbf {\bibinfo {volume} {362}},\ \bibinfo
  {pages} {714 -- 725} (\bibinfo {year} {2015})}\BibitemShut {NoStop}%
\bibitem [{\citenamefont {Smith}\ \emph
  {et~al.}(2017{\natexlab{a}})\citenamefont {Smith}, \citenamefont {Knolle},
  \citenamefont {Kovrizhin},\ and\ \citenamefont {Moessner}}]{Smith01}%
  \BibitemOpen
  \bibfield  {author} {\bibinfo {author} {\bibfnamefont {A.}~\bibnamefont
  {Smith}}, \bibinfo {author} {\bibfnamefont {J.}~\bibnamefont {Knolle}},
  \bibinfo {author} {\bibfnamefont {D.~L.}\ \bibnamefont {Kovrizhin}}, \ and\
  \bibinfo {author} {\bibfnamefont {R.}~\bibnamefont {Moessner}},\ }\bibfield
  {title} {\enquote {\bibinfo {title} {Disorder-free localization},}\ }\href
  {\doibase 10.1103/PhysRevLett.118.266601} {\bibfield  {journal} {\bibinfo
  {journal} {Phys. Rev. Lett.}\ }\textbf {\bibinfo {volume} {118}},\ \bibinfo
  {pages} {266601} (\bibinfo {year} {2017}{\natexlab{a}})}\BibitemShut
  {NoStop}%
\bibitem [{\citenamefont {Smith}\ \emph
  {et~al.}(2017{\natexlab{b}})\citenamefont {Smith}, \citenamefont {Knolle},
  \citenamefont {Moessner},\ and\ \citenamefont {Kovrizhin}}]{Smith02}%
  \BibitemOpen
  \bibfield  {author} {\bibinfo {author} {\bibfnamefont {A.}~\bibnamefont
  {Smith}}, \bibinfo {author} {\bibfnamefont {J.}~\bibnamefont {Knolle}},
  \bibinfo {author} {\bibfnamefont {R.}~\bibnamefont {Moessner}}, \ and\
  \bibinfo {author} {\bibfnamefont {D.~L.}\ \bibnamefont {Kovrizhin}},\
  }\bibfield  {title} {\enquote {\bibinfo {title} {Absence of ergodicity
  without quenched disorder: From quantum disentangled liquids to many-body
  localization},}\ }\href {\doibase 10.1103/PhysRevLett.119.176601} {\bibfield
  {journal} {\bibinfo  {journal} {Phys. Rev. Lett.}\ }\textbf {\bibinfo
  {volume} {119}},\ \bibinfo {pages} {176601} (\bibinfo {year}
  {2017}{\natexlab{b}})}\BibitemShut {NoStop}%
\bibitem [{\citenamefont {Brenes}\ \emph {et~al.}(2018)\citenamefont {Brenes},
  \citenamefont {Dalmonte}, \citenamefont {Heyl},\ and\ \citenamefont
  {Scardicchio}}]{Brenes18}%
  \BibitemOpen
  \bibfield  {author} {\bibinfo {author} {\bibfnamefont {Marlon}\ \bibnamefont
  {Brenes}}, \bibinfo {author} {\bibfnamefont {Marcello}\ \bibnamefont
  {Dalmonte}}, \bibinfo {author} {\bibfnamefont {Markus}\ \bibnamefont {Heyl}},
  \ and\ \bibinfo {author} {\bibfnamefont {Antonello}\ \bibnamefont
  {Scardicchio}},\ }\bibfield  {title} {\enquote {\bibinfo {title} {Many-body
  localization dynamics from gauge invariance},}\ }\href {\doibase
  10.1103/PhysRevLett.120.030601} {\bibfield  {journal} {\bibinfo  {journal}
  {Phys. Rev. Lett.}\ }\textbf {\bibinfo {volume} {120}},\ \bibinfo {pages}
  {030601} (\bibinfo {year} {2018})}\BibitemShut {NoStop}%
\bibitem [{\citenamefont {van Nieuwenburg}\ \emph {et~al.}(2018)\citenamefont
  {van Nieuwenburg}, \citenamefont {Baum},\ and\ \citenamefont
  {Refael}}]{Refael18}%
  \BibitemOpen
  \bibfield  {author} {\bibinfo {author} {\bibfnamefont {Evert P.~L.}\
  \bibnamefont {van Nieuwenburg}}, \bibinfo {author} {\bibfnamefont {Yuval}\
  \bibnamefont {Baum}}, \ and\ \bibinfo {author} {\bibfnamefont {Gil}\
  \bibnamefont {Refael}},\ }\href@noop {} {\enquote {\bibinfo {title} {From
  bloch oscillations to many body localization in clean interacting systems},}\
  } (\bibinfo {year} {2018}),\ \Eprint {http://arxiv.org/abs/arXiv:1808.00471}
  {arXiv:1808.00471} \BibitemShut {NoStop}%
\bibitem [{\citenamefont {Schulz}\ \emph {et~al.}(2019)\citenamefont {Schulz},
  \citenamefont {Hooley}, \citenamefont {Moessner},\ and\ \citenamefont
  {Pollmann}}]{Schulz19}%
  \BibitemOpen
  \bibfield  {author} {\bibinfo {author} {\bibfnamefont {M.}~\bibnamefont
  {Schulz}}, \bibinfo {author} {\bibfnamefont {C.~A.}\ \bibnamefont {Hooley}},
  \bibinfo {author} {\bibfnamefont {R.}~\bibnamefont {Moessner}}, \ and\
  \bibinfo {author} {\bibfnamefont {F.}~\bibnamefont {Pollmann}},\ }\bibfield
  {title} {\enquote {\bibinfo {title} {Stark many-body localization},}\ }\href
  {\doibase 10.1103/PhysRevLett.122.040606} {\bibfield  {journal} {\bibinfo
  {journal} {Phys. Rev. Lett.}\ }\textbf {\bibinfo {volume} {122}},\ \bibinfo
  {pages} {040606} (\bibinfo {year} {2019})}\BibitemShut {NoStop}%
\bibitem [{\citenamefont {Biroli}\ \emph {et~al.}(2010)\citenamefont {Biroli},
  \citenamefont {Kollath},\ and\ \citenamefont {L\"auchli}}]{Biroli2010}%
  \BibitemOpen
  \bibfield  {author} {\bibinfo {author} {\bibfnamefont {Giulio}\ \bibnamefont
  {Biroli}}, \bibinfo {author} {\bibfnamefont {Corinna}\ \bibnamefont
  {Kollath}}, \ and\ \bibinfo {author} {\bibfnamefont {Andreas~M.}\
  \bibnamefont {L\"auchli}},\ }\bibfield  {title} {\enquote {\bibinfo {title}
  {Effect of rare fluctuations on the thermalization of isolated quantum
  systems},}\ }\href {\doibase 10.1103/PhysRevLett.105.250401} {\bibfield
  {journal} {\bibinfo  {journal} {Phys. Rev. Lett.}\ }\textbf {\bibinfo
  {volume} {105}},\ \bibinfo {pages} {250401} (\bibinfo {year}
  {2010})}\BibitemShut {NoStop}%
\bibitem [{\citenamefont {Mori}(2016)}]{MoriWeakETH}%
  \BibitemOpen
  \bibfield  {author} {\bibinfo {author} {\bibfnamefont {Takashi}\ \bibnamefont
  {Mori}},\ }\href@noop {} {\enquote {\bibinfo {title} {Weak eigenstate
  thermalization with large deviation bound},}\ } (\bibinfo {year} {2016}),\
  \Eprint {http://arxiv.org/abs/arXiv:1609.09776} {arXiv:1609.09776}
  \BibitemShut {NoStop}%
\bibitem [{\citenamefont {Shiraishi}\ and\ \citenamefont
  {Mori}(2017)}]{ShiraishiMori}%
  \BibitemOpen
  \bibfield  {author} {\bibinfo {author} {\bibfnamefont {Naoto}\ \bibnamefont
  {Shiraishi}}\ and\ \bibinfo {author} {\bibfnamefont {Takashi}\ \bibnamefont
  {Mori}},\ }\bibfield  {title} {\enquote {\bibinfo {title} {Systematic
  construction of counterexamples to the eigenstate thermalization
  hypothesis},}\ }\href {\doibase 10.1103/PhysRevLett.119.030601} {\bibfield
  {journal} {\bibinfo  {journal} {Phys. Rev. Lett.}\ }\textbf {\bibinfo
  {volume} {119}},\ \bibinfo {pages} {030601} (\bibinfo {year}
  {2017})}\BibitemShut {NoStop}%
\bibitem [{\citenamefont {Moudgalya}\ \emph
  {et~al.}(2018{\natexlab{a}})\citenamefont {Moudgalya}, \citenamefont
  {Rachel}, \citenamefont {Bernevig},\ and\ \citenamefont
  {Regnault}}]{Moudgalya01}%
  \BibitemOpen
  \bibfield  {author} {\bibinfo {author} {\bibfnamefont {Sanjay}\ \bibnamefont
  {Moudgalya}}, \bibinfo {author} {\bibfnamefont {Stephan}\ \bibnamefont
  {Rachel}}, \bibinfo {author} {\bibfnamefont {B.~Andrei}\ \bibnamefont
  {Bernevig}}, \ and\ \bibinfo {author} {\bibfnamefont {Nicolas}\ \bibnamefont
  {Regnault}},\ }\bibfield  {title} {\enquote {\bibinfo {title} {Exact excited
  states of nonintegrable models},}\ }\href {\doibase
  10.1103/PhysRevB.98.235155} {\bibfield  {journal} {\bibinfo  {journal} {Phys.
  Rev. B}\ }\textbf {\bibinfo {volume} {98}},\ \bibinfo {pages} {235155}
  (\bibinfo {year} {2018}{\natexlab{a}})}\BibitemShut {NoStop}%
\bibitem [{\citenamefont {Moudgalya}\ \emph
  {et~al.}(2018{\natexlab{b}})\citenamefont {Moudgalya}, \citenamefont
  {Regnault},\ and\ \citenamefont {Bernevig}}]{Moudgalya02}%
  \BibitemOpen
  \bibfield  {author} {\bibinfo {author} {\bibfnamefont {Sanjay}\ \bibnamefont
  {Moudgalya}}, \bibinfo {author} {\bibfnamefont {Nicolas}\ \bibnamefont
  {Regnault}}, \ and\ \bibinfo {author} {\bibfnamefont {B.~Andrei}\
  \bibnamefont {Bernevig}},\ }\bibfield  {title} {\enquote {\bibinfo {title}
  {Entanglement of exact excited states of affleck-kennedy-lieb-tasaki models:
  Exact results, many-body scars, and violation of the strong eigenstate
  thermalization hypothesis},}\ }\href {\doibase 10.1103/PhysRevB.98.235156}
  {\bibfield  {journal} {\bibinfo  {journal} {Phys. Rev. B}\ }\textbf {\bibinfo
  {volume} {98}},\ \bibinfo {pages} {235156} (\bibinfo {year}
  {2018}{\natexlab{b}})}\BibitemShut {NoStop}%
\bibitem [{\citenamefont {Iadecola}\ and\ \citenamefont
  {Znidaric}(2018)}]{Iadecola2018}%
  \BibitemOpen
  \bibfield  {author} {\bibinfo {author} {\bibfnamefont {Thomas}\ \bibnamefont
  {Iadecola}}\ and\ \bibinfo {author} {\bibfnamefont {Marko}\ \bibnamefont
  {Znidaric}},\ }\href@noop {} {\enquote {\bibinfo {title} {Exact localized and
  ballistic eigenstates in disordered chaotic spin ladders and the
  fermi-hubbard model},}\ } (\bibinfo {year} {2018}),\ \Eprint
  {http://arxiv.org/abs/arXiv:1811.07903} {arXiv:1811.07903} \BibitemShut
  {NoStop}%
\bibitem [{\citenamefont {Iadecola}\ \emph {et~al.}(2019)\citenamefont
  {Iadecola}, \citenamefont {Schecter},\ and\ \citenamefont
  {Xu}}]{Iadecola2019}%
  \BibitemOpen
  \bibfield  {author} {\bibinfo {author} {\bibfnamefont {Thomas}\ \bibnamefont
  {Iadecola}}, \bibinfo {author} {\bibfnamefont {Michael}\ \bibnamefont
  {Schecter}}, \ and\ \bibinfo {author} {\bibfnamefont {Shenglong}\
  \bibnamefont {Xu}},\ }\href@noop {} {\enquote {\bibinfo {title} {Quantum
  many-body scars and space-time crystalline order from magnon condensation},}\
  } (\bibinfo {year} {2019}),\ \Eprint {http://arxiv.org/abs/arXiv:1903.10517}
  {arXiv:1903.10517} \BibitemShut {NoStop}%
\bibitem [{\citenamefont {Ok}\ \emph {et~al.}(2019)\citenamefont {Ok},
  \citenamefont {Choo}, \citenamefont {Mudry}, \citenamefont {Castelnovo},
  \citenamefont {Chamon},\ and\ \citenamefont {Neupert}}]{Neupert2019}%
  \BibitemOpen
  \bibfield  {author} {\bibinfo {author} {\bibfnamefont {Seulgi}\ \bibnamefont
  {Ok}}, \bibinfo {author} {\bibfnamefont {Kenny}\ \bibnamefont {Choo}},
  \bibinfo {author} {\bibfnamefont {Christopher}\ \bibnamefont {Mudry}},
  \bibinfo {author} {\bibfnamefont {Claudio}\ \bibnamefont {Castelnovo}},
  \bibinfo {author} {\bibfnamefont {Claudio}\ \bibnamefont {Chamon}}, \ and\
  \bibinfo {author} {\bibfnamefont {Titus}\ \bibnamefont {Neupert}},\
  }\href@noop {} {\enquote {\bibinfo {title} {Topological many-body scar states
  in dimensions 1, 2, and 3},}\ } (\bibinfo {year} {2019}),\ \Eprint
  {http://arxiv.org/abs/arXiv:1901.01260} {arXiv:1901.01260} \BibitemShut
  {NoStop}%
\bibitem [{\citenamefont {van Horssen}\ \emph {et~al.}(2015)\citenamefont {van
  Horssen}, \citenamefont {Levi},\ and\ \citenamefont
  {Garrahan}}]{Garrahan2015}%
  \BibitemOpen
  \bibfield  {author} {\bibinfo {author} {\bibfnamefont {Merlijn}\ \bibnamefont
  {van Horssen}}, \bibinfo {author} {\bibfnamefont {Emanuele}\ \bibnamefont
  {Levi}}, \ and\ \bibinfo {author} {\bibfnamefont {Juan~P.}\ \bibnamefont
  {Garrahan}},\ }\bibfield  {title} {\enquote {\bibinfo {title} {Dynamics of
  many-body localization in a translation-invariant quantum glass model},}\
  }\href {\doibase 10.1103/PhysRevB.92.100305} {\bibfield  {journal} {\bibinfo
  {journal} {Phys. Rev. B}\ }\textbf {\bibinfo {volume} {92}},\ \bibinfo
  {pages} {100305} (\bibinfo {year} {2015})}\BibitemShut {NoStop}%
\bibitem [{\citenamefont {Lan}\ \emph {et~al.}(2018)\citenamefont {Lan},
  \citenamefont {van Horssen}, \citenamefont {Powell},\ and\ \citenamefont
  {Garrahan}}]{Garrahan2018}%
  \BibitemOpen
  \bibfield  {author} {\bibinfo {author} {\bibfnamefont {Zhihao}\ \bibnamefont
  {Lan}}, \bibinfo {author} {\bibfnamefont {Merlijn}\ \bibnamefont {van
  Horssen}}, \bibinfo {author} {\bibfnamefont {Stephen}\ \bibnamefont
  {Powell}}, \ and\ \bibinfo {author} {\bibfnamefont {Juan~P.}\ \bibnamefont
  {Garrahan}},\ }\bibfield  {title} {\enquote {\bibinfo {title} {Quantum slow
  relaxation and metastability due to dynamical constraints},}\ }\href
  {\doibase 10.1103/PhysRevLett.121.040603} {\bibfield  {journal} {\bibinfo
  {journal} {Phys. Rev. Lett.}\ }\textbf {\bibinfo {volume} {121}},\ \bibinfo
  {pages} {040603} (\bibinfo {year} {2018})}\BibitemShut {NoStop}%
\bibitem [{\citenamefont {Turner}\ \emph
  {et~al.}(2018{\natexlab{a}})\citenamefont {Turner}, \citenamefont
  {Michailidis}, \citenamefont {Abanin}, \citenamefont {Serbyn},\ and\
  \citenamefont {Papic}}]{TurnerNatPhys}%
  \BibitemOpen
  \bibfield  {author} {\bibinfo {author} {\bibfnamefont {CJ}~\bibnamefont
  {Turner}}, \bibinfo {author} {\bibfnamefont {AA}~\bibnamefont {Michailidis}},
  \bibinfo {author} {\bibfnamefont {Dmitry}\ \bibnamefont {Abanin}}, \bibinfo
  {author} {\bibfnamefont {Maksym}\ \bibnamefont {Serbyn}}, \ and\ \bibinfo
  {author} {\bibfnamefont {Zlatko}\ \bibnamefont {Papic}},\ }\bibfield  {title}
  {\enquote {\bibinfo {title} {Weak ergodicity breaking from quantum many-body
  scars},}\ }\href {\doibase 10.1038/s41567-018-0137-5} {\bibfield  {journal}
  {\bibinfo  {journal} {Nature Physics}\ }\textbf {\bibinfo {volume} {14}}
  (\bibinfo {year} {2018}{\natexlab{a}}),\
  10.1038/s41567-018-0137-5}\BibitemShut {NoStop}%
\bibitem [{\citenamefont {Turner}\ \emph
  {et~al.}(2018{\natexlab{b}})\citenamefont {Turner}, \citenamefont
  {Michailidis}, \citenamefont {Abanin}, \citenamefont {Serbyn},\ and\
  \citenamefont {Papi\ifmmode~\acute{c}\else \'{c}\fi{}}}]{TurnerPRB}%
  \BibitemOpen
  \bibfield  {author} {\bibinfo {author} {\bibfnamefont {C.~J.}\ \bibnamefont
  {Turner}}, \bibinfo {author} {\bibfnamefont {A.~A.}\ \bibnamefont
  {Michailidis}}, \bibinfo {author} {\bibfnamefont {D.~A.}\ \bibnamefont
  {Abanin}}, \bibinfo {author} {\bibfnamefont {M.}~\bibnamefont {Serbyn}}, \
  and\ \bibinfo {author} {\bibfnamefont {Z.}~\bibnamefont
  {Papi\ifmmode~\acute{c}\else \'{c}\fi{}}},\ }\bibfield  {title} {\enquote
  {\bibinfo {title} {Quantum scarred eigenstates in a rydberg atom chain:
  Entanglement, breakdown of thermalization, and stability to perturbations},}\
  }\href {\doibase 10.1103/PhysRevB.98.155134} {\bibfield  {journal} {\bibinfo
  {journal} {Phys. Rev. B}\ }\textbf {\bibinfo {volume} {98}},\ \bibinfo
  {pages} {155134} (\bibinfo {year} {2018}{\natexlab{b}})}\BibitemShut
  {NoStop}%
\bibitem [{\citenamefont {Choi}\ \emph {et~al.}(2018)\citenamefont {Choi},
  \citenamefont {Turner}, \citenamefont {Pichler}, \citenamefont {Ho},
  \citenamefont {Michailidis}, \citenamefont {Papić}, \citenamefont {Serbyn},
  \citenamefont {Lukin},\ and\ \citenamefont {Abanin}}]{Choi2018}%
  \BibitemOpen
  \bibfield  {author} {\bibinfo {author} {\bibfnamefont {Soonwon}\ \bibnamefont
  {Choi}}, \bibinfo {author} {\bibfnamefont {Christopher~J.}\ \bibnamefont
  {Turner}}, \bibinfo {author} {\bibfnamefont {Hannes}\ \bibnamefont
  {Pichler}}, \bibinfo {author} {\bibfnamefont {Wen~Wei}\ \bibnamefont {Ho}},
  \bibinfo {author} {\bibfnamefont {Alexios~A.}\ \bibnamefont {Michailidis}},
  \bibinfo {author} {\bibfnamefont {Zlatko}\ \bibnamefont {Papić}}, \bibinfo
  {author} {\bibfnamefont {Maksym}\ \bibnamefont {Serbyn}}, \bibinfo {author}
  {\bibfnamefont {Mikhail~D.}\ \bibnamefont {Lukin}}, \ and\ \bibinfo {author}
  {\bibfnamefont {Dmitry~A.}\ \bibnamefont {Abanin}},\ }\href@noop {} {\enquote
  {\bibinfo {title} {Emergent su(2) dynamics and perfect quantum many-body
  scars},}\ } (\bibinfo {year} {2018}),\ \Eprint
  {http://arxiv.org/abs/arXiv:1812.05561} {arXiv:1812.05561} \BibitemShut
  {NoStop}%
\bibitem [{\citenamefont {Lin}\ and\ \citenamefont
  {Motrunich}(2018)}]{Motrunich2018}%
  \BibitemOpen
  \bibfield  {author} {\bibinfo {author} {\bibfnamefont {Cheng-Ju}\
  \bibnamefont {Lin}}\ and\ \bibinfo {author} {\bibfnamefont {Olexei~I.}\
  \bibnamefont {Motrunich}},\ }\href@noop {} {\enquote {\bibinfo {title} {Exact
  quantum many-body scar states in the rydberg-blockaded atom chain},}\ }
  (\bibinfo {year} {2018}),\ \Eprint {http://arxiv.org/abs/arXiv:1810.00888}
  {arXiv:1810.00888} \BibitemShut {NoStop}%
\bibitem [{\citenamefont {Feldmeier}\ \emph {et~al.}(2019)\citenamefont
  {Feldmeier}, \citenamefont {Pollmann},\ and\ \citenamefont
  {Knap}}]{Feldmeier19}%
  \BibitemOpen
  \bibfield  {author} {\bibinfo {author} {\bibfnamefont {Johannes}\
  \bibnamefont {Feldmeier}}, \bibinfo {author} {\bibfnamefont {Frank}\
  \bibnamefont {Pollmann}}, \ and\ \bibinfo {author} {\bibfnamefont {Michael}\
  \bibnamefont {Knap}},\ }\href@noop {} {\enquote {\bibinfo {title} {Dynamical
  phase transitions in the quantum dimer model on a square lattice},}\ }
  (\bibinfo {year} {2019}),\ \Eprint {http://arxiv.org/abs/arXiv:1901.07597}
  {arXiv:1901.07597} \BibitemShut {NoStop}%
\bibitem [{\citenamefont {Bernien}\ \emph {et~al.}(2017)\citenamefont
  {Bernien}, \citenamefont {Schwartz}, \citenamefont {Keesling}, \citenamefont
  {Levine}, \citenamefont {Omran}, \citenamefont {Pichler}, \citenamefont
  {Choi}, \citenamefont {Zibrov}, \citenamefont {Endres}, \citenamefont
  {Greiner}, \citenamefont {Vuletic},\ and\ \citenamefont
  {D.~Lukin}}]{Bernien2017}%
  \BibitemOpen
  \bibfield  {author} {\bibinfo {author} {\bibfnamefont {Hannes}\ \bibnamefont
  {Bernien}}, \bibinfo {author} {\bibfnamefont {Sylvain}\ \bibnamefont
  {Schwartz}}, \bibinfo {author} {\bibfnamefont {Alexander}\ \bibnamefont
  {Keesling}}, \bibinfo {author} {\bibfnamefont {Harry}\ \bibnamefont
  {Levine}}, \bibinfo {author} {\bibfnamefont {Ahmed}\ \bibnamefont {Omran}},
  \bibinfo {author} {\bibfnamefont {Hannes}\ \bibnamefont {Pichler}}, \bibinfo
  {author} {\bibfnamefont {Soonwon}\ \bibnamefont {Choi}}, \bibinfo {author}
  {\bibfnamefont {A}~\bibnamefont {Zibrov}}, \bibinfo {author} {\bibfnamefont
  {Manuel}\ \bibnamefont {Endres}}, \bibinfo {author} {\bibfnamefont {Markus}\
  \bibnamefont {Greiner}}, \bibinfo {author} {\bibfnamefont {Vladan}\
  \bibnamefont {Vuletic}}, \ and\ \bibinfo {author} {\bibfnamefont {Mikhail}\
  \bibnamefont {D.~Lukin}},\ }\bibfield  {title} {\enquote {\bibinfo {title}
  {Probing many-body dynamics on a 51-atom quantum simulator},}\ }\href
  {\doibase 10.1038/nature24622} {\bibfield  {journal} {\bibinfo  {journal}
  {Nature}\ }\textbf {\bibinfo {volume} {551}} (\bibinfo {year} {2017}),\
  10.1038/nature24622}\BibitemShut {NoStop}%
\bibitem [{\citenamefont {Haah}(2011)}]{Haah11}%
  \BibitemOpen
  \bibfield  {author} {\bibinfo {author} {\bibfnamefont {Jeongwan}\
  \bibnamefont {Haah}},\ }\bibfield  {title} {\enquote {\bibinfo {title} {Local
  stabilizer codes in three dimensions without string logical operators},}\
  }\href {\doibase 10.1103/PhysRevA.83.042330} {\bibfield  {journal} {\bibinfo
  {journal} {Phys. Rev. A}\ }\textbf {\bibinfo {volume} {83}},\ \bibinfo
  {pages} {042330} (\bibinfo {year} {2011})}\BibitemShut {NoStop}%
\bibitem [{\citenamefont {Vijay}\ \emph {et~al.}(2015)\citenamefont {Vijay},
  \citenamefont {Haah},\ and\ \citenamefont {Fu}}]{Vijay15}%
  \BibitemOpen
  \bibfield  {author} {\bibinfo {author} {\bibfnamefont {Sagar}\ \bibnamefont
  {Vijay}}, \bibinfo {author} {\bibfnamefont {Jeongwan}\ \bibnamefont {Haah}},
  \ and\ \bibinfo {author} {\bibfnamefont {Liang}\ \bibnamefont {Fu}},\
  }\bibfield  {title} {\enquote {\bibinfo {title} {A new kind of topological
  quantum order: A dimensional hierarchy of quasiparticles built from
  stationary excitations},}\ }\href {\doibase 10.1103/PhysRevB.92.235136}
  {\bibfield  {journal} {\bibinfo  {journal} {Phys. Rev. B}\ }\textbf {\bibinfo
  {volume} {92}},\ \bibinfo {pages} {235136} (\bibinfo {year}
  {2015})}\BibitemShut {NoStop}%
\bibitem [{\citenamefont {Chamon}(2005)}]{Chamon05}%
  \BibitemOpen
  \bibfield  {author} {\bibinfo {author} {\bibfnamefont {Claudio}\ \bibnamefont
  {Chamon}},\ }\bibfield  {title} {\enquote {\bibinfo {title} {Quantum
  glassiness in strongly correlated clean systems: An example of topological
  overprotection},}\ }\href {\doibase 10.1103/PhysRevLett.94.040402} {\bibfield
   {journal} {\bibinfo  {journal} {Phys. Rev. Lett.}\ }\textbf {\bibinfo
  {volume} {94}},\ \bibinfo {pages} {040402} (\bibinfo {year}
  {2005})}\BibitemShut {NoStop}%
\bibitem [{\citenamefont {Vijay}\ \emph {et~al.}(2016)\citenamefont {Vijay},
  \citenamefont {Haah},\ and\ \citenamefont {Fu}}]{Vijay16}%
  \BibitemOpen
  \bibfield  {author} {\bibinfo {author} {\bibfnamefont {Sagar}\ \bibnamefont
  {Vijay}}, \bibinfo {author} {\bibfnamefont {Jeongwan}\ \bibnamefont {Haah}},
  \ and\ \bibinfo {author} {\bibfnamefont {Liang}\ \bibnamefont {Fu}},\
  }\bibfield  {title} {\enquote {\bibinfo {title} {Fracton topological order,
  generalized lattice gauge theory, and duality},}\ }\href {\doibase
  10.1103/PhysRevB.94.235157} {\bibfield  {journal} {\bibinfo  {journal} {Phys.
  Rev. B}\ }\textbf {\bibinfo {volume} {94}},\ \bibinfo {pages} {235157}
  (\bibinfo {year} {2016})}\BibitemShut {NoStop}%
\bibitem [{\citenamefont {Hsieh}\ and\ \citenamefont
  {Hal\'asz}(2017)}]{Hsieh17}%
  \BibitemOpen
  \bibfield  {author} {\bibinfo {author} {\bibfnamefont {Timothy~H.}\
  \bibnamefont {Hsieh}}\ and\ \bibinfo {author} {\bibfnamefont {G\'abor~B.}\
  \bibnamefont {Hal\'asz}},\ }\bibfield  {title} {\enquote {\bibinfo {title}
  {Fractons from partons},}\ }\href {\doibase 10.1103/PhysRevB.96.165105}
  {\bibfield  {journal} {\bibinfo  {journal} {Phys. Rev. B}\ }\textbf {\bibinfo
  {volume} {96}},\ \bibinfo {pages} {165105} (\bibinfo {year}
  {2017})}\BibitemShut {NoStop}%
\bibitem [{\citenamefont {Hal\'asz}\ \emph {et~al.}(2017)\citenamefont
  {Hal\'asz}, \citenamefont {Hsieh},\ and\ \citenamefont {Balents}}]{Gabor17}%
  \BibitemOpen
  \bibfield  {author} {\bibinfo {author} {\bibfnamefont {G\'abor~B.}\
  \bibnamefont {Hal\'asz}}, \bibinfo {author} {\bibfnamefont {Timothy~H.}\
  \bibnamefont {Hsieh}}, \ and\ \bibinfo {author} {\bibfnamefont {Leon}\
  \bibnamefont {Balents}},\ }\bibfield  {title} {\enquote {\bibinfo {title}
  {Fracton topological phases from strongly coupled spin chains},}\ }\href
  {\doibase 10.1103/PhysRevLett.119.257202} {\bibfield  {journal} {\bibinfo
  {journal} {Phys. Rev. Lett.}\ }\textbf {\bibinfo {volume} {119}},\ \bibinfo
  {pages} {257202} (\bibinfo {year} {2017})}\BibitemShut {NoStop}%
\bibitem [{\citenamefont {Shirley}\ \emph {et~al.}(2018)\citenamefont
  {Shirley}, \citenamefont {Slagle}, \citenamefont {Wang},\ and\ \citenamefont
  {Chen}}]{Kevin18}%
  \BibitemOpen
  \bibfield  {author} {\bibinfo {author} {\bibfnamefont {Wilbur}\ \bibnamefont
  {Shirley}}, \bibinfo {author} {\bibfnamefont {Kevin}\ \bibnamefont {Slagle}},
  \bibinfo {author} {\bibfnamefont {Zhenghan}\ \bibnamefont {Wang}}, \ and\
  \bibinfo {author} {\bibfnamefont {Xie}\ \bibnamefont {Chen}},\ }\bibfield
  {title} {\enquote {\bibinfo {title} {Fracton models on general
  three-dimensional manifolds},}\ }\href {\doibase 10.1103/PhysRevX.8.031051}
  {\bibfield  {journal} {\bibinfo  {journal} {Phys. Rev. X}\ }\textbf {\bibinfo
  {volume} {8}},\ \bibinfo {pages} {031051} (\bibinfo {year}
  {2018})}\BibitemShut {NoStop}%
\bibitem [{\citenamefont {Pretko}(2017{\natexlab{a}})}]{PretkoSub}%
  \BibitemOpen
  \bibfield  {author} {\bibinfo {author} {\bibfnamefont {Michael}\ \bibnamefont
  {Pretko}},\ }\bibfield  {title} {\enquote {\bibinfo {title} {Subdimensional
  particle structure of higher rank $u(1)$ spin liquids},}\ }\href {\doibase
  10.1103/PhysRevB.95.115139} {\bibfield  {journal} {\bibinfo  {journal} {Phys.
  Rev. B}\ }\textbf {\bibinfo {volume} {95}},\ \bibinfo {pages} {115139}
  (\bibinfo {year} {2017}{\natexlab{a}})}\BibitemShut {NoStop}%
\bibitem [{\citenamefont {{Pretko}}(2018)}]{PretkoPrin}%
  \BibitemOpen
  \bibfield  {author} {\bibinfo {author} {\bibfnamefont {M.}~\bibnamefont
  {{Pretko}}},\ }\bibfield  {title} {\enquote {\bibinfo {title} {{The Fracton
  Gauge Principle}},}\ }\href@noop {} {\bibfield  {journal} {\bibinfo
  {journal} {ArXiv e-prints}\ } (\bibinfo {year} {2018})},\ \Eprint
  {http://arxiv.org/abs/1807.11479} {arXiv:1807.11479 [cond-mat.str-el]}
  \BibitemShut {NoStop}%
\bibitem [{\citenamefont {Pretko}(2017{\natexlab{b}})}]{Pretko17b}%
  \BibitemOpen
  \bibfield  {author} {\bibinfo {author} {\bibfnamefont {Michael}\ \bibnamefont
  {Pretko}},\ }\bibfield  {title} {\enquote {\bibinfo {title} {Higher-spin
  witten effect and two-dimensional fracton phases},}\ }\href {\doibase
  10.1103/PhysRevB.96.125151} {\bibfield  {journal} {\bibinfo  {journal} {Phys.
  Rev. B}\ }\textbf {\bibinfo {volume} {96}},\ \bibinfo {pages} {125151}
  (\bibinfo {year} {2017}{\natexlab{b}})}\BibitemShut {NoStop}%
\bibitem [{\citenamefont {{Williamson}}\ \emph {et~al.}(2018)\citenamefont
  {{Williamson}}, \citenamefont {{Bi}},\ and\ \citenamefont
  {{Cheng}}}]{Williamson18}%
  \BibitemOpen
  \bibfield  {author} {\bibinfo {author} {\bibfnamefont {Dominic~J.}\
  \bibnamefont {{Williamson}}}, \bibinfo {author} {\bibfnamefont {Zhen}\
  \bibnamefont {{Bi}}}, \ and\ \bibinfo {author} {\bibfnamefont {Meng}\
  \bibnamefont {{Cheng}}},\ }\bibfield  {title} {\enquote {\bibinfo {title}
  {{Fractonic Matter in Symmetry-Enriched U(1) Gauge Theory}},}\ }\href@noop {}
  {\bibfield  {journal} {\bibinfo  {journal} {arXiv e-prints}\ ,\ \bibinfo
  {eid} {arXiv:1809.10275}} (\bibinfo {year} {2018})},\ \Eprint
  {http://arxiv.org/abs/1809.10275} {arXiv:1809.10275 [cond-mat.str-el]}
  \BibitemShut {NoStop}%
\bibitem [{\citenamefont {{Bulmash}}\ and\ \citenamefont
  {{Barkeshli}}(2018)}]{Bulmash18b}%
  \BibitemOpen
  \bibfield  {author} {\bibinfo {author} {\bibfnamefont {Daniel}\ \bibnamefont
  {{Bulmash}}}\ and\ \bibinfo {author} {\bibfnamefont {Maissam}\ \bibnamefont
  {{Barkeshli}}},\ }\bibfield  {title} {\enquote {\bibinfo {title}
  {{Generalized $U(1)$ Gauge Field Theories and Fractal Dynamics}},}\
  }\href@noop {} {\bibfield  {journal} {\bibinfo  {journal} {arXiv e-prints}\
  ,\ \bibinfo {eid} {arXiv:1806.01855}} (\bibinfo {year} {2018})},\ \Eprint
  {http://arxiv.org/abs/1806.01855} {arXiv:1806.01855 [cond-mat.str-el]}
  \BibitemShut {NoStop}%
\bibitem [{\citenamefont {{Marcos}}\ \emph {et~al.}(2014)\citenamefont
  {{Marcos}}, \citenamefont {{Widmer}}, \citenamefont {{Rico}}, \citenamefont
  {{Hafezi}}, \citenamefont {{Rabl}}, \citenamefont {{Wiese}},\ and\
  \citenamefont {{Zoller}}}]{DimertoQLM}%
  \BibitemOpen
  \bibfield  {author} {\bibinfo {author} {\bibfnamefont {D.}~\bibnamefont
  {{Marcos}}}, \bibinfo {author} {\bibfnamefont {P.}~\bibnamefont {{Widmer}}},
  \bibinfo {author} {\bibfnamefont {E.}~\bibnamefont {{Rico}}}, \bibinfo
  {author} {\bibfnamefont {M.}~\bibnamefont {{Hafezi}}}, \bibinfo {author}
  {\bibfnamefont {P.}~\bibnamefont {{Rabl}}}, \bibinfo {author} {\bibfnamefont
  {U.-J.}\ \bibnamefont {{Wiese}}}, \ and\ \bibinfo {author} {\bibfnamefont
  {P.}~\bibnamefont {{Zoller}}},\ }\bibfield  {title} {\enquote {\bibinfo
  {title} {{Two-dimensional lattice gauge theories with superconducting quantum
  circuits}},}\ }\href {\doibase 10.1016/j.aop.2014.09.011} {\bibfield
  {journal} {\bibinfo  {journal} {Annals of Physics}\ }\textbf {\bibinfo
  {volume} {351}},\ \bibinfo {pages} {634--654} (\bibinfo {year} {2014})},\
  \Eprint {http://arxiv.org/abs/1407.6066} {arXiv:1407.6066 [quant-ph]}
  \BibitemShut {NoStop}%
\bibitem [{\citenamefont {{Gromov}}(2018)}]{Gromov18}%
  \BibitemOpen
  \bibfield  {author} {\bibinfo {author} {\bibfnamefont {Andrey}\ \bibnamefont
  {{Gromov}}},\ }\bibfield  {title} {\enquote {\bibinfo {title} {{Towards
  classification of Fracton phases: the multipole algebra}},}\ }\href@noop {}
  {\bibfield  {journal} {\bibinfo  {journal} {arXiv e-prints}\ ,\ \bibinfo
  {eid} {arXiv:1812.05104}} (\bibinfo {year} {2018})},\ \Eprint
  {http://arxiv.org/abs/1812.05104} {arXiv:1812.05104 [cond-mat.str-el]}
  \BibitemShut {NoStop}%
\bibitem [{\citenamefont {Bulmash}\ and\ \citenamefont
  {Barkeshli}(2018)}]{Bulmash18}%
  \BibitemOpen
  \bibfield  {author} {\bibinfo {author} {\bibfnamefont {Daniel}\ \bibnamefont
  {Bulmash}}\ and\ \bibinfo {author} {\bibfnamefont {Maissam}\ \bibnamefont
  {Barkeshli}},\ }\bibfield  {title} {\enquote {\bibinfo {title} {Higgs
  mechanism in higher-rank symmetric u(1) gauge theories},}\ }\href {\doibase
  10.1103/PhysRevB.97.235112} {\bibfield  {journal} {\bibinfo  {journal} {Phys.
  Rev. B}\ }\textbf {\bibinfo {volume} {97}},\ \bibinfo {pages} {235112}
  (\bibinfo {year} {2018})}\BibitemShut {NoStop}%
\bibitem [{\citenamefont {Ma}\ \emph {et~al.}(2018)\citenamefont {Ma},
  \citenamefont {Hermele},\ and\ \citenamefont {Chen}}]{Han18}%
  \BibitemOpen
  \bibfield  {author} {\bibinfo {author} {\bibfnamefont {Han}\ \bibnamefont
  {Ma}}, \bibinfo {author} {\bibfnamefont {Michael}\ \bibnamefont {Hermele}}, \
  and\ \bibinfo {author} {\bibfnamefont {Xie}\ \bibnamefont {Chen}},\
  }\bibfield  {title} {\enquote {\bibinfo {title} {Fracton topological order
  from the higgs and partial-confinement mechanisms of rank-two gauge
  theory},}\ }\href {\doibase 10.1103/PhysRevB.98.035111} {\bibfield  {journal}
  {\bibinfo  {journal} {Phys. Rev. B}\ }\textbf {\bibinfo {volume} {98}},\
  \bibinfo {pages} {035111} (\bibinfo {year} {2018})}\BibitemShut {NoStop}%
\bibitem [{\citenamefont {Prem}\ \emph {et~al.}(2017)\citenamefont {Prem},
  \citenamefont {Haah},\ and\ \citenamefont {Nandkishore}}]{Abhinav17}%
  \BibitemOpen
  \bibfield  {author} {\bibinfo {author} {\bibfnamefont {Abhinav}\ \bibnamefont
  {Prem}}, \bibinfo {author} {\bibfnamefont {Jeongwan}\ \bibnamefont {Haah}}, \
  and\ \bibinfo {author} {\bibfnamefont {Rahul}\ \bibnamefont {Nandkishore}},\
  }\bibfield  {title} {\enquote {\bibinfo {title} {Glassy quantum dynamics in
  translation invariant fracton models},}\ }\href {\doibase
  10.1103/PhysRevB.95.155133} {\bibfield  {journal} {\bibinfo  {journal} {Phys.
  Rev. B}\ }\textbf {\bibinfo {volume} {95}},\ \bibinfo {pages} {155133}
  (\bibinfo {year} {2017})}\BibitemShut {NoStop}%
\bibitem [{\citenamefont {Pretko}(2017{\natexlab{c}})}]{Pretko17}%
  \BibitemOpen
  \bibfield  {author} {\bibinfo {author} {\bibfnamefont {Michael}\ \bibnamefont
  {Pretko}},\ }\bibfield  {title} {\enquote {\bibinfo {title}
  {Finite-temperature screening of $u$(1) fractons},}\ }\href {\doibase
  10.1103/PhysRevB.96.115102} {\bibfield  {journal} {\bibinfo  {journal} {Phys.
  Rev. B}\ }\textbf {\bibinfo {volume} {96}},\ \bibinfo {pages} {115102}
  (\bibinfo {year} {2017}{\natexlab{c}})}\BibitemShut {NoStop}%
\bibitem [{\citenamefont {Castelnovo}\ and\ \citenamefont
  {Chamon}(2012)}]{Castelnovo12}%
  \BibitemOpen
  \bibfield  {author} {\bibinfo {author} {\bibfnamefont {Claudio}\ \bibnamefont
  {Castelnovo}}\ and\ \bibinfo {author} {\bibfnamefont {Claudio}\ \bibnamefont
  {Chamon}},\ }\bibfield  {title} {\enquote {\bibinfo {title} {Topological
  quantum glassiness},}\ }\href {\doibase 10.1080/14786435.2011.609152}
  {\bibfield  {journal} {\bibinfo  {journal} {Philosophical Magazine}\ }\textbf
  {\bibinfo {volume} {92}},\ \bibinfo {pages} {304--323} (\bibinfo {year}
  {2012})},\ \Eprint
  {http://arxiv.org/abs/https://doi.org/10.1080/14786435.2011.609152}
  {https://doi.org/10.1080/14786435.2011.609152} \BibitemShut {NoStop}%
\bibitem [{\citenamefont {Rezayi}\ and\ \citenamefont
  {Haldane}(1994)}]{PhysRevB.50.17199}%
  \BibitemOpen
  \bibfield  {author} {\bibinfo {author} {\bibfnamefont {E.~H.}\ \bibnamefont
  {Rezayi}}\ and\ \bibinfo {author} {\bibfnamefont {F.~D.~M.}\ \bibnamefont
  {Haldane}},\ }\bibfield  {title} {\enquote {\bibinfo {title} {Laughlin state
  on stretched and squeezed cylinders and edge excitations in the quantum hall
  effect},}\ }\href {\doibase 10.1103/PhysRevB.50.17199} {\bibfield  {journal}
  {\bibinfo  {journal} {Phys. Rev. B}\ }\textbf {\bibinfo {volume} {50}},\
  \bibinfo {pages} {17199--17207} (\bibinfo {year} {1994})}\BibitemShut
  {NoStop}%
\bibitem [{\citenamefont {{Bergholtz}}\ and\ \citenamefont
  {{Karlhede}}(2008)}]{Bergholtz08}%
  \BibitemOpen
  \bibfield  {author} {\bibinfo {author} {\bibfnamefont {E.~J.}\ \bibnamefont
  {{Bergholtz}}}\ and\ \bibinfo {author} {\bibfnamefont {A.}~\bibnamefont
  {{Karlhede}}},\ }\bibfield  {title} {\enquote {\bibinfo {title} {{Quantum
  Hall system in Tao-Thouless limit}},}\ }\href {\doibase
  10.1103/PhysRevB.77.155308} {\bibfield  {journal} {\bibinfo  {journal}
  {\prb}\ }\textbf {\bibinfo {volume} {77}},\ \bibinfo {eid} {155308} (\bibinfo
  {year} {2008})},\ \Eprint {http://arxiv.org/abs/0712.1927} {arXiv:0712.1927}
  \BibitemShut {NoStop}%
\bibitem [{\citenamefont {Bergholtz}\ \emph {et~al.}(2011)\citenamefont
  {Bergholtz}, \citenamefont {Nakamura},\ and\ \citenamefont
  {Suorsa}}]{BERGHOLTZ2011755}%
  \BibitemOpen
  \bibfield  {author} {\bibinfo {author} {\bibfnamefont {Emil~J.}\ \bibnamefont
  {Bergholtz}}, \bibinfo {author} {\bibfnamefont {Masaaki}\ \bibnamefont
  {Nakamura}}, \ and\ \bibinfo {author} {\bibfnamefont {Juha}\ \bibnamefont
  {Suorsa}},\ }\bibfield  {title} {\enquote {\bibinfo {title} {Effective spin
  chains for fractional quantum hall states},}\ }\href {\doibase
  https://doi.org/10.1016/j.physe.2010.07.044} {\bibfield  {journal} {\bibinfo
  {journal} {Physica E: Low-dimensional Systems and Nanostructures}\ }\textbf
  {\bibinfo {volume} {43}},\ \bibinfo {pages} {755 -- 760} (\bibinfo {year}
  {2011})},\ \bibinfo {note} {nanoPHYS 09}\BibitemShut {NoStop}%
\bibitem [{\citenamefont {Wang}\ \emph {et~al.}(2012)\citenamefont {Wang},
  \citenamefont {Takayoshi},\ and\ \citenamefont
  {Nakamura}}]{PhysRevB.86.155104}%
  \BibitemOpen
  \bibfield  {author} {\bibinfo {author} {\bibfnamefont {Zheng-Yuan}\
  \bibnamefont {Wang}}, \bibinfo {author} {\bibfnamefont {Shintaro}\
  \bibnamefont {Takayoshi}}, \ and\ \bibinfo {author} {\bibfnamefont {Masaaki}\
  \bibnamefont {Nakamura}},\ }\bibfield  {title} {\enquote {\bibinfo {title}
  {Spin-chain description of fractional quantum hall states in the jain
  series},}\ }\href {\doibase 10.1103/PhysRevB.86.155104} {\bibfield  {journal}
  {\bibinfo  {journal} {Phys. Rev. B}\ }\textbf {\bibinfo {volume} {86}},\
  \bibinfo {pages} {155104} (\bibinfo {year} {2012})}\BibitemShut {NoStop}%
\bibitem [{\citenamefont {Nakamura}\ \emph {et~al.}(2012)\citenamefont
  {Nakamura}, \citenamefont {Wang},\ and\ \citenamefont
  {Bergholtz}}]{PhysRevLett.109.016401}%
  \BibitemOpen
  \bibfield  {author} {\bibinfo {author} {\bibfnamefont {Masaaki}\ \bibnamefont
  {Nakamura}}, \bibinfo {author} {\bibfnamefont {Zheng-Yuan}\ \bibnamefont
  {Wang}}, \ and\ \bibinfo {author} {\bibfnamefont {Emil~J.}\ \bibnamefont
  {Bergholtz}},\ }\bibfield  {title} {\enquote {\bibinfo {title} {Exactly
  solvable fermion chain describing a $\ensuremath{\nu}=1/3$ fractional quantum
  hall state},}\ }\href {\doibase 10.1103/PhysRevLett.109.016401} {\bibfield
  {journal} {\bibinfo  {journal} {Phys. Rev. Lett.}\ }\textbf {\bibinfo
  {volume} {109}},\ \bibinfo {pages} {016401} (\bibinfo {year}
  {2012})}\BibitemShut {NoStop}%
\bibitem [{\citenamefont {Pai}\ \emph {et~al.}(2019)\citenamefont {Pai},
  \citenamefont {Pretko},\ and\ \citenamefont {Nandkishore}}]{Pai18}%
  \BibitemOpen
  \bibfield  {author} {\bibinfo {author} {\bibfnamefont {Shriya}\ \bibnamefont
  {Pai}}, \bibinfo {author} {\bibfnamefont {Michael}\ \bibnamefont {Pretko}}, \
  and\ \bibinfo {author} {\bibfnamefont {Rahul~M.}\ \bibnamefont
  {Nandkishore}},\ }\bibfield  {title} {\enquote {\bibinfo {title}
  {Localization in fractonic random circuits},}\ }\href {\doibase
  10.1103/PhysRevX.9.021003} {\bibfield  {journal} {\bibinfo  {journal} {Phys.
  Rev. X}\ }\textbf {\bibinfo {volume} {9}},\ \bibinfo {pages} {021003}
  (\bibinfo {year} {2019})}\BibitemShut {NoStop}%
\bibitem [{\citenamefont {Lesanovsky}\ and\ \citenamefont
  {Katsura}(2012)}]{Lesanovsky12}%
  \BibitemOpen
  \bibfield  {author} {\bibinfo {author} {\bibfnamefont {Igor}\ \bibnamefont
  {Lesanovsky}}\ and\ \bibinfo {author} {\bibfnamefont {Hosho}\ \bibnamefont
  {Katsura}},\ }\bibfield  {title} {\enquote {\bibinfo {title} {Interacting
  fibonacci anyons in a rydberg gas},}\ }\href {\doibase
  10.1103/PhysRevA.86.041601} {\bibfield  {journal} {\bibinfo  {journal} {Phys.
  Rev. A}\ }\textbf {\bibinfo {volume} {86}},\ \bibinfo {pages} {041601}
  (\bibinfo {year} {2012})}\BibitemShut {NoStop}%
\bibitem [{Zla()}]{Zlatko}%
  \BibitemOpen
  \href@noop {} {}\bibinfo {note} {We thank Zlatko Papic for bringing this to
  our attention.}\BibitemShut {Stop}%
\bibitem [{Gra()}]{GralPcons}%
  \BibitemOpen
  \href@noop {} {}\bibinfo {note} {A general dipole conserving Hamiltonian is
  invariant under the transformation $S_n^{\pm}\to e^{\pm i\alpha \cdot
  n}S_n^{\pm}$ generated by the unitary $\exp\big(i\alpha \sum_n nS_n^z \big)$.
  Thus, for any spin representation, a local term $(S_{n^+_1}^+)^{d_1^+}\cdots
  (S_{n^+_k}^+)^{d_k^+}(S_{n^-_1}^-)^{d_1^-}\cdots (S_{n^-_k}^-)^{d_k^-} $ is
  dipole conserving if the following condition holds: \\$\sum_{n_i^+}
  d_i^+n_i^+-\sum_{n_i^-} d_i^-n_i^-=0 $.}\BibitemShut {Stop}%
\bibitem [{PBC()}]{PBC_Note}%
  \BibitemOpen
  \href@noop {} {}\bibinfo {note} {For periodic boundary conditions, the
  discussion would be similar to that of the position operator on a
  ring~\cite{Resta98}, and one would have to define the dipole moment through
  the unitary operator $\exp\big(i\frac{2\pi}{N}P \big)$. However, the choice
  of boundary conditions cannot effect the dynamics at finite times in the
  thermodynamic limit.}\BibitemShut {Stop}%
\bibitem [{\citenamefont {Reimann}(2007)}]{Reimann07}%
  \BibitemOpen
  \bibfield  {author} {\bibinfo {author} {\bibfnamefont {Peter}\ \bibnamefont
  {Reimann}},\ }\bibfield  {title} {\enquote {\bibinfo {title} {Typicality for
  generalized microcanonical ensembles},}\ }\href {\doibase
  10.1103/PhysRevLett.99.160404} {\bibfield  {journal} {\bibinfo  {journal}
  {Phys. Rev. Lett.}\ }\textbf {\bibinfo {volume} {99}},\ \bibinfo {pages}
  {160404} (\bibinfo {year} {2007})}\BibitemShut {NoStop}%
\bibitem [{\citenamefont {Jochen~Gemmer}\ and\ \citenamefont
  {Mahler}(2009)}]{QuantumThermo}%
  \BibitemOpen
  \bibfield  {author} {\bibinfo {author} {\bibfnamefont {M.~Michel}\
  \bibnamefont {Jochen~Gemmer}}\ and\ \bibinfo {author} {\bibfnamefont
  {Günter}\ \bibnamefont {Mahler}},\ }\href {\doibase
  10.1017/CBO9780511813467} {\emph {\bibinfo {title} {Quantum Thermodynamics:
  Emergence of Thermodynamic Behavior Within Composite Quantum Systems}}}\
  (\bibinfo  {publisher} {Springer, Berlin, Heidelberg},\ \bibinfo {year}
  {2009})\BibitemShut {NoStop}%
\bibitem [{\citenamefont {Steinigeweg}\ \emph {et~al.}(2015)\citenamefont
  {Steinigeweg}, \citenamefont {Gemmer},\ and\ \citenamefont
  {Brenig}}]{Steinigeweg15}%
  \BibitemOpen
  \bibfield  {author} {\bibinfo {author} {\bibfnamefont {Robin}\ \bibnamefont
  {Steinigeweg}}, \bibinfo {author} {\bibfnamefont {Jochen}\ \bibnamefont
  {Gemmer}}, \ and\ \bibinfo {author} {\bibfnamefont {Wolfram}\ \bibnamefont
  {Brenig}},\ }\bibfield  {title} {\enquote {\bibinfo {title} {Spin and energy
  currents in integrable and nonintegrable spin-$\frac{1}{2}$ chains: A
  typicality approach to real-time autocorrelations},}\ }\href {\doibase
  10.1103/PhysRevB.91.104404} {\bibfield  {journal} {\bibinfo  {journal} {Phys.
  Rev. B}\ }\textbf {\bibinfo {volume} {91}},\ \bibinfo {pages} {104404}
  (\bibinfo {year} {2015})}\BibitemShut {NoStop}%
\bibitem [{Czt()}]{Czth_Note}%
  \BibitemOpen
  \href@noop {} {}\bibinfo {note} {Since $C_0^z(0)=2/3 $,
  $\sum_n\avg{S_n^z(t)S_{0}^z(0)}_{T=\infty}$ is conserved and the system is
  thermalizing and translational invariant, the initial value $2/3$ will be
  homogeneously distributed over the chain.}\BibitemShut {Stop}%
\bibitem [{\citenamefont {Saad}(1992)}]{Saad92}%
  \BibitemOpen
  \bibfield  {author} {\bibinfo {author} {\bibfnamefont {Y.}~\bibnamefont
  {Saad}},\ }\bibfield  {title} {\enquote {\bibinfo {title} {Analysis of some
  krylov subspace approximations to the matrix exponential operator},}\
  }\href@noop {} {\bibfield  {journal} {\bibinfo  {journal} {SIAM Journal on
  Numerical Analysis}\ }\textbf {\bibinfo {volume} {29}},\ \bibinfo {pages}
  {209} (\bibinfo {year} {1992})}\BibitemShut {NoStop}%
\bibitem [{\citenamefont {Pauling}(1931)}]{Pauling}%
  \BibitemOpen
  \bibfield  {author} {\bibinfo {author} {\bibfnamefont {Linus.}\ \bibnamefont
  {Pauling}},\ }\bibfield  {title} {\enquote {\bibinfo {title} {The nature of
  the chemical bond. application of results obtained from the quantum mechanics
  and from a theory of paramagnetic susceptibility to the structure of
  molecules},}\ }\href {\doibase 10.1021/ja01355a027} {\bibfield  {journal}
  {\bibinfo  {journal} {Journal of the American Chemical Society}\ }\textbf
  {\bibinfo {volume} {53}},\ \bibinfo {pages} {1367--1400} (\bibinfo {year}
  {1931})},\ \Eprint {http://arxiv.org/abs/https://doi.org/10.1021/ja01355a027}
  {https://doi.org/10.1021/ja01355a027} \BibitemShut {NoStop}%
\bibitem [{\citenamefont {Onsager}\ and\ \citenamefont
  {Dupuis}(1959)}]{Dupuis59}%
  \BibitemOpen
  \bibfield  {author} {\bibinfo {author} {\bibfnamefont {L.}~\bibnamefont
  {Onsager}}\ and\ \bibinfo {author} {\bibfnamefont {M.}~\bibnamefont
  {Dupuis}},\ }\bibfield  {title} {\enquote {\bibinfo {title} {The electric
  properties of ice(*)},}\ }\href@noop {} {\bibfield  {journal} {\bibinfo
  {journal} {in Rend. Sc. Int. Fis. ``Enrico Fermi", Corso X, Varenna}\ ,\
  \bibinfo {pages} {pp. 294–315}} (\bibinfo {year} {1959})}\BibitemShut
  {NoStop}%
\bibitem [{Dif()}]{Diffusion_Note}%
  \BibitemOpen
  \href@noop {} {}\bibinfo {note} {Although $H_3$ acts as free hopping on a
  \emph{single} dipole, at finite dipole density---relevant for calculating the
  infinite temperature quantities we consider---transport of dipoles can still
  be diffusive, as is the case of usual charge/energy transport at high
  temperatures~\cite{Kadanoff1963,Rosch13,Bohrdt16}.}\BibitemShut {Stop}%
\bibitem [{\citenamefont {{Moudgalya}}\ \emph {et~al.}(2019)\citenamefont
  {{Moudgalya}}, \citenamefont {{Bernevig}},\ and\ \citenamefont
  {{Regnault}}}]{Sanjay19}%
  \BibitemOpen
  \bibfield  {author} {\bibinfo {author} {\bibfnamefont {Sanjay}\ \bibnamefont
  {{Moudgalya}}}, \bibinfo {author} {\bibfnamefont {B.~Andrei}\ \bibnamefont
  {{Bernevig}}}, \ and\ \bibinfo {author} {\bibfnamefont {Nicolas}\
  \bibnamefont {{Regnault}}},\ }\bibfield  {title} {\enquote {\bibinfo {title}
  {{Quantum Many-body Scars in a Landau Level on a Thin Torus}},}\ }\href@noop
  {} {\bibfield  {journal} {\bibinfo  {journal} {arXiv e-prints}\ ,\ \bibinfo
  {eid} {arXiv:1906.05292}} (\bibinfo {year} {2019})},\ \Eprint
  {http://arxiv.org/abs/1906.05292} {arXiv:1906.05292 [cond-mat.str-el]}
  \BibitemShut {NoStop}%
\bibitem [{\citenamefont {Mazur}(1969)}]{Mazur69}%
  \BibitemOpen
  \bibfield  {author} {\bibinfo {author} {\bibfnamefont {P.}~\bibnamefont
  {Mazur}},\ }\bibfield  {title} {\enquote {\bibinfo {title} {Non-ergodicity of
  phase functions in certain systems},}\ }\href {\doibase
  https://doi.org/10.1016/0031-8914(69)90185-2} {\bibfield  {journal} {\bibinfo
   {journal} {Physica}\ }\textbf {\bibinfo {volume} {43}},\ \bibinfo {pages}
  {533 -- 545} (\bibinfo {year} {1969})}\BibitemShut {NoStop}%
\bibitem [{\citenamefont {Suzuki}(1971)}]{Suzuki71}%
  \BibitemOpen
  \bibfield  {author} {\bibinfo {author} {\bibfnamefont {M.}~\bibnamefont
  {Suzuki}},\ }\bibfield  {title} {\enquote {\bibinfo {title} {Ergodicity,
  constants of motion, and bounds for susceptibilities},}\ }\href {\doibase
  https://doi.org/10.1016/0031-8914(71)90226-6} {\bibfield  {journal} {\bibinfo
   {journal} {Physica}\ }\textbf {\bibinfo {volume} {51}},\ \bibinfo {pages}
  {277 -- 291} (\bibinfo {year} {1971})}\BibitemShut {NoStop}%
\bibitem [{\citenamefont {{Caux}}\ and\ \citenamefont
  {{Mossel}}(2011)}]{Caux10}%
  \BibitemOpen
  \bibfield  {author} {\bibinfo {author} {\bibfnamefont {Jean-S{\'e}bastien}\
  \bibnamefont {{Caux}}}\ and\ \bibinfo {author} {\bibfnamefont {Jorn}\
  \bibnamefont {{Mossel}}},\ }\bibfield  {title} {\enquote {\bibinfo {title}
  {{Remarks on the notion of quantum integrability}},}\ }\href {\doibase
  10.1088/1742-5468/2011/02/P02023} {\bibfield  {journal} {\bibinfo  {journal}
  {Journal of Statistical Mechanics: Theory and Experiment}\ }\textbf {\bibinfo
  {volume} {2011}},\ \bibinfo {pages} {02023} (\bibinfo {year} {2011})},\
  \Eprint {http://arxiv.org/abs/1012.3587} {arXiv:1012.3587 [cond-mat.str-el]}
  \BibitemShut {NoStop}%
\bibitem [{H5()}]{H5}%
  \BibitemOpen
  \href@noop {} {}\bibinfo {note} {The Hamiltonian $H_5$ appearing in Fig. 4 is
  given by $H_5=\sum_n\big[ S_{n-2}^+S_{n-1}^-S_{n+1}^-S_{n+2}^+ + H.c.
  \big]$.}\BibitemShut {Stop}%
\bibitem [{Thi()}]{ThirdRef}%
  \BibitemOpen
  \href@noop {} {}\bibinfo {note} {We thank the anonymous referee for pointing
  out this observation to us.}\BibitemShut {Stop}%
\bibitem [{Dim()}]{Dimratio_Note}%
  \BibitemOpen
  \href@noop {} {}\bibinfo {note} {Note that the ratio of the estimated number
  of frozen states, $(2S+3)^{N/\ell}$, and the total Hilbert space dimension
  $(2S+1)^N$, decreases with $S$ for $\ell>1$.}\BibitemShut {Stop}%
\bibitem [{\citenamefont {Santos}\ and\ \citenamefont
  {Rigol}(2010)}]{Santos2010}%
  \BibitemOpen
  \bibfield  {author} {\bibinfo {author} {\bibfnamefont {Lea~F.}\ \bibnamefont
  {Santos}}\ and\ \bibinfo {author} {\bibfnamefont {Marcos}\ \bibnamefont
  {Rigol}},\ }\bibfield  {title} {\enquote {\bibinfo {title} {Onset of quantum
  chaos in one-dimensional bosonic and fermionic systems and its relation to
  thermalization},}\ }\href {\doibase 10.1103/PhysRevE.81.036206} {\bibfield
  {journal} {\bibinfo  {journal} {Phys. Rev. E}\ }\textbf {\bibinfo {volume}
  {81}},\ \bibinfo {pages} {036206} (\bibinfo {year} {2010})}\BibitemShut
  {NoStop}%
\bibitem [{\citenamefont {Sorg}\ \emph {et~al.}(2014)\citenamefont {Sorg},
  \citenamefont {Vidmar}, \citenamefont {Pollet},\ and\ \citenamefont
  {Heidrich-Meisner}}]{Sorg2014}%
  \BibitemOpen
  \bibfield  {author} {\bibinfo {author} {\bibfnamefont {S.}~\bibnamefont
  {Sorg}}, \bibinfo {author} {\bibfnamefont {L.}~\bibnamefont {Vidmar}},
  \bibinfo {author} {\bibfnamefont {L.}~\bibnamefont {Pollet}}, \ and\ \bibinfo
  {author} {\bibfnamefont {F.}~\bibnamefont {Heidrich-Meisner}},\ }\bibfield
  {title} {\enquote {\bibinfo {title} {Relaxation and thermalization in the
  one-dimensional bose-hubbard model: A case study for the interaction quantum
  quench from the atomic limit},}\ }\href {\doibase 10.1103/PhysRevA.90.033606}
  {\bibfield  {journal} {\bibinfo  {journal} {Phys. Rev. A}\ }\textbf {\bibinfo
  {volume} {90}},\ \bibinfo {pages} {033606} (\bibinfo {year}
  {2014})}\BibitemShut {NoStop}%
\bibitem [{\citenamefont {Mondaini}\ \emph {et~al.}(2016)\citenamefont
  {Mondaini}, \citenamefont {Fratus}, \citenamefont {Srednicki},\ and\
  \citenamefont {Rigol}}]{Rigol2DIsing}%
  \BibitemOpen
  \bibfield  {author} {\bibinfo {author} {\bibfnamefont {Rubem}\ \bibnamefont
  {Mondaini}}, \bibinfo {author} {\bibfnamefont {Keith~R.}\ \bibnamefont
  {Fratus}}, \bibinfo {author} {\bibfnamefont {Mark}\ \bibnamefont
  {Srednicki}}, \ and\ \bibinfo {author} {\bibfnamefont {Marcos}\ \bibnamefont
  {Rigol}},\ }\bibfield  {title} {\enquote {\bibinfo {title} {Eigenstate
  thermalization in the two-dimensional transverse field ising model},}\ }\href
  {\doibase 10.1103/PhysRevE.93.032104} {\bibfield  {journal} {\bibinfo
  {journal} {Phys. Rev. E}\ }\textbf {\bibinfo {volume} {93}},\ \bibinfo
  {pages} {032104} (\bibinfo {year} {2016})}\BibitemShut {NoStop}%
\bibitem [{ETH()}]{ETHnote}%
  \BibitemOpen
  \href@noop {} {}\bibinfo {note} {In any case, the existence of exponentially
  many subspaces indicates that one would need to fix an extensively large
  number of symmetries to target a specific one.}\BibitemShut {Stop}%
\bibitem [{\citenamefont {Pai}\ and\ \citenamefont {Pretko}(2019)}]{Pai19}%
  \BibitemOpen
  \bibfield  {author} {\bibinfo {author} {\bibfnamefont {Shriya}\ \bibnamefont
  {Pai}}\ and\ \bibinfo {author} {\bibfnamefont {Michael}\ \bibnamefont
  {Pretko}},\ }\href@noop {} {\enquote {\bibinfo {title} {Manifestation of
  quantum many-body scars in fracton systems},}\ } (\bibinfo {year} {2019}),\
  \Eprint {http://arxiv.org/abs/arXiv:1903.06173} {arXiv:1903.06173}
  \BibitemShut {NoStop}%
\bibitem [{\citenamefont {Ritort}\ and\ \citenamefont
  {Sollich}(2003)}]{Ritort03}%
  \BibitemOpen
  \bibfield  {author} {\bibinfo {author} {\bibfnamefont {F.}~\bibnamefont
  {Ritort}}\ and\ \bibinfo {author} {\bibfnamefont {P.}~\bibnamefont
  {Sollich}},\ }\bibfield  {title} {\enquote {\bibinfo {title} {Glassy dynamics
  of kinetically constrained models},}\ }\href {\doibase
  10.1080/0001873031000093582} {\bibfield  {journal} {\bibinfo  {journal}
  {Advances in Physics}\ }\textbf {\bibinfo {volume} {52}},\ \bibinfo {pages}
  {219--342} (\bibinfo {year} {2003})},\ \Eprint
  {http://arxiv.org/abs/https://doi.org/10.1080/0001873031000093582}
  {https://doi.org/10.1080/0001873031000093582} \BibitemShut {NoStop}%
\bibitem [{\citenamefont {Basko}\ \emph
  {et~al.}(2006{\natexlab{b}})\citenamefont {Basko}, \citenamefont {Aleiner},\
  and\ \citenamefont {Altshuler}}]{BASKO20061126}%
  \BibitemOpen
  \bibfield  {author} {\bibinfo {author} {\bibfnamefont {D.M.}\ \bibnamefont
  {Basko}}, \bibinfo {author} {\bibfnamefont {I.L.}\ \bibnamefont {Aleiner}}, \
  and\ \bibinfo {author} {\bibfnamefont {B.L.}\ \bibnamefont {Altshuler}},\
  }\bibfield  {title} {\enquote {\bibinfo {title} {Metal–insulator transition
  in a weakly interacting many-electron system with localized single-particle
  states},}\ }\href {\doibase https://doi.org/10.1016/j.aop.2005.11.014}
  {\bibfield  {journal} {\bibinfo  {journal} {Annals of Physics}\ }\textbf
  {\bibinfo {volume} {321}},\ \bibinfo {pages} {1126 -- 1205} (\bibinfo {year}
  {2006}{\natexlab{b}})}\BibitemShut {NoStop}%
\bibitem [{\citenamefont {Moudgalya}\ \emph
  {et~al.}(2018{\natexlab{c}})\citenamefont {Moudgalya}, \citenamefont
  {Rachel}, \citenamefont {Bernevig},\ and\ \citenamefont
  {Regnault}}]{Sanjay18}%
  \BibitemOpen
  \bibfield  {author} {\bibinfo {author} {\bibfnamefont {Sanjay}\ \bibnamefont
  {Moudgalya}}, \bibinfo {author} {\bibfnamefont {Stephan}\ \bibnamefont
  {Rachel}}, \bibinfo {author} {\bibfnamefont {B.~Andrei}\ \bibnamefont
  {Bernevig}}, \ and\ \bibinfo {author} {\bibfnamefont {Nicolas}\ \bibnamefont
  {Regnault}},\ }\bibfield  {title} {\enquote {\bibinfo {title} {Exact excited
  states of nonintegrable models},}\ }\href {\doibase
  10.1103/PhysRevB.98.235155} {\bibfield  {journal} {\bibinfo  {journal} {Phys.
  Rev. B}\ }\textbf {\bibinfo {volume} {98}},\ \bibinfo {pages} {235155}
  (\bibinfo {year} {2018}{\natexlab{c}})}\BibitemShut {NoStop}%
\bibitem [{\citenamefont {Pretko}(2017{\natexlab{d}})}]{PretkoMach}%
  \BibitemOpen
  \bibfield  {author} {\bibinfo {author} {\bibfnamefont {Michael}\ \bibnamefont
  {Pretko}},\ }\bibfield  {title} {\enquote {\bibinfo {title} {Emergent gravity
  of fractons: Mach's principle revisited},}\ }\href {\doibase
  10.1103/PhysRevD.96.024051} {\bibfield  {journal} {\bibinfo  {journal} {Phys.
  Rev. D}\ }\textbf {\bibinfo {volume} {96}},\ \bibinfo {pages} {024051}
  (\bibinfo {year} {2017}{\natexlab{d}})}\BibitemShut {NoStop}%
\bibitem [{\citenamefont {van Nieuwenburg}\ \emph {et~al.}(2019)\citenamefont
  {van Nieuwenburg}, \citenamefont {Baum},\ and\ \citenamefont
  {Refael}}]{vanNieuwenburg19}%
  \BibitemOpen
  \bibfield  {author} {\bibinfo {author} {\bibfnamefont {Evert}\ \bibnamefont
  {van Nieuwenburg}}, \bibinfo {author} {\bibfnamefont {Yuval}\ \bibnamefont
  {Baum}}, \ and\ \bibinfo {author} {\bibfnamefont {Gil}\ \bibnamefont
  {Refael}},\ }\bibfield  {title} {\enquote {\bibinfo {title} {From bloch
  oscillations to many-body localization in clean interacting systems},}\
  }\href {\doibase 10.1073/pnas.1819316116} {\bibfield  {journal} {\bibinfo
  {journal} {Proceedings of the National Academy of Sciences}\ }\textbf
  {\bibinfo {volume} {116}},\ \bibinfo {pages} {9269--9274} (\bibinfo {year}
  {2019})},\ \Eprint
  {http://arxiv.org/abs/https://www.pnas.org/content/116/19/9269.full.pdf}
  {https://www.pnas.org/content/116/19/9269.full.pdf} \BibitemShut {NoStop}%
\bibitem [{\citenamefont {Kim}\ and\ \citenamefont {Haah}(2016)}]{Kim16}%
  \BibitemOpen
  \bibfield  {author} {\bibinfo {author} {\bibfnamefont {Isaac~H.}\
  \bibnamefont {Kim}}\ and\ \bibinfo {author} {\bibfnamefont {Jeongwan}\
  \bibnamefont {Haah}},\ }\bibfield  {title} {\enquote {\bibinfo {title}
  {Localization from superselection rules in translationally invariant
  systems},}\ }\href {\doibase 10.1103/PhysRevLett.116.027202} {\bibfield
  {journal} {\bibinfo  {journal} {Phys. Rev. Lett.}\ }\textbf {\bibinfo
  {volume} {116}},\ \bibinfo {pages} {027202} (\bibinfo {year}
  {2016})}\BibitemShut {NoStop}%
\bibitem [{\citenamefont {Page}(1993)}]{Page93}%
  \BibitemOpen
  \bibfield  {author} {\bibinfo {author} {\bibfnamefont {Don~N.}\ \bibnamefont
  {Page}},\ }\bibfield  {title} {\enquote {\bibinfo {title} {Average entropy of
  a subsystem},}\ }\href {\doibase 10.1103/PhysRevLett.71.1291} {\bibfield
  {journal} {\bibinfo  {journal} {Phys. Rev. Lett.}\ }\textbf {\bibinfo
  {volume} {71}},\ \bibinfo {pages} {1291--1294} (\bibinfo {year}
  {1993})}\BibitemShut {NoStop}%
\bibitem [{\citenamefont {{Shiraishi}}(2019)}]{Shiraishi19}%
  \BibitemOpen
  \bibfield  {author} {\bibinfo {author} {\bibfnamefont {Naoto}\ \bibnamefont
  {{Shiraishi}}},\ }\bibfield  {title} {\enquote {\bibinfo {title} {{Connection
  between quantum-many-body scars and the AKLT model from the viewpoint of
  embedded Hamiltonians}},}\ }\href@noop {} {\bibfield  {journal} {\bibinfo
  {journal} {arXiv e-prints}\ ,\ \bibinfo {eid} {arXiv:1904.05182}} (\bibinfo
  {year} {2019})},\ \Eprint {http://arxiv.org/abs/1904.05182} {arXiv:1904.05182
  [cond-mat.stat-mech]} \BibitemShut {NoStop}%
\bibitem [{OBC()}]{OBCPXP}%
  \BibitemOpen
  \href@noop {} {}\bibinfo {note} {In the case of open boundary conditions, the
  two additional terms $X_1P_2$ and $X_{L-1}P_L$ were added to
  Eq.~\eqref{eq:PXP} in the analysis of Ref.~\cite{TurnerPRB} that we follow in
  this appendix.}\BibitemShut {Stop}%
\bibitem [{\citenamefont {Roberts}\ \emph {et~al.}(2015)\citenamefont
  {Roberts}, \citenamefont {Stanford},\ and\ \citenamefont
  {Susskind}}]{Roberts2015}%
  \BibitemOpen
  \bibfield  {author} {\bibinfo {author} {\bibfnamefont {Daniel~A.}\
  \bibnamefont {Roberts}}, \bibinfo {author} {\bibfnamefont {Douglas}\
  \bibnamefont {Stanford}}, \ and\ \bibinfo {author} {\bibfnamefont {Leonard}\
  \bibnamefont {Susskind}},\ }\bibfield  {title} {\enquote {\bibinfo {title}
  {Localized shocks},}\ }\href {\doibase 10.1007/JHEP03(2015)051} {\bibfield
  {journal} {\bibinfo  {journal} {Journal of High Energy Physics}\ }\textbf
  {\bibinfo {volume} {2015}},\ \bibinfo {pages} {51} (\bibinfo {year}
  {2015})}\BibitemShut {NoStop}%
\bibitem [{\citenamefont {von Keyserlingk}\ \emph {et~al.}(2018)\citenamefont
  {von Keyserlingk}, \citenamefont {Rakovszky}, \citenamefont {Pollmann},\ and\
  \citenamefont {Sondhi}}]{Keyserlingk2018}%
  \BibitemOpen
  \bibfield  {author} {\bibinfo {author} {\bibfnamefont {C.~W.}\ \bibnamefont
  {von Keyserlingk}}, \bibinfo {author} {\bibfnamefont {Tibor}\ \bibnamefont
  {Rakovszky}}, \bibinfo {author} {\bibfnamefont {Frank}\ \bibnamefont
  {Pollmann}}, \ and\ \bibinfo {author} {\bibfnamefont {S.~L.}\ \bibnamefont
  {Sondhi}},\ }\bibfield  {title} {\enquote {\bibinfo {title} {Operator
  hydrodynamics, otocs, and entanglement growth in systems without conservation
  laws},}\ }\href {\doibase 10.1103/PhysRevX.8.021013} {\bibfield  {journal}
  {\bibinfo  {journal} {Phys. Rev. X}\ }\textbf {\bibinfo {volume} {8}},\
  \bibinfo {pages} {021013} (\bibinfo {year} {2018})}\BibitemShut {NoStop}%
\bibitem [{\citenamefont {Nahum}\ \emph {et~al.}(2018)\citenamefont {Nahum},
  \citenamefont {Vijay},\ and\ \citenamefont {Haah}}]{Nahum2018}%
  \BibitemOpen
  \bibfield  {author} {\bibinfo {author} {\bibfnamefont {Adam}\ \bibnamefont
  {Nahum}}, \bibinfo {author} {\bibfnamefont {Sagar}\ \bibnamefont {Vijay}}, \
  and\ \bibinfo {author} {\bibfnamefont {Jeongwan}\ \bibnamefont {Haah}},\
  }\bibfield  {title} {\enquote {\bibinfo {title} {Operator spreading in random
  unitary circuits},}\ }\href {\doibase 10.1103/PhysRevX.8.021014} {\bibfield
  {journal} {\bibinfo  {journal} {Phys. Rev. X}\ }\textbf {\bibinfo {volume}
  {8}},\ \bibinfo {pages} {021014} (\bibinfo {year} {2018})}\BibitemShut
  {NoStop}%
\bibitem [{\citenamefont {Rakovszky}\ \emph {et~al.}(2018)\citenamefont
  {Rakovszky}, \citenamefont {Pollmann},\ and\ \citenamefont {von
  Keyserlingk}}]{Rakovszky2018}%
  \BibitemOpen
  \bibfield  {author} {\bibinfo {author} {\bibfnamefont {Tibor}\ \bibnamefont
  {Rakovszky}}, \bibinfo {author} {\bibfnamefont {Frank}\ \bibnamefont
  {Pollmann}}, \ and\ \bibinfo {author} {\bibfnamefont {C.~W.}\ \bibnamefont
  {von Keyserlingk}},\ }\bibfield  {title} {\enquote {\bibinfo {title}
  {Diffusive hydrodynamics of out-of-time-ordered correlators with charge
  conservation},}\ }\href {\doibase 10.1103/PhysRevX.8.031058} {\bibfield
  {journal} {\bibinfo  {journal} {Phys. Rev. X}\ }\textbf {\bibinfo {volume}
  {8}},\ \bibinfo {pages} {031058} (\bibinfo {year} {2018})}\BibitemShut
  {NoStop}%
\bibitem [{\citenamefont {Khemani}\ \emph {et~al.}(2018)\citenamefont
  {Khemani}, \citenamefont {Vishwanath},\ and\ \citenamefont
  {Huse}}]{Khemani2018}%
  \BibitemOpen
  \bibfield  {author} {\bibinfo {author} {\bibfnamefont {Vedika}\ \bibnamefont
  {Khemani}}, \bibinfo {author} {\bibfnamefont {Ashvin}\ \bibnamefont
  {Vishwanath}}, \ and\ \bibinfo {author} {\bibfnamefont {David~A.}\
  \bibnamefont {Huse}},\ }\bibfield  {title} {\enquote {\bibinfo {title}
  {Operator spreading and the emergence of dissipative hydrodynamics under
  unitary evolution with conservation laws},}\ }\href {\doibase
  10.1103/PhysRevX.8.031057} {\bibfield  {journal} {\bibinfo  {journal} {Phys.
  Rev. X}\ }\textbf {\bibinfo {volume} {8}},\ \bibinfo {pages} {031057}
  (\bibinfo {year} {2018})}\BibitemShut {NoStop}%
\bibitem [{\citenamefont {Verstraete}\ \emph {et~al.}(2008)\citenamefont
  {Verstraete}, \citenamefont {Murg},\ and\ \citenamefont
  {Cirac}}]{MurgReview}%
  \BibitemOpen
  \bibfield  {author} {\bibinfo {author} {\bibfnamefont {F.}~\bibnamefont
  {Verstraete}}, \bibinfo {author} {\bibfnamefont {V.}~\bibnamefont {Murg}}, \
  and\ \bibinfo {author} {\bibfnamefont {J.I.}\ \bibnamefont {Cirac}},\
  }\bibfield  {title} {\enquote {\bibinfo {title} {Matrix product states,
  projected entangled pair states, and variational renormalization group
  methods for quantum spin systems},}\ }\href {\doibase
  10.1080/14789940801912366} {\bibfield  {journal} {\bibinfo  {journal}
  {Advances in Physics}\ }\textbf {\bibinfo {volume} {57}},\ \bibinfo {pages}
  {143--224} (\bibinfo {year} {2008})},\ \Eprint
  {http://arxiv.org/abs/https://doi.org/10.1080/14789940801912366}
  {https://doi.org/10.1080/14789940801912366} \BibitemShut {NoStop}%
\bibitem [{\citenamefont {Vidal}(2003)}]{VidalTEBD}%
  \BibitemOpen
  \bibfield  {author} {\bibinfo {author} {\bibfnamefont {Guifr\'e}\
  \bibnamefont {Vidal}},\ }\bibfield  {title} {\enquote {\bibinfo {title}
  {Efficient classical simulation of slightly entangled quantum
  computations},}\ }\href {\doibase 10.1103/PhysRevLett.91.147902} {\bibfield
  {journal} {\bibinfo  {journal} {Phys. Rev. Lett.}\ }\textbf {\bibinfo
  {volume} {91}},\ \bibinfo {pages} {147902} (\bibinfo {year}
  {2003})}\BibitemShut {NoStop}%
\bibitem [{\citenamefont {{Resta}}(1998)}]{Resta98}%
  \BibitemOpen
  \bibfield  {author} {\bibinfo {author} {\bibfnamefont {R.}~\bibnamefont
  {{Resta}}},\ }\bibfield  {title} {\enquote {\bibinfo {title}
  {{Quantum-Mechanical Position Operator in Extended Systems}},}\ }\href
  {\doibase 10.1103/PhysRevLett.80.1800} {\bibfield  {journal} {\bibinfo
  {journal} {Physical Review Letters}\ }\textbf {\bibinfo {volume} {80}},\
  \bibinfo {pages} {1800--1803} (\bibinfo {year} {1998})},\ \Eprint
  {http://arxiv.org/abs/cond-mat/9709306} {cond-mat/9709306} \BibitemShut
  {NoStop}%
\bibitem [{\citenamefont {Kadanoff}\ and\ \citenamefont
  {Martin}(1963)}]{Kadanoff1963}%
  \BibitemOpen
  \bibfield  {author} {\bibinfo {author} {\bibfnamefont {Leo~P}\ \bibnamefont
  {Kadanoff}}\ and\ \bibinfo {author} {\bibfnamefont {Paul~C}\ \bibnamefont
  {Martin}},\ }\bibfield  {title} {\enquote {\bibinfo {title} {Hydrodynamic
  equations and correlation functions},}\ }\href {\doibase
  https://doi.org/10.1016/0003-4916(63)90078-2} {\bibfield  {journal} {\bibinfo
   {journal} {Annals of Physics}\ }\textbf {\bibinfo {volume} {24}},\ \bibinfo
  {pages} {419 -- 469} (\bibinfo {year} {1963})}\BibitemShut {NoStop}%
\bibitem [{\citenamefont {Lux}\ \emph {et~al.}(2014)\citenamefont {Lux},
  \citenamefont {M\"uller}, \citenamefont {Mitra},\ and\ \citenamefont
  {Rosch}}]{Rosch13}%
  \BibitemOpen
  \bibfield  {author} {\bibinfo {author} {\bibfnamefont {Jonathan}\
  \bibnamefont {Lux}}, \bibinfo {author} {\bibfnamefont {Jan}\ \bibnamefont
  {M\"uller}}, \bibinfo {author} {\bibfnamefont {Aditi}\ \bibnamefont {Mitra}},
  \ and\ \bibinfo {author} {\bibfnamefont {Achim}\ \bibnamefont {Rosch}},\
  }\bibfield  {title} {\enquote {\bibinfo {title} {Hydrodynamic long-time tails
  after a quantum quench},}\ }\href {\doibase 10.1103/PhysRevA.89.053608}
  {\bibfield  {journal} {\bibinfo  {journal} {Phys. Rev. A}\ }\textbf {\bibinfo
  {volume} {89}},\ \bibinfo {pages} {053608} (\bibinfo {year}
  {2014})}\BibitemShut {NoStop}%
\bibitem [{\citenamefont {Bohrdt}\ \emph {et~al.}(2017)\citenamefont {Bohrdt},
  \citenamefont {Mendl}, \citenamefont {Endres},\ and\ \citenamefont
  {Knap}}]{Bohrdt16}%
  \BibitemOpen
  \bibfield  {author} {\bibinfo {author} {\bibfnamefont {A}~\bibnamefont
  {Bohrdt}}, \bibinfo {author} {\bibfnamefont {C~B}\ \bibnamefont {Mendl}},
  \bibinfo {author} {\bibfnamefont {M}~\bibnamefont {Endres}}, \ and\ \bibinfo
  {author} {\bibfnamefont {M}~\bibnamefont {Knap}},\ }\bibfield  {title}
  {\enquote {\bibinfo {title} {Scrambling and thermalization in a diffusive
  quantum many-body system},}\ }\href
  {http://stacks.iop.org/1367-2630/19/i=6/a=063001} {\bibfield  {journal}
  {\bibinfo  {journal} {New Journal of Physics}\ }\textbf {\bibinfo {volume}
  {19}},\ \bibinfo {pages} {063001} (\bibinfo {year} {2017})}\BibitemShut
  {NoStop}%
\end{thebibliography}%

\end{document}